\documentclass[11pt]{article}
\usepackage{graphicx,deleq}

\newcommand{\bfit}[1]{\textit{\textbf{#1}\/}}
\newcommand{\bfsl}[1]{\textsl{\textbf{#1}\/}}

\begin{document}

\vspace*{3 cm}

\centerline{\LARGE\bf Multimode gravitational wave detection:}

\centerline{\LARGE\bf the spherical detector theory}

\vspace{2 cm}

\begin{center}
{\large\bf Jos\'e Alberto Lobo} \\[0.5 em]
Departament de F\'\i sica Fonamental \\
Universitat de Barcelona, Spain \\
e-mail:\ {\tt lobo@hermes.ffn.ub.es}
\vspace{2.1 em}
\end{center}

\begin{abstract}
Gravitational waves (GW) are propagated perturbations of the space-time
geometry, and they show up locally as {\em tides\/}, i.e., local gravity
gradients which vary with time. These gradients are determined by {\em six\/}
often called {\em electric\/} components of the Riemann tensor, which in
a local coordinate frame are $R_{0i0j}$, ($i,j=1,2,3$). A GW detector is
a device designed to generate information on those quantities on the basis
of suitable measurements. Both Weber bars and laser interferometers are
{\em single mode\/} antennas, i.e., they generate a {\em single\/} readout
which is the result of the combined action of those six GW's Riemann
tensor amplitudes on the system sensor's output port.

A {\em spherical\/} detector is {\em not\/} limited in that fashion: this
is because its {\em degenerate\/} oscillation eigenmodes are uniquely matched
to the structure of the above Riemann tensor components. This means that
a solid spherical body is a natural {\em multimode\/} GW detector, i.e., it
is fully capable of delivering six channel outputs, precise combinations
of which completely deconvolve the six GW amplitudes.

The present article is concerned with the theoretical reasons of the
remarkable properties of a spherical GW detector. The analysis proceeds
from {\em first principles\/} and is based on essentially no {\em ad hoc\/}
hypotheses. The mathematical beauty of the theory is outstanding, and
abundant detail is given for a thorough understanding of the fundamental
facts and ideas. Experimental evidence of the accuracy of the model is
also provided, where possible.
\end{abstract}

\newpage


\section*{Summary}

The following pages contain the fusion of two published papers:
{\em What can we learn about gravitational wave physics with an elastic
spherical antenna?\/}, published in the {\sl Physical Review\/} (PRD
{\bf 52}, 591-604 (1995)), and {\em  Multiple mode gravitational wave
detection with a spherical antenna\/}, scheduled to appear in the July-2000
issue of {\sl Monthly Notices of the Royal Astronomical Society\/} (MNRAS
{\bf 316}, 173-194 (2000))\footnote{
The research contained in this second paper was actually complete by early
1996, and its main results first presented in a Winter School at Warsaw
(Poland) in march 1996 \protect\cite{mini}, then in a plenary lecture
at the {\em Second Edoardo Amaldi Conference on Gravitational Waves\/}
at {\sl CERN\/} in July~1997~\protect\cite{lobo2}. Essentially in its
present form ---see acknowledgements in page~\pageref{ack2} below---,
the paper was submitted to PRD in 1997. It was rejected {\em after\/}
the Editor, {\sf Dennis Nordstrom}, had already accepted it, on the basis
of a positive report. The reason for this opinion switch was a later,
extremely vague yet demolishingly negative and utterly arrogant judgement
by a second referee, to whom Nordstrom grants unlimited credit.

An {\em unchanged\/} version of the manuscript was, again, positively
reported on for MNRAS, and so the article will at long last see the light,
three years too late.}.
In them, I first developed an analytic model to describe the interaction
of a solid elastic sphere with an incoming flux of gravitational waves of
the general {\em metric\/} class, then extended it to also address the
multimode signal deconvolution problem by means of suitable layouts of
{\em resonant\/} motion sensors.

I think the model is quite complete as regards the major {\em conceptual\/}
issues in a spherical~GW detector, therefore think merging the two papers
together makes positive sense. Let me however stress that I do not mean
to understate the fundamental problem of {\em noise\/}, which is here left
aside; rather, it is now being actively investigated in detail within the
very convenient general framework set up below.

{\em What can we learn about GW Physics\,\ldots} is already five years old,
and certain parts of it have been the subject of further research since.
More specifically, this happens with Brans-Dicke absorption cross sections,
see section~4, where reference has been added to such newer work. In
addition,  two new tables and more mathematical detail have been added to
its Appendix B. The latter should help the interested reader with certain
technicalities,  while the tables and attendant new formulas are meant to
provide explicit numerical values of the frequency spectrum of the sphere
rather than the original {\em only graphics\/} information. I expect this
to be useful reference in numerical and/or experimental determinations of
these quantities.

No changes have been included in the second paper relative to its MNRAS
version. In merging the two articles together, however, I have found
expedient to renumber the equations and bibliographic references into
corresponding single streams. The bibliography has been updated and
made into a unique list at the end of the file in alphabetical order.
Finally, I have included minor notation changes in order to reconcile
otherwise small mistunings between both articles.

\vspace{2 cm}

\noindent\hspace*{11 cm} Alberto Lobo \\
\noindent\hspace*{11 cm} Barcelona, June 2000

\newpage


\begin{flushleft}
{\LARGE\sf What can we learn about GW Physics with an elastic
spherical antenna?} \\[0.6 em]
PRD {\bf 52}, 591-604 (1995) \\[1 em]
{\large\bf Jos\'e Alberto Lobo} \\[0.5 em]
Departament de F\'\i sica Fonamental \\
Universitat de Barcelona, Spain \\
e-mail:\ {\tt lobo@hermes.ffn.ub.es}
\end{flushleft}
\vspace{2.1 em}

\begin{abstract}

A general formalism is set up to analyse the response of an
{\it arbitrary\/} solid elastic body to an {\it arbitrary metric\/}
Gravitational Wave perturbation, which fully displays the details of
the interaction wave-antenna. The formalism is applied to the spherical
detector, whose sensitivity parameters are thereby scrutinised. A
{\it multimode\/} transfer function is defined to study the amplitude
sensitivity, and absorption cross sections are calculated for a general
metric theory of GW physics. Their {\it scaling\/} properties are shown
to be {\it independent\/} of the underlying theory, with interesting
consequences for future detector design. The GW incidence direction
deconvolution problem is also discussed, always within the context
of a general metric theory of the gravitational field.

\end{abstract}

\section{Introduction}

Spherical antennae are considered by many to be the natural next step
in the development of resonant GW detectors. The reasons for this new
trend essentially derive from the {\it improved sensitivity\/} of a sphere
---which can be nearly an order of magnitude better than a cylinder
having the same resonance frequency, see below and~\cite{clo}---, and
from its {\it multimode capabilities\/}, first recognised by
Forward~\cite{fo71} and further elaborated in~\cite{wp77,jm93}.

Although some of the most relevant aspects of detector sensitivity
have already received attention in the literature, it seems to me
that a sufficiently general and flexible analysis of the interaction
between GW and detector has not been satisfactorily developed to date.
This {\it theoretical\/} shortage has a number of {\it practical\/}
negative consequences, too. Traditional analysis, to mention but an
example, is almost invariably restricted to General Relativity or
scalar--tensor theories of gravity; while it may be argued that this
is already very general, any such argument is, as a matter of fact,
understating the potentialities actually offered by a spherical GW
antenna to help decide for or against any one specific theory of the
gravitational field on the basis of {\it experimental observation\/}.

I thus propose to develop in this paper a full fledged mathematical
formalism which will enable analysis of the antenna's response to a
completely general GW, i.e., making no {\it a priori\/} assumptions
about which is the correct theory underlying GW physics (other than,
indeed, that it is a {\it metric\/} theory), {\it and\/} also making
no assumptions about detector shape, structure or boundary conditions.
Considering things in such generality is not only ``theoretically nice''
---it also brings about new results and a {\it better understanding\/}
of older ones. For example, it will be proved that the sphere is the
{\it most efficient\/} GW elastic detector shape, and that higher mode
absorption cross sections {\it scale independently of GW physics\/}.
I will also discuss the direction of incidence deconvolution problem
in the context of a general metric theory of gravity.

The paper is organised as follows: section~2 is devoted to the development
of the general mathematical framework, leading to a formula in which an
elastic solid's response is related to the action of an arbitrary metric
GW impinging on it. In section~3 the general equations are applied to the
homogeneous spherical body, and a discussion of the {\it deconvolution\/}
problem is presented as well. Section~4 contains the description of the
sphere's sensitivity parameters, specifically leading to the concept of
{\it multimode}, or vector, {\it transfer function}, and to an analysis
of the {\it absorption cross section\/} presented by this detector to a
passing by GW. Conclusions and prospects are summarised in section~5,
and two appendices are added which include mathematical derivations.

\section{General mathematical framework}

In the mathematical model, I shall be assuming that the antenna is a solid
elastic body which responds to GW perturbations according to the equations
of classical non-relativistic linear Elasticity Theory~\cite{ll70}. This
is fully justified because GW-induced displacements will be very small
indeed, and the speed of such displacements much smaller than that of
light for any foreseeable frequencies. Although our primary interest is
a spherical antenna, the considerations which follow in the remainder of
this section {\em have general validity for arbitrarily shaped, isotropic
elastic solids\/}.

Let ${\bf u}({\bf x},t)$ be the displacement vector of the infinitesimal
mass element sitting at point {\bf x} relative to the solid's centre of
mass in its unperturbed state, whose density distribution in that state
is $\varrho$. Let $\lambda$ and $\mu$ be the material's elastic Lam\'e
coefficients. If a volume force density ${\bf f}({\bf x},t)$ acts on such
solid, the displacement field ${\bf u}({\bf x},t)$ is the solution to the
system of partial differential equations~\cite{ll70}

\begin{equation}
 \varrho \frac{\partial^2 {\bf u}}{\partial t^2} - \mu\nabla^2 {\bf u} -
   (\lambda+\mu)\,\nabla(\nabla{\bf\cdot}{\bf u}) = {\bf f}({\bf x},t)
   \label{2.1}
\end{equation}
with the appropriate initial and boundary conditions. A summary of
notation and general results regarding the solution to that system
is briefly outlined in the ensuing subsection, as they are necessary
for the subsequent developments in this paper, and also in future work
---e.g. page~\pageref{sec:intro} and ss.\ below.

\subsection{Separable driving force}

For reasons which will become clear later on, we shall only be interested
in driving forces of the separable type

\begin{equation}
   {\bf f}({\bf x},t) = {\bf f}({\bf x})\,g(t)    \label{2.2}
\end{equation}
or, indeed, linear combinations thereof. The solution to~(\ref{2.1})
does not require us to specify the precise boundary conditions on
${\bf u}({\bf x},t)$ at this stage, but we need to set the initial
conditions. We adopt the following:

\begin{equation}
   {\bf u}({\bf x},0) = {\bf \dot{u}}({\bf x},0) = 0
\end{equation}
where \ ${\bf\dot{}}\equiv\partial/\partial t$, implying that the
antenna is at complete rest before observation begins at $t$=0. The
structure of the force field~(\ref{2.2}) is such that the displacements
${\bf u}({\bf x},t)$ can be expressed by means of a {\it Green function\/}
integral of the form

\begin{equation}
   {\bf u}({\bf x},t) = \int_0^\infty{\bf S}({\bf x};t-t')\,g(t')\,dt'
   \label{2.6}
\end{equation}

The deductive procedure whereby ${\bf S}({\bf x};t-t')$ is calculated
can be found in many standard textbooks ---see e.g.~\cite{tm87}. The
result is

\begin{equation}
  {\bf S}({\bf x};t) = \left\{\begin{array}{ll}
    0 & \ \ {\rm if}\ \ t\leq 0 \\[1 em]
    \sum_N\,\mbox{\large $\frac{f_N}{\omega_N}$}\,
    {\bf u}_N({\bf x})\sin\omega_Nt
    & \ \ {\rm if}\ \ t\geq 0 \end{array}\right.   \label{2.15}
\end{equation}
where

\begin{equation}
  f_N\equiv\frac{1}{M}\,\int_{\rm Solid}{\bf u}_{N}^*({\bf x})\cdot
  {\bf f}({\bf x})\,d^3x      \label{2.14}
\end{equation}
and ${\bf u}_N({\bf x})$ are the normalised {\it eigen--solutions\/} to

\begin{equation}
   \mu\nabla^2{\bf u}_N + (\lambda+\mu)\,\nabla(\nabla{\bf \cdot}{\bf u}_N)
      = - \omega_N^2\varrho{\bf u}_N        \label{2.9}
\end{equation}
with suitable {\it boundary conditions\/}. Here $N\/$ represents an index,
or set of indices, labelling the {\it eigenmode\/} of frequency~$\omega_N$.
The normalisation condition is (arbitrarily) chosen so that

\begin{equation}
  \int_{\rm Solid}{\bf u}_{N'}^*({\bf x})\cdot{\bf u}_{N}({\bf x})
  \,\varrho({\bf x})\,d^3x = M\,\delta_{N'N}     
  \label{2.10}
\end{equation}
where $M\/$ is the total mass of the solid, and the asterisk denotes
complex conjugation. Replacing now~(\ref{2.15}) into~(\ref{2.6}) we
can write the solution to our problem as a series expansion:

\begin{equation}
   {\bf u}({\bf x},t) =
   \sum_N\,\frac{f_N}{\omega_N}\,{\bf u}_N({\bf x})\,g_N (t)
   \label{2.16}
\end{equation}
where

\begin{equation}
   g_N(t)\equiv\int_0^t g(t')\,\sin\omega_N (t-t')\,dt'   \label{2.17}
\end{equation}

Equation~(\ref{2.16}) is the {\it formal\/} solution to our problem; it
has the standard form of an orthogonal expansion and is valid for
{\it any\/} solid driven by a separable force like~(\ref{2.2}) and
{\it any\/} boundary conditions. It is therefore {\it completely
general\/}, given that type of force.

Before we go on, it is perhaps interesting to quote a simple but useful
example. It is the case of a solid hit by a {\it hammer blow\/}, i.e.,
receiving a sudden stroke at a point on its surface. Exam of the response
of a GW antenna to such perturbation is being used for correct tuning and
monitoring of the device~\cite{wjpc}. If the driving force density is
represented by the simple model 

\begin{equation}
   {\bf f}^{({\rm hb})}({\bf x},t) = {\bf f}_0\,
   \delta^{(3)}({\bf x}-{\bf x}_0)\,\delta(t)   \label{2.18}
\end{equation}

where ${\bf x}_0$ is the surface point hit, and ${\bf f}_0$ is a constant
vector, then the system's response is immediately seen to be

\begin{equation}
   {\bf u}^{({\rm hb})}({\bf x},t) =
   \sum_N\,\frac{f_N^0}{\omega_N}\,{\bf u}_N({\bf x})\sin\omega_N t
   \label{2.19}
\end{equation}
with $f_N^0=M^{-1}\,{\bf f}_0\!\cdot\!{\bf u}_N^*({\bf x}_0)$. A hammer
blow thus excites {\it all\/} the solid's normal modes, except those
{\it perpendicular\/} to ${\bf f}_0$, with amplitudes which are
{\it inversely proportional to the mode's frequency\/}. This is seen to
be a rather general result in the theory of sound waves in isotropic
elastic solids.

\subsection{The GW tidal forces}

An incoming GW manifests itself as a {\it tidal\/} force density; in the
long wavelength linear approximation~\cite{mtw} it only depends on the
``electric'' components of the Riemann tensor:

\begin{equation}
   f_i({\bf x},t) = \varrho c^2\,R_{0i0j}(t)\,x_j    \label{3.1}
\end{equation}
where $c\/$ is the speed of light, and sum over the repeated index {\it j\/}
is understood. In~(\ref{3.1}) tidal forces are referred to the antenna's
centre of mass, and thus {\bf x} is a vector originating there. Note that
I have omitted any dependence of $R_{0i0j}$ on spatial coordinates, since
it only needs to be evaluated at the solid's centre. The Riemann tensor
is only required to first order at this stage~\cite{we72}:

\begin{equation}
   R_{0i0j} = \frac{1}{2}\left(h_{ij,00}-h_{0i,0j}-h_{0j,0i}+h_{00,ij}
                         \right)   \label{3.2}
\end{equation}
where $h_{\mu\nu}$ are the perturbations to flat geometry\footnote{
Throughout this paper, {\it Greek\/} indices ($\mu,\nu,\ldots$) will run
through space-time values 0,1,2,3; {\it Latin\/} indices ($i,j,\ldots$)
will run through space values 1,2,3 only.}, always at the centre of mass
of the detector.

The form~(\ref{3.1}) is seen to be a sum of three terms like~(\ref{2.2})
---but this three term ``straightforward'' splitting is not the most
convenient, due to lack of invariance and symmetry. A better choice is
now outlined.

An {\it arbitrary symmetric\/} tensor ${\cal S}_{ij}$ admits the
following decomposition:

\begin{equation}
   {\cal S}_{ij}(t) = {\cal S}^{(00)}(t)\,E_{ij}^{(00)}\ +\ 
        \sum_{m=-2}^2\,{\cal S}^{(2m)}(t)\,E_{ij}^{(2m)}  \label{3.3}
\end{equation}
where $E_{ij}^{(2m)}$ are 5 linearly independent {\it symmetric\/} and
{\it traceless\/} tensors, and $E_{ij}^{00)}$ is a multiple of the
{\it unit\/} \mbox{tensor $\delta_{ij}$.} ${\cal S}^{(lm)}(t)$ are
uniquely defined functions, whose explicit form depends on the particular
representation of the \mbox{$E\/$-matrices} chosen. A convenient one is
the following:

\begin{deqarr}
\arrlabel{3.4}
  & E_{ij}^{(00)} = \left(\mbox{\large $\frac{1}{4\pi}$}
                 \right)^{\frac{1}{2}}\left(\begin{array}{ccc}
       1 & 0 & 0 \\ 0 & 1 & 0 \\ 0 & 0 & 1 \end{array}\right) & 
                 \label{3.4a}      \\*[1.2 em]
  & E_{ij}^{(20)} = \left(\mbox{\large $\frac{5}{16\pi}$}
                 \right)^{\frac{1}{2}}\left(\begin{array}{ccc}
      -1 & 0 & 0 \\ 0 & -1 & 0 \\ 0 & 0 & 2 \end{array}\right)\ ,\ \ 
  E_{ij}^{(2\pm 1)} = \left(\mbox{\large $\frac{15}{32\pi}$}
                 \right)^{\frac{1}{2}}\left(\begin{array}{ccc}
       0 & 0 & \mp 1 \\ 0 & 0 & -i \\ \mp 1 & -i & 0 \end{array}\right) &
		 \label{3.4b}      \\*[1.2 em]
  & E_{ij}^{(2\pm 2)} = \left(\mbox{\large $\frac{15}{32\pi}$}
                 \right)^{\frac{1}{2}}\left(\begin{array}{ccc}
       1 & \pm i & 0 \\ \pm i & -1 & 0 \\ 0 & 0 & 0 \end{array}\right) &
                       \label{3.4c}
\end{deqarr}

The excellence of this representation stems from its ability to display
the {\it spin features\/} of the driving terms in~(\ref{3.1}). Such
features are characterised by the relationships

\begin{equation}
  E_{ij}^{(lm)}\,n_i n_j = Y_{lm}({\bf n})\ ,\qquad
  l=0,2\ ;\ \ \ \ m= -l,\ldots,l  \label{3.6}
\end{equation}
where {\bf n}$\,\equiv\,${\bf x}/$|{\bf x}|$ is the radial unit vector,
and $Y_{lm}({\bf n})$ are spherical harmonics~\cite{Ed60}. Details about
the above $E\/$-matrices are given in Appendix A. In particular, the
orthogonality relations~(\ref{A.9}) can be used to invert~(\ref{3.3}):

\begin{deqarr}
\arrlabel{3.77}
   {\cal S}^{(00)}(t) & = & \frac{4\pi}{3}\,E_{ij}^{(00)}\,{\cal S}_{ij}(t)
                      \label{3.77a}  \\[1 em]
   {\cal S}^{(2m)}(t) & = & \frac{8\pi}{15}\,E_{ij}^{*(2m)}\,{\cal S}_{ij}(t)
                     \ ,\qquad m=-2,\ldots,2    \label{3.77b}
\end{deqarr}
where an asterisk denotes complex conjugation. Note that
${\cal S}^{(00)}(t) = \sqrt{4\pi}{\cal S}(t)/3$, where
${\cal S}(t)\equiv\delta_{ij}\,{\cal S}_{ij}(t)$ is the tensor's trace.

We now take advantage of~(\ref{3.3}) to express the GW tidal
force~(\ref{3.1}) as a sum of split terms like~(\ref{2.2}):

\begin{equation}
   {\bf f}({\bf x},t) = {\bf f}^{(00)}({\bf x})\,g^{(00)}(t)\ +\ 
       \sum_{m=-2}^2\,{\bf f}^{(2m)}({\bf x})\,g^{(2m)}(t)      \label{3.7}
\end{equation}
with

\begin{deqarr}
\arrlabel{3.8}
 f_i^{(00)}({\bf x}) = \varrho\,E_{ij}^{(00)}\,x_j\ \ & , & \ \ \ 
 g^{(00)}(t) = \frac{4\pi}{3}\,E_{ij}^{*(00)}\,R_{0i0j}(t)\,c^2
 \label{3.8a} \\*[1 em]
 f_i^{(2m)}({\bf x}) = \varrho\,E_{ij}^{(2m)}\,x_j\ \ & , & \ \ \ 
 g^{(2m)}(t) = \frac{8\pi}{15}\,E_{ij}^{*(2m)}\,R_{0i0j}(t)\,c^2
     \qquad (m = -2,\ldots,2)  \label{3.8b}
\end{deqarr}


Straightforward application of~(\ref{2.16}) yields the formal solution
of the antenna response to a GW perturbation:

\begin{equation}
   {\bf u}({\bf x},t) = \sum_N\,\omega_N^{-1}{\bf u}_N({\bf x})\,\left[
   f_N^{(00)}\,g_N^{(00)}(t)\ +\ \sum_{m=-2}^2\,f_N^{(2m)}\,g_N^{(2m)}(t)
   \right]         \label{3.9}
\end{equation}
with the notation of~(\ref{2.14}) and~(\ref{2.17}) applied
{\it mutatis mutandi\/} to the terms in~(\ref{3.8}).

Equation~(\ref{3.9}) gives the response of an arbitrary elastic solid
to an incoming weak GW, {\it independently of the underlying gravity
theory\/}, be it General Relativity (GR) or indeed any other {\it metric\/}
theory of the gravitational interaction. It is also valid for
{\it any antenna shape\/} and {\it any boundary conditions\/}, thus giving
the formalism, in particular, the capability of being used to study the
response of a detector which is {\it suspended\/} by means of a mechanical
device in the laboratory site ---a situation of much practical importance.
It is therefore {\it very\/} general.

Equation~(\ref{3.9}) also tells us that that only {\it monopole\/} and
{\it quadrupole detector modes\/} can possibly be excited by a metric GW.
The nice thing about~(\ref{3.9}) is that it {\it fully\/} displays the
monopole-quadrupole {\it structure\/} of the solution to the fundamental
differential equations~(\ref{2.1}).

In a non-symmetric body, all (or {\it nearly\/} all) the modes have
monopole and quadrupole moments, and~(\ref{3.9}) precisely shows how much
each of them contributes to the detector's response. A {\it homogeneous
spherical\/} antenna, which is very symmetric, has a set of vibrational
eigenmodes which are particularly well matched to the form~(\ref{3.9}):
it only possesses {\it one\/} series of monopole modes and {\it one\/}
(five-fold degenerate) series of quadrupole modes ---see next section
and Appendix B for details. The existence of so {\it few\/} modes which
couple to GWs means that {\it all the absorbed incoming radiation energy
will be distributed amongst those few modes only}, thereby making the
sphere the {\it most efficient\/} detector, even from the sensitivity
point of view. The higher energy cross section {\it per unit mass\/}
reported for spheres on the basis of \mbox{GR \cite{clo},} for example,
finds here its qualitative explanation. The generality of~(\ref{3.9}),
on the other hand, means that {\it this excellence of the spherical
detector is there independently of which is the correct GW theory}.

Before going further, let me mention another potentially useful
application of the formalism so far. Cylindrical antennas, for instance,
are usually studied in the {\it thin rod\/} approximation; although this
is generally quite satisfactory, equation~(\ref{3.9}) offers the
possibility of eventually considering corrections to such simplifying
hypothesis by use of more realistic eigenfunctions, such as those given
in~\cite{ric,Ras}. Recent new proposals for stumpy cylinder arrays~\cite{maf}
may well benefit from the above approach, too.

\section{The spherical antenna}

To explore the consequences of~(\ref{3.9}) in a particular case, the mode
amplitudes ${\bf u}_N({\bf x})$ and frequencies $\omega_N\/$ must be
specified. From now on I will focus on a {\it homogeneous sphere\/} whose
surface is free of tractions and/or tensions; the latter happens to be
quite a good approximation, even if the sphere is suspended in the static
gravitational field~\cite{ol}.

The normal modes of the free sphere fall into two families: so called
{\it toroidal\/} ---where the sphere only undergoes twisting which
keep its shape unchanged throughout the volume--- and
{\it spheroidal\/}~\cite{mzno}, where radial as well as tangential
displacements take place. I use the notation

\begin{equation}
  {\bf u}_{nlm}^T({\bf x})\,e^{\pm i\omega_{nl}^T t}\qquad ,\qquad
  {\bf u}_{nlm}^P({\bf x})\,e^{\pm i\omega_{nl}^P t}  \label{3.10}
\end{equation}
for them, respectively; note that the index $N\/$ of the previous section is
a {\it multiple\/} index $\{nlm\}$ for {\it each\/} family; $l\/$ and $m\/$
are the usual {\it multipole\/} indices, and $n\/$ numbers from 1 to
$\infty$ each of the $l\/$-pole modes. The frequencies happen to be
independent of $m$, and so every one mode~(\ref{3.10}) is (2$l\/$+1)-fold
degenerate. Further details about these eigenmodes are given in Appendix B.

In order to see what~(\ref{3.9}) looks like in this case, integrals of
the form~(\ref{2.14}) ought to be evaluated. It is straightforward to
prove that they all vanish for the toroidal modes, the spheroidal modes
contributing the only non-vanishing terms; after some algebra one finds

\begin{equation}
\label{3.11}
  f_{nlm}^{(l'm')}\equiv\frac{1}{M}\,\int_{\rm Sphere}
  {\bf u}_{nlm}^{P\,*}({\bf x})\cdot{\bf f}^{(l'm')}({\bf x})\,d^3x =
  a_{nl}\,\delta_{l'l}\,\delta_{m'm}\ ,\qquad l'=0,2,\ m'=-l',\ldots,l'
\end{equation}
where

\begin{deqarr}
\arrlabel{3.12}
  a_{n0} & = & -\frac{1}{M}\,\int_0^R A_{n0}(r)\,\varrho\,r^3\,dr
  \label{3.12a}    \\*[1 em]
  a_{n2} & = & -\frac{1}{M}\,\int_0^R \left[A_{n2}(r) +
  3\,B_{n2}(r)\right]\,\varrho\,r^3\,dr         \label{3.12b}
\end{deqarr}

The functions $A_{nl}(r)$, $B_{nl}(r)$ are given in Appendix B, and $R\/$
is the sphere's radius. To our reassurance, only the monopole and quadrupole
{\it sphere modes\/} survive, as seen by the presence of the factors
$\delta_{l'l}\/$ in~(\ref{3.11}). The final series is thus a relatively
simple one, even in spite of its generality\footnote{
From now on I will drop the label $P\/$, meaning {\it spheroidal\/}
mode, to ease the notation since {\it toroidal\/} modes no longer appear in
the formulae.}:

\begin{equation}
   {\bf u}({\bf x},t) = \sum_{n=1}^\infty\,\frac{a_{n0}}{\omega_{n0}}
   \,{\bf u}_{n00}({\bf x})\,g_{n0}^{(00)}(t)\ +\ 
   \sum_{n=1}^\infty\,\frac{a_{n2}}{\omega_{n2}}\,\left[\sum_{m=-2}^2
   \,{\bf u}_{n2m}({\bf x})\,g_{n2}^{(2m)}(t)\right]\ ,\qquad (t>0)
   \label{3.16}
\end{equation}
where, it is recalled,

\begin{equation}
 g_{nl}^{(lm)}(t)=\int_0^t g^{(lm)}(t')\,\sin\omega_{nl}(t-t')\,dt'
 \ ,\qquad (l=0,2;\ \ \ \ m=-l,\ldots,l)     \label{3.6b}
\end{equation}

Equation~(\ref{3.16}) constitutes the sphere's response to an arbitrary
tidal GW perturbation, and will be used to analyse the sensitivity of the
spherical detector in the next section. Before doing so, however, a few
comments on the antenna's signal {\it deconvolution capabilities\/}, within
the context of a completely general metric theory of GWs, are in order.

\subsection{The deconvolution problem}

Let us first of all take the Fourier transform of~(\ref{3.16}):

\begin{equation}
  {\bf U}({\bf x},\omega)\equiv\int_{-\infty}^\infty
  {\bf u}({\bf x},t)\,e^{-i\omega t}\,dt            \label{3.17}
\end{equation}

This is seen to be

\begin{eqnarray}
  {\bf U}({\bf x},\omega) & = & \frac{\pi}{i}\,\sum_{n=1}^\infty\,
  \frac{a_{n0}}{\omega_{n0}}\,{\bf u}_{n00}({\bf x})\,G^{(00)}(\omega)
  \left[\delta(\omega\!-\!\omega_{n0}) - \delta(\omega\!+\!
  \omega_{n0})\right]
  \ + \nonumber \\*[1 em]
   & + & \!\frac{\pi}{i}\,\sum_{n=1}^\infty\,\frac{a_{n2}}{\omega_{n2}}
   \left[\sum_{m=-2}^2\,{\bf u}_{n2m}({\bf x})\,G^{(2m)}(\omega)\right]
   \left[\delta(\omega\!-\!\omega_{n2}) - \delta(\omega\!+\!
   \omega_{n2})\right]
   \label{3.18}
\end{eqnarray}
where $G^{(lm)}(\omega)$ are the Fourier transforms of $g^{(lm)}(t)$,
respectively:

\begin{equation}
  G^{(lm)}(\omega)\equiv\int_0^\infty g^{(lm)}(t)\,e^{-i\omega t}\,dt
  \label{3.19}
\end{equation}

The $\delta\/$-function factors are of course idealisations corresponding
to infinitely long integration times and infinitely narrow resonance
line-widths ---but the essentials of the ensuing discussion will not be
affected by those idealisations.

If the measuring system were (ideally) sensitive to all frequencies,
filters could be applied to examine the antenna's oscillations at each
monopole and quadrupole frequency: a single transducer would suffice to
reveal $G^{(00)}(\omega)$ around the monopole frequencies $\omega_{n0}$,
whilst {\it five\/} (placed at suitable positions) would be required to
calculate the five degenerate amplitudes $G^{(2m)}(\omega)$ around the
quadrupole frequencies $\omega_{n2}$. Once the {\it six\/} functions
$G^{(lm)}(\omega)$ would have thus been determined, inverse Fourier
transforms would give us the functions $g^{(lm)}(t)$, and thereby the
six Riemann tensor components $R_{0i0j}(t)$ through inversion of the
second of equations~(\ref{3.8}), i.e., as an expansion like~(\ref{3.3})
---only with $g\/$'s instead of ${\cal S}\/$'s. Deconvolution would
then be complete.

Well, not quite\ldots\ Knowledge of the Riemann tensor in the
{\it laboratory\/} frame coordinates is not really sufficient to say the
waveform has been completely deconvolved, unless we {\it also\/} know the
{\it source position\/} in the sky. There clearly are two possibilities:

\begin{enumerate}
\item[{\sf i)}] The source position {\it is\/} known ahead of time by some
other astronomical observation methods. Let me rush to emphasise that, far
from trivial or uninteresting, this is a {\it very important\/} case to
consider, specially during the first stages of GW Astronomy, when any
reported GW event will have to be thoroughly checked by all possible means.

If the incidence direction is known, then a rotation must be applied to
the just obtained quantities $R_{0i0j}(t)$, which takes the laboratory
\mbox{$z\/$-axis} into coincidence with the incoming wave propagation
vector. A classification procedure must thereafter be applied to the so
transformed Riemann tensor in order to see which is the theory (or class
of theories) compatible with the actual observations. Such classification
procedure has been described in detail in~\cite{el73}; see also~\cite{m98}
for an updated discussion.

The spherical antenna is thus seen to have the {\it capability of furnishing
the analyst sufficient information to discern amongst different competing
theories of GW physics, whenever the wave incidence direction is known
prior to detection}.

\item[{\sf ii)}] The source position is {\it not\/} known at detection
time. This makes things more complex, since the above rotation between the
laboratory and GW frames cannot be performed.

In order to deconvolve the incidence direction in this case, a specific
theory of the GWs {\it must\/} be assumed ---a given choice being made
on the basis of whatever prior information is available or, simply,
dictated by the the decision to probe a particular theory. Wagoner and
Paik~\cite{wp77} propose a method which is useful both for GR and BD
theory, their idea being simple and elegant at the same time: since
neither of these theories predicts the excitation of the $m\/$=$\pm$1
quadrupole modes {\it of the wave}, the source position is determined
precisely by the rotation angles which, when applied to the laboratory
axes, cause the amplitudes of those {\it antenna\/} modes to vanish;
the rotated frame is thereby associated to the GW natural frame.

A generalisation of this idea can conceivably be found on the basis of a
detailed ---and possibly rather casuistic--- analysis of the canonical
forms of of the Riemann tensor for a list of theories of gravity, along
the following line of argument: any one particular theory will be
characterised by certain (homogeneous) canonical relationships amongst
the monopole and quadrupole components of the Riemann tensor,
$g^{(lm)}(t)$, and so enforcement of those relations upon rotation of
the laboratory frame axes should enable determination of the rotation
angles or, equivalently, of the incoming radiation incidence direction.
Scalar-tensor theories e.g. have $g^{(2\pm 1)}(t)=0$ in their canonical
forms, hence Wagoner and Paik's proposal for this particular case.

Before any deconvolution procedure is triggered, however, it is very
important to make sure that it will be {\it viable}. More precisely,
since the transformation from the laboratory to the ultimate canonical
frame is going to be linear, {\it invariants\/} must be preserved. This
means that, even if the source position is unknown, certain theories will
forthrightly be {\it vetoed\/} by the observed $R_{0i0j}(t)$ if their
predicted invariants are incompatible with the observed ones. To give but
an easy example, if $R_{0i0j}(t)$ is {\it observed\/} to have a non-null
trace $R_{0i0i}(t)$, then a veto on GR will be readily served, and
therefore no algorithm based on that theory should be applied~\cite{lms}.

I would like to make a final remark here. Assume a direction deconvolution
procedure has been successfully carried through to the end on the basis
of certain GW theory, so that the analyst comes up with a pair of numbers
$(\theta,\varphi)$ expressing the source's coordinates in the sky. Of
course, these numbers will represent the {\it actual\/} source position
{\it only if the assumed theory is correct}. Now, how do we know it
{\it is\/} correct? Strictly speaking, ``correctness'' of a scientific
theory is an {\it asymptotic\/} concept ---in the sense that the
possibility always remains open that new facts be eventually discovered
which contradict the theory---, and so {\it reliability\/} of the
estimate $(\theta,\varphi)$ of the source position can only be assessed
in practice in terms of the {\it consistency\/} between the assumed
theory and whatever experimental evidence is available {\it to date},
including, indeed, GW measurements themselves. It is thus very important
to have a method to verify that the estimate $(\theta,\varphi)$ does
not contradict the theory which enabled its very determination.

Such verification is a {\it logical\/} absurdity if only {\it one\/}
measurement of position is available; this happens for instance if the
recorded signal is a {\it short burst\/} of radiation, and so {\it two
antennas\/} are at least necessary to check consistency in that case.
The test would proceed as a check that the time delay between reception
of the signal at both detectors is consistent with the calculated
$(\theta,\varphi)$\footnote{
Note that the two detectors will agree on the same $(\theta,\varphi)$,
even if the assumed theory is wrong, since the sphere deformations will
be the same if caused by the same signal.},
given their relative position and the wave propagation speed predicted by
the assumed theory. If, on the other hand, the signal being tracked is
a {\it long duration\/} signal, then a single antenna may be sufficient
to perform the test by looking at the observed Doppler patterns and
checking them against those expected with the given $(\theta,\varphi)$.

\end{enumerate}

The above considerations have been made ignoring noise in the detector
and monitor systems. A fundamental constraint introduced by noise is
that it makes the antenna {\it bandwidth limited\/} in sensitivity. As
a consequence, any deconvolution procedure is deemed to be incomplete
or, rather, {\it ambiguous\/}~\cite{als}, since information about the
signal can possibly be retrieved only within a reduced bandwidth,
whilst the rest will be lost. I thus come to a detailed discussion of
the sensitivity of the spherical GW antenna in the next section.

\section{The sensitivity parameters}

I will consider successively {\it amplitude\/} and {\it energy\/}
sensitivities; the first leads to the concept of {\it transfer
function}, while the second to that of absorption {\it cross
section}. I devote separate subsections to analyse each of them
in some detail.

\subsection{The transfer function}

A widely used and useful concept in linear system theory is that of
{\it transfer function\/}~\cite{he68}. It is defined as the Fourier
transform of the system's impulse response, or as the system's
impedance/admittance, and can be inferred from the frequency response
function~(\ref{3.18}).

We recall from the previous section that the sphere is a multimode device
---due to its monopole and five-fold degenerate quadrupole modes. It is
expedient to define a {\it multimode\/} or {\it vector transfer function\/}
as a useful construct which encompasses all six different modes into a
single conceptual block, according to

\begin{equation}
   {\bf U}({\bf x},\omega) = \sum_\alpha\,{\bf Z}^{(\alpha)}({\bf x},\omega)
               \,G^{(\alpha)}(\omega)          \label{4.1}
\end{equation}
where $G^{(\alpha)}(\omega)$ are the six driving terms $G^{(lm)}(\omega)$
given in~(\ref{3.19}). The transfer function is
${\bf Z}^{(\alpha)}({\bf x},\omega)$, and its ``vector'' character alluded
above is reflected by the {\it multimode index\/} $\alpha$. Looking
at~(\ref{3.18}) it is readily seen that

\begin{deqarr}
\arrlabel{4.2}
  {\bf Z}^{(00)}({\bf x},\omega) & = & \frac{\pi}{i}\,\sum_{n=1}^\infty\,
  \frac{a_{n0}}{\omega_{n0}}\,{\bf u}_{n00}({\bf x})\,\left[
  \delta(\omega\!-\!\omega_{n0}) - \delta(\omega\!+\!\omega_{n0})
  \right]	\label{4.2a}    \\[1.3 em]
  {\bf Z}^{(2m)}({\bf x},\omega) & = & \frac{\pi}{i}\,\sum_{n=1}^\infty\,
  \frac{a_{n2}}{\omega_{n2}}\,{\bf u}_{n2m}({\bf x})\,\left[
  \delta(\omega\!-\!\omega_{n2}) - \delta(\omega\!+\!\omega_{n2})
  \right]  \qquad (m=-2,\ldots,2)     \label{4.2b}
\end{deqarr}

As we observe in these formulae, the sphere's sensitivity to monopole
excitations is governed by $a_{n0}/\omega_{n0}$, and to quadrupole ones by
$a_{n2}/\omega_{n2}$. Closed expressions happen to exist for $a_{n0}$ and
$a_{n2}$; using the notation of Appendix B, they are

\begin{deqarr}
\arrlabel{4.4}
   \frac{a_{n0}}{R} & = & \frac{3\,C(n,0)}{8\pi}\;\frac{j_2(q_{n0}R)}{q_{n0}R}
   \label{4.4a}  \\[1 em]
   \frac{a_{n2}}{R} & = & -\frac{3\,C(n,2)}{8\pi}\,\left[\beta_3(k_{n2}R)\,
   \frac{j_2(q_{n2}R)}{q_{n2}R} - 3\,\frac{q_{n2}}{k_{n2}}\,
   \beta_1(q_{n2}R)\,\frac{j_2(k_{n2}R)}{k_{n2}R}
   \right]     \label{4.4b}
\end{deqarr}

Numerical investigation of the behaviour of these coefficients shows that
they decay asymptotically as $n^{-2}$:

\begin{equation}
    a_{nl} \stackrel{n\rightarrow\infty}{\longrightarrow}
    {\rm const}\times n^{-2}          \label{4.6}
\end{equation}

Likewise, it is found that the frequencies $\omega_{n0}\/$ and
$\omega_{n2}\/$ diverge like $n\/$ for large $n$, so that
${\bf Z}^{(\alpha)}({\bf x},\omega)$ drops as $\omega^{-3}$ for
large $\omega$. Figures 6 and 7 display a symbolic plot of
$\omega^3\,{\bf Z}^{(00)}({\bf x},\omega)$ and
$\omega^3\,{\bf Z}^{(2m)}({\bf x},\omega)$, respectively, which illustrates
the situation: monopole modes soon reach the asymptotic regime, while
there appear to be 3 sub-families of quadrupole modes regularly intertwined;
the asymptotic regime for these sub-families is more irregularly reached.
Note also the perfectly regular alternate changes of phase (by $\pi\/$
radians) in both monopole and each quadrupole family.

The sharp fall in sensitivity of a sphere for higher frequency modes
($n^{-3}$) indicates that only the lowest ones stand a chance of being
observable in an actual GW antenna. I report in Table I the numerical
values of the relevant parameters for the first few monopole and
quadrupole modes. The reason for the last (fourth) columns will become
clear later.

\begin{table}
\label{tab1}
\caption{First few monopole (left) and quadrupole (right) sphere
parameters, for a $\sigma\/$=\,0.33 material. First and second columns
on either side of the central line number the modes and give the
corresponding eigenvalue; rows are intertwined in order of ascending
frequency, which is proportional to $kR\/$ ---see (\protect\ref{B.4})
below. Third columns contain the $a_{n0}$ and $a_{n2}$ coefficients
defined in equations (\protect\ref{3.12}); the fourth columns display
the cross section {\it ratios\/} $(k_{10}a_{10}/k_{n0}a_{n0})^2$ and
$(k_{12}a_{12}/k_{n2}a_{n2})^2$ for higher frequency modes, respectively,
taking as reference the lowest in each family ---cf. equations
(\protect\ref{4.23}).}

\begin{center}
\begin{tabular}{cccc|cccc}
$n\ \ $ & $k_{n0}R$ & $a_{n0}/R$ & $\sigma_{10}/\sigma_{n0}\ $ &
$\ n\ \ $ & $k_{n2}R$ & $a_{n2}/R$ & $\sigma_{12}/\sigma_{n2}$
\\ \hline \\
  & &  &  & \ 1\ \  & 2.650 &  0.328 & 1                         \\
  & &  &  & \ 2\ \  & 5.088 &  0.106 & 2.61    \\
1\ \  & 5.432 & 0.214  & 1\     &       &       &         &     \\
  & &  &  & \ 3\ \  & 8.617  & $-$1.907$\times\! 10^{-2}$ & 27.95 \\
  & &  &  & \ 4\ \  & 10.917 & $-$9.101$\times\! 10^{-3}$ & 76.42 \\
2\ \  & 12.138 & $-$3.772$\times\! 10^{-2}$ & 6.46\  & & & &    \\
  & &  &  & \ 5\ \  & 12.280 &  1.387$\times\! 10^{-2}$ & 25.99   \\
  & &  &  & \ 6\ \  & 15.347 &  6.879$\times\! 10^{-3}$ & 67.87   \\
3\ \  & 18.492 & 1.600$\times\! 10^{-2}$  & 15.49\    & & & &   \\
\end{tabular}
\end{center}
\end{table}

\subsection{The absorption cross section}

Let us calculate now the energy of the oscillating sphere. We
first define the {\it spectral energy density\/} at frequency $\omega$,
which is naturally given by\footnote{
$T\/$ is the integration time ---assumed very large. The peaks in the
$\delta\/$-functions diverge like $T/\pi$.}

\begin{equation}
  W(\omega) = \frac{1}{T}\;\int_{\rm Solid}\frac{1}{2}\,\omega^2\,
        \left|{\bf U}({\bf x},\omega)\right|^2\,\varrho\,d^3x  \label{4.7}
\end{equation}
and can be easily evaluated:

\begin{equation}
  W(\omega) =
  \frac{1}{2}\pi M\,\sum_{n=1}^\infty\,\sum_{l=0,2}\,\sum_{m=-l}^{l}\,
  a_{nl}^2\,\left|G^{(lm)}(\omega)\right|^2\,
  \left[\delta(\omega\!-\!\omega_{nl}) + \delta(\omega\!+\!
  \omega_{nl})\right]
  \label{4.8}
\end{equation}

The {\it energy\/} at any one spectral frequency $\omega_{nl}\/$ is
obtained by {\it integration\/} of the spectral density in a narrow
interval around $\omega = \pm\omega_{nl}\/$:

\begin{equation}
  E(\omega_{nl})=\int_{-\omega_{nl}-\varepsilon}^{-\omega_{nl}+\varepsilon}
               + \int_{\omega_{nl}-\varepsilon}^{\omega_{nl}+\varepsilon}
                 \;W(\omega)\,\frac{d\omega}{2\pi}      \label{4.9}
\end{equation}

In this case,

\begin{equation}
  E(\omega_{nl}) = \frac{1}{2}\,M\,a_{nl}^2\,
  \sum_{m=-l}^l\,\left|G^{(lm)}(\omega_{n2})\right|^2\ ,\qquad l=0,2
  \label{4.10}
\end{equation}

The sensitivity parameter associated with the vibrational energy of the
modes is the detector's {\it absorption cross section\/}, defined as the
energy it absorbs per unit incident GW spectral flux density, or

\begin{equation}
   \sigma_{\rm abs}(\omega) = \frac{E(\omega)}{\Phi(\omega)}  \label{4.12}
\end{equation}
where $\Phi(\omega)$ is the number of joules per square metre and Hz
carried by the GW at frequency $\omega\/$ as it passes by the antenna.
Thus, for the frequencies of interest,

\begin{deqarr}
\arrlabel{4.13}
  \sigma_{\rm abs}(\omega_{n0}) & = &
  \frac{1}{2}Ma_{n0}^2\,\frac{\left|G^{(00)}(\omega_{n0})\right|^2}
  {\Phi(\omega_{n0})}  \label{4.13a}    \\*[0.7 em]
  \sigma_{\rm abs}(\omega_{n2}) & = &
  \frac{1}{2}Ma_{n2}^2\,\frac{\sum_{m=-2}^2\,\left|G^{(2m)}(\omega_{n2})
  \right|^2}{\Phi(\omega_{n2})}  \label{4.13b}
\end{deqarr}

These quantities have very precise values, but such values can only be
calculated on the basis of a {\it specific underlying theory of the GW
physics}. In the absence of such theory, neither $\Phi(\omega)$ nor
$G^{(lm)}(\omega)$ can possibly be calculated, since they are {\em not\/}
theory independent quantities. To date, only GR calculations have been
reported in the literature~\cite{clo,MZ94,wp77}. As I will now show,
even though the fractions in the rhs of~(\ref{4.13}) are {\it not\/}
theory independent, some very general results can still be obtained
about the sphere's cross section within the context of metric theories
of the gravitational interaction. To do so, it will be necessary to go
into a short digression on the general nature of weak metric GWs.

No matter which is the (metric) theory which happens to be the ``correct
one'' to describe gravitation, it is beyond reasonable doubt that any
GWs reaching the Earth ought to be {\it very weak}. The linear approximation
should therefore be an extremely good one to describe the propagating field
variables in the neighbourhood of the detector. In such circumstances, the
field equations can be derived from a Poincar\'e invariant variational
principle based on an action integral of the type

\begin{equation}
   \int\,{\cal L}(\psi_A,\psi_{A,\mu})\,d^4x      \label{4.15}
\end{equation}

where the Lagrangian density ${\cal L}\/$ is a {\it quadratic\/} functional
of the field variables $\psi_A(x)$ and their space-time derivatives
$\psi_{A,\mu}(x)$; these variables include the metric perturbations
$h_{\mu\nu}\/$, plus any other fields required by the specific theory
under consideration ---e.g. a scalar field in the theory of Brans--Dicke,
etc. The requirement that ${\cal L}\/$ be quadratic ensures that the
Euler--Lagrange equations of motion are {\it linear}.

The energy and momentum transported by the waves can be calculated in
this formalism in terms of the components $\tau^{\mu\nu}$ of the canonical
energy-momentum tensor\footnote{
This tensor is {\it not\/} symmetric in general, but can be
symmetrized by a standard method due to Belinfante~\cite{Barut,ll85}. For
the considerations which follow in this paper it is unnecessary to go
into those details, and the {\it canonical\/} form~(\ref{4.16}) will be
sufficient.}

\begin{equation}
   \tau^{\mu\nu}({\bf x},t) = \sum_A\,\frac{\partial{\cal L}}
    {\partial\psi_{A,\mu}}\,\psi_A^{,\nu} - {\cal L}\,\eta^{\mu\nu}
    \label{4.16}
\end{equation}

The flux energy density, or Poynting, vector is given by
$S_i = c^2\,\tau^{0i}$, i.e.,

\begin{equation}
   {\bf S}({\bf x},t) = c^3\,\sum_A\,
   \frac{\partial{\cal L}}{\partial\dot\psi_A}\,\nabla\psi_A
   \label{4.17}
\end{equation}
where \ ${\bf\dot{}}\equiv\partial/\partial t$. Any GW hitting the antenna
will be seen plane, due to the enormous distance to the source. If {\bf k}
is the incidence direction (normal to the wave front), then the fields
will depend on the variable $ct\,$$-${\bf k$\cdot$x}, so that the GW
energy reaching the detector per unit time and area is

\begin{equation}
  \phi(t)\equiv {\bf k\!\cdot\!S}({\bf x},t) = -c^2\,\sum_A\,
   \frac{\partial{\cal L}}{\partial\dot\psi_A}\,\dot\psi_A
   \label{4.18}
\end{equation}
where {\bf x} is the sphere's centre position relative to the source
---which is {\it fixed\/}, and so its dependence can be safely dropped
in the lhs of the above expression. The important thing to note in
equation~(\ref{4.18}) is that it tells us that $\phi(t)$ {\it can be
written as a quadratic form in the time derivatives of the
fields~$\psi_A$}. As a consequence, the spectral density $\Phi(\omega)$,
defined by

\begin{equation}
   \int_{-\infty}^\infty\,\phi(t)\,dt = \int_0^\infty\,
   \Phi(\omega)\,\frac{d\omega}{2\pi}      \label{4.20}
\end{equation}
can be ascertained to factorise as

\begin{equation}
   \Phi(\omega) = \omega^2\,\Phi_0(\omega)    \label{4.21}
\end{equation}
where $\Phi_0(\omega)$ is again a {\it quadratic\/} function of the
Fourier transforms $\Psi_A(\omega)$ of the fields $\psi_A$. On the
other hand, the functions $G^{(lm)}(\omega)$ in~(\ref{4.13}) which,
it is recalled, are the Fourier transforms of $g^{(lm)}(t)$ in~(\ref{3.8}),
contain {\it second\/} order derivatives of the {\it metric\/}
fields~$h_{\mu\nu}$, and therefore of {\it all\/} the
fields~$\psi_A$ as a result of the theory's field equations. Since we
are considering {\it plane wave\/} solutions to those equations, all
derivatives can be reduced to {\it time\/} derivatives ---just like
in~(\ref{4.18}) above. We can thus write

\begin{equation}
   G^{(lm)}(\omega) = -\omega^2\,\Psi^{(lm)}(\omega)   \label{4.22}
\end{equation}
with $\Psi^{(lm)}(\omega)$ suitable {\it linear\/} combinations of the
$\Psi_A(\omega)$. Replacing the last two equations into~(\ref{4.13})
and manipulating dimensions expediently, we come to the remarkable
result that

\begin{deqarr}
\arrlabel{4.23}
  \sigma_{\rm abs}(\omega_{n0}) & = &
  K_S(\aleph)\,\frac{GMv_t^2}{c^3}\,(k_{n0}a_{n0})^2
  \label{4.23a}     \\*[0.5 em]
  \sigma_{\rm abs}(\omega_{n2}) & = &
  K_Q(\aleph)\,\frac{GMv_t^2}{c^3}\,(k_{n2}a_{n2})^2  \label{4.23b}
\end{deqarr}
where $v_t^2\,$$\equiv\,$(2+2$\sigma)^{-1}$$\,v_{\rm s}^2$, $v_{\rm s}\/$
being the speed of sound in the detector's material, and $\sigma\/$ its
Poisson ratio; $G\/$ is the Gravitational constant. The ``remarkable''
about the above is that the coefficients $K_S(\aleph)$ and $K_Q(\aleph)$
{\it are independent of frequency\/}: they {\it exclusively depend on the
underlying gravitation theory}, which I symbolically denote by $\aleph$.
To see that this is the case, it is enough to consider a monochromatic
incident wave: since the coefficients $K_S(\aleph)$ and $K_Q(\aleph)$
happen to be {\it invariant\/} with respect to field amplitude
{\it scalings}, this means they will {\it only\/} depend on the amplitudes'
relative weights, i.e., on the field equations' {\it specific structure}.

By way of example, it is interesting to see what the results for General
Relativity (GR) and Brans--Dicke (BD) theory are. After somehow lengthy
algebra it is found that

\begin{equation}
   \aleph = {\rm GR} \Rightarrow \left\{\begin{array}{l}
          K_S(\aleph) = 0 \\[0.7 em]
          K_Q(\aleph) = \mbox{\large $\frac{16\,\pi^2}{15}$}
          \end{array}\right.   \label{4.25}
\end{equation}
and

\begin{equation}
   \aleph = {\rm BD} \Rightarrow \left\{\begin{array}{l}
          K_S(\aleph) = \mbox{\large $\frac{8\,\pi^2}{9}$}\,
          (3+2\Omega)^{-2}\,k\,\left[1+\mbox{\large $\frac
          {k\Omega}{(3+2\Omega)^2}$}\right]^{-1}  \\*[1.3 em]
          K_Q(\aleph) = \mbox{\large $\frac{16\,\pi^2}{15}$}\,\left[1 +
          \frac{1}{6}\,(3+2\Omega)^{-2}\,k\right]
          \left[1+\mbox{\large $\frac{k\Omega}{(3+2\Omega)^2}$}\right]^{-1}
          \end{array}\right.   \label{4.26}
\end{equation}

In the latter formulae, $\Omega$ is the usual Brans--Dicke parameter
$\omega\/$~\cite{bd61}, renamed here to avoid confusion with
{\it frequency}, and $k\/$ is a dimensionless parameter, generally
of order one, depending on the source's properties~\cite{lobonotes}. As
is well known, GR is obtained in the limit $\Omega\rightarrow\infty$
of BD~\cite{we72}; the quoted results are of course in agreement with
that limit.

Incidentally, an interesting consequence of the above equations is
that the presence of a scalar field in the theory of Brans and Dicke
causes {\it not only the monopole\/} sphere's modes to be excited,
{\it but also the {\rm $m\/$=0} quadrupole ones}; what we see in
equations~(\ref{4.26}) is that {\it precisely\/} 5/6 of the total
energy extracted from the scalar wave goes into the antenna's monopole
modes, whilst there is still a remaining 1/6 which is communicated to
the quadrupoles, independently of the values of $\Omega$ and $k$\footnote{
Note however that since monopole and quadrupole detector modes occur at
different frequencies, this particular {\it distribution\/} of energy
may not be seen if the sphere's vibrations are monitored at a single
resonance.}.

This somehow non-intuitive result finds its explanation in the structure
of the Riemann tensor in BD theory, in which the {\it excess\/}
$R_{0i0j}\/$ with respect to General Relativity happens {\it not\/} to be
proportional to the scalar part $E_{ij}^{(00)}$, but to a combination of
$E_{ij}^{(00)}$ \mbox{and $E_{ij}^{(20)}$.} More in-depth analysis of
these facts has been investigated in reference~\cite{maura}.

Equations~(\ref{4.23}) show that, no matter which is the gravity theory
assumed, the sphere's absorption cross sections for higher modes
{\it scale\/} as the successive coefficients $(k_{n0}a_{n0})^2$ and
$(k_{n2}a_{n2})^2$ for monopole and quadrupole modes, respectively. In
particular, the result quoted in~\cite{clo} that cross section for the
second quadrupole mode is 2.61 times less than that for the first,
assuming GR, is in fact valid, as we now see, {\it independently of
which is the\/ {\rm (metric)} theory of gravity actually governing GW
physics}. The fourth columns in Table~\ref{tab1} display these scaling
properties. It is seen that the drop in cross section from the first to
the second monopole mode is as high as 6.46. It should however be stressed
that the frequency of such mode would be over 4 kHz for a (likely) sphere
whose fundamental {\it quadrupole\/} frequency be 900 Hz~\cite{clo}. Note
finally the asymptotic cross section drop as $n^{-2}\/$ for large
$n\/$ ---cf. equation~(\ref{4.6}) and the ensuing paragraph.

\section{Conclusion}

The main purpose of this paper has been to set up a {\it sound\/}
mathematical formalism to address with as much generality as possible any
questions related to the interaction between a resonant antenna and a weak
incoming GW, with much special emphasis on the homogeneous {\it sphere}.
New results have been found along this line, such as the {\it scaling\/}
properties of cross sections for higher frequency modes, or the sensitivity
of the antenna to arbitrary metric GWs; also, new ideas have been put
forward regarding the {\it direction deconvolution\/} problem within the
context of an arbitrary metric theory of GW physics. Less spectacularly,
the full machinery has also been applied to produce {\it independent\/}
checks of previously published results.

The whole investigation reported herein has been developed with no
{\it a priori\/} assumptions about any specific (metric) theory of the
GWs, and is therefore {\it very\/} general. ``Too general solutions'' are
often impractical in science; here, however, the ``very general'' appears
to be rather ``cheap'', as seen in the results expressed by the equations
of section 3 above. An immediate consequence is that solid elastic detectors
of GWs (and, in particular, spheres) offer, as a matter of principle, the
possibility of probing {\it any\/} given theory of GW physics with just as
much effort as it would take, e.g., to probe General Relativity: the vector
transfer function of section 4 supplies the requisite theoretical vehicle
for the purpose.

An important question, however, has not been considered in this paper.
This is the {\it transducer\/} problem: the sphere's oscillations can
only be revealed to the observer by means of suitable (usually
electromechanical) transducers. These devices, however, are not
{\it neutral\/}, i.e., they {\it couple\/} to the antenna's motions,
thereby exercising a back action on it which must be taken into
consideration if one is to correctly interpret the system's readout.
Preliminary studies and proposals have already been published~\cite{jm93},
but further work is clearly needed for a more thorough understanding of
the problems involved.

Progress in this direction is currently being made ---which I expect to
report on shortly. The formalism developed in this paper provides basic
support to that further work\footnote{
This {\em was\/} underway in 1995, and is now complete. The results are
presented right below, from page~\pageref{sec:intro} on.}.

\section*{Acknowledgments}

It is a pleasure for me to thank Eugenio Coccia for his critical reading
and comments on the manuscript. I also want to express gratitude to M
Montero and JA Ortega for their assistance during the first stages of
this work, and JMM Senovilla for some useful discussions. I have received
support from the Spanish Ministry of Education through contract number
PB93--1050.

\section*{Appendices}

\renewcommand{\thesection}{\Alph{section}}
\setcounter{section}{1}	   

\subsection{Algebra of the \bfit{E}-matrices}

Let ${\bf e}_x,{\bf e}_y,{\bf e}_z$ be three orthonormal Cartesian vectors
defining the sphere's laboratory reference frame. We define the equivalent
triad

\begin{equation}
  {\bf e}^{(0)}={\bf e}_z \qquad,\qquad {\bf e}^{(\pm 1)}=
  \mbox{\large $\frac{1}{\sqrt{2}}$}\,({\bf e}_x\pm i{\bf e}_y)
  \label{A.1}
\end{equation}
having the properties

\begin{equation}
   {\bf e}^{*(m')}\cdot{\bf e}^{(m)} = \delta_{m'm}
   \qquad ,\qquad m,m'=-1,0,1   \label{A.2}
\end{equation}

We say that the vectors~(\ref{A.1}) are the {\it natural basis\/} for the
$l\/$=1 irreducible representation of the rotation group; they behave
under arbitrary rotations precisely like the spherical harmonics
$Y_{1m}({\bf n})$. In particular, if a rotation of angle $\alpha\/$ around
the \mbox{$z\/$-axis} is applied to the original frame then

\begin{equation}
  {\bf e}^{(\pm 1)}\rightarrow\exp(\pm i\alpha)\,{\bf e}^{(\pm 1)}
  \qquad,\qquad {\bf e}^{(0)}\rightarrow{\bf e}^{(0)}   \label{A.3}
\end{equation}

Higher rank tensors have specific multipole characteristics depending on
the number of tensor indices, and the above basis lends itself to reveal
those characteristics, too. For example, the five dimensional linear space
of traceless symmetric tensors supports the $l\/$=2 irreducible
representation of the rotation group, while a tensor's trace is an
invariant. A general symmetric tensor can be expressed as an ``orthogonal''
sum of a traceless symmetric tensor and a multiple of the unit tensor.
A convenient basis to expand any such tensor is the following:

\begin{deqarr}
\arrlabel{A.4}
  & {\bf e}^{(1)}\otimes{\bf e}^{(1)}\qquad,\qquad
    {\bf e}^{(-1)}\otimes{\bf e}^{(-1)} & \label{A.4a} \\*[0.5 em]
  & {\bf e}^{(0)}\otimes{\bf e}^{(1)}+{\bf e}^{(1)}\otimes{\bf e}^{(0)}
                               \ \ ,\ \ 
    {\bf e}^{(0)}\otimes{\bf e}^{(-1)}+{\bf e}^{(-1)}\otimes{\bf e}^{(0)} & 
                             \label{A.4b} \\*[0.5 em]
  & {\bf e}^{(1)}\otimes{\bf e}^{(-1)}+{\bf e}^{(-1)}\otimes{\bf e}^{(1)} -
    2\,{\bf e}^{(0)}\otimes{\bf e}^{(0)} & \label{A.4c} \\*[0.5 em]
  & {\bf e}^{(1)}\otimes{\bf e}^{(-1)}+{\bf e}^{(-1)}\otimes{\bf e}^{(1)} +
    {\bf e}^{(0)}\otimes{\bf e}^{(0)} & \label{A.4d}
\end{deqarr}

The elements~(\ref{A.4a}) get multiplied by e$^{\pm 2i\alpha}\/$ in a
rotation of angle $\alpha\/$ around the \mbox{$z\/$-axis}, respectively,
the~(\ref{A.4b}) by e$^{\pm i\alpha}\/$, and~(\ref{A.4c}) and
(\ref{A.4d}) are invariant, as is readily seen. These properties define
the ``spin characteristics'' of the corresponding tensors. Also, the
five elements~(\ref{A.4a})--(\ref{A.4c}) are {\it traceless\/} tensors,
while~(\ref{A.4d}) is the {\it unit\/} tensor. Any symmetric tensor can
be expressed as a linear combination of the six~(\ref{A.4}), and the
respective coefficients carry the information about the weights of the
different monopole and quadrupole components of the tensor.

Equations~(\ref{3.4}) in the text are the matrix representation of the
above tensors in the Cartesian basis ${\bf e}_x,{\bf e}_y,{\bf e}_z$,
except that they are multiplied by suitable coefficients to ensure that
the conditions

\begin{equation}
  E_{ij}^{(lm)}\,n_i n_j = Y_{lm}({\bf n})\ ,\qquad
  l=0,2;\ \ \ \ m=-l,\ldots,l\label{A.8}
\end{equation}
where {\bf n}$\,\equiv\,${\bf x}/$|{\bf x}|$ is the radial unit vector,
hold. They are arbitrary, but expedient for the calculations in this
paper. The following {\it orthogonality\/} relations can be easily
established:

\begin{equation}
  E_{ij}^{*(2m')}\,E_{ij}^{(2m)} = 
  \mbox{\large $\frac{15}{8\pi}$}\,\delta_{m'm}\ \ ,\ \ 
  E_{ij}^{*(00)}\,E_{ij}^{(2m)} = 0\ \ ,\ \ 
  E_{ij}^{*(00)}\,E_{ij}^{(00)} = \mbox{\large $\frac{3}{4\pi}$}
  \label{A.9}
\end{equation}
with the indices $m$,$m'$ running from $-2$ to 2, and with an understood
sum over the repeated $i\/$ and $j$. It is also easy to prove the
{\it closure\/} properties

\begin{equation}
  E_{ij}^{*(00)}\,E_{kl}^{(00)}\ +\ \frac{2}{5}\,
  \sum_{m=-2}^2\,E_{ij}^{*(2m)}\,E_{kl}^{(2m)} = 
  \mbox{\large $\frac{3}{8\pi}$}\,(\delta_{ik}\delta_{jl}+
  \delta_{il}\delta_{jk})   \label{A.10}
\end{equation}

Equations~(\ref{A.9}) and~(\ref{A.10}) constitute the {\it completeness\/}
equations of the \mbox{$E\/$-matrix} basis of Euclidean symmetric tensors.

\subsection{The sphere's spectrum and wave-functions}

This Appendix is intended to give a rather complete summary of the
frequency spectrum and eigenmodes of a uniform elastic sphere. Although
this is a classical problem in Elasticity Theory~\cite{lo44}, some of
the results which follow have never been published so far. Also, its
scope is to serve as reference for notation, etc., in future work
---see ensuing article in page~\pageref{sec:intro}.

The uniform\footnote{
By {\it uniform\/} I mean its density $\varrho\/$ is constant
throughout the solid in the unperturbed state.}
elastic sphere's normal modes are obtained as the solutions to the
eigenvalue equation 

\begin{equation}
   \mu\nabla^2{\bf u} + (\lambda+\mu)\,\nabla(\nabla{\bf \cdot}{\bf u})
      = - \omega^2\varrho{\bf u}     \label{B.0}
\end{equation}
with the boundary conditions that its surface be free of any tensions
and/or tractions; this is expressed by the equations~\cite{ll70}

\begin{equation}
   \sigma_{ij}\,n_j = 0 \qquad {\rm at}\ \ r\!=\!R     \label{B.1}
\end{equation}
where $R\/$ is the sphere's radius, {\bf n} the outward normal, and
$\sigma_{ij}$ the {\it stress\/} tensor

\begin{equation}
   \sigma_{ij} = \lambda\,u_{kk}\delta_{ij} + 2\mu\,u_{ij}   \label{B.1a}
\end{equation}
with $u_{ij}\equiv{\mbox{\normalsize $\frac{1}{2}$}}(u_{i,j}+u_{j,i})$,
the {\it strain\/} tensor, and $\lambda,\mu\/$ the Lam\'e
coefficients~\cite{ll70}.

Like any differentiable vector field, {\bf u}({\bf x}) can be expressed as
a sum of an irrotational vector and a divergence-free vector,

\begin{equation}
  {\bf u}({\bf x}) = {\bf u}_{\rm irrot.}({\bf x}) +
  {\bf u}_{\rm div-free}({\bf x})\ ,
  \label{B.1b}
\end{equation}
say; on substituting this into equation~(\ref{B.0}), and after a few
easy manipulations, one can see that

\begin{equation}
  (\nabla^2 + k^2)\,{\bf u}_{\rm div-free}({\bf x}) = 0\ ,\qquad
  (\nabla^2 + q^2)\,{\bf u}_{\rm irrot.}({\bf x}) = 0
  \label{B.1c}
\end{equation}
where

\begin{equation}
   k^2\equiv\frac{\varrho\omega^2}{\mu} \ ,\qquad
   q^2\equiv\frac{\varrho\omega^2}{\lambda+2\mu}
   \label{B.4}
\end{equation}

Now the irrotational component can generically be expressed as
the {\em gradient\/} of a scalar function, i.e.,

\begin{equation}
   {\bf u}_{\rm irrot.}({\bf x}) = \nabla\phi({\bf x})
  \label{B.1d}
\end{equation}
while there are {\em two\/} linearly independent divergence-free components
which, as can be readily verified, are

\begin{equation}
  {\bf u}_{\rm div-free}^{(1)}({\bf x}) = {\bf L}\psi^{(1)}({\bf x})
  \ ,\ \ \ \mbox{and}\ \ \ \ 
  {\bf u}_{\rm div-free}^{(2)}({\bf x}) =
          \nabla\!\times\!{\bf L}\psi^{(2)}({\bf x})
  \label{B.1e}
\end{equation}
where ${\bf L}\equiv-i{\bf x}\!\times\!\nabla$ is the ``angular momentum''
operator, cf.~\cite{Ed60}, and $\psi^{(1)}$ and $\psi^{(2)}$ are also
scalar functions. If~(\ref{B.1d}) and~(\ref{B.1e}) are now respectively
substituted in~(\ref{B.1c}), it is found that $\phi({\bf x})$,
$\psi^{(1)}({\bf x})$, and $\psi^{(2)}({\bf x})$ satisfy Helmholtz equations:

\begin{equation}
  (\nabla^2 + k^2)\,\psi({\bf x}) = 0\ ,\qquad
  (\nabla^2 + q^2)\,\phi({\bf x}) = 0
   \label{B.1f}
\end{equation}
where $\psi({\bf x})$ stands for either $\psi^{(1)}({\bf x})$ or
$\psi^{(2)}({\bf x})$. Therefore

\begin{equation}
  \phi({\bf x})=j_l(qr)\,Y_{lm}({\bf n}) \qquad,\qquad
  \psi({\bf x})=j_l(kr)\,Y_{lm}({\bf n})     \label{B.3}
\end{equation}
in order to ensure regularity at the centre of the sphere, $r\/$=0. Here,
$j_l\/$ is a {\it spherical\/} Bessel function ---see~\cite{ab72} for
general conventions on these functions---, and $Y_{lm}\/$ a spherical
harmonic~\cite{Ed60}. Finally thus,

\begin{equation}
  {\bf u}({\bf x}) = \frac{C_0}{q^2}\,\nabla\phi({\bf x})
		   + \frac{iC_1}{k}\,{\bf L}\psi({\bf x})
		   + \frac{iC_2}{k^2}\,\nabla\!\times\!{\bf L}\psi({\bf x})
  \label{B.2}
\end{equation}
where $C_0,C_1,C_2$ are three constants which will be determined by the
boundary conditions~(\ref{B.1}) (the denominators under them have been
included for notational convenience). After lengthy algebra, those
conditions can be expressed as the following system of linear equations:

\begin{deqarr}
\arrlabel{B.5}
  \left[\beta_2(qR)-\mbox{\large $\frac{\lambda}{2\mu}$}\,
  q^2R^2\,\beta_0(qR)\right]\,C_0
  - l(l+1)\,\beta_1(kR)\,C_2 & = & 0   \label{B.5a} \\*[0.5 em]
  \beta_1(kR)\,C_1 & = & 0             \label{B.5b} \\*[0.5 em]
  \beta_1(qR)\,C_0 - \left[\mbox{\large $\frac{1}{2}$}\,\beta_2(kR)+
  \left\{\mbox{\large $\frac{l(l+1)}{2}$}-1\right\}\,
  \beta_0(kR)\right]\,C_2 & = & 0
  \label{B.5c}
\end{deqarr}
where

\begin{equation}
  \beta_0(z)\equiv\frac{j_l(z)}{z^2}\ ,\ \ \
  \beta_1(z)\equiv\frac{d}{dz}\left[\frac{j_l(z)}{z}\right]\ ,\ \ \
  \beta_2(z)\equiv\frac{d^2}{dz^2}\left[j_l(z)\right]  \label{B.8}
\end{equation}

There are clearly {\it two\/} families of solutions to~(\ref{B.5}):

\begin{enumerate}
\item[{\sf i)}] {\it Toroidal\/} modes. These are characterised by

\begin{equation}
   \beta_1(kR)=0\ ,\qquad C_0=C_2=0      \label{B.11}
\end{equation}

The frequencies of these modes are {\it independent\/} of $\lambda$, and
thence independent of the material's Poisson ratio. Their amplitudes are

\begin{equation}
   {\bf u}_{nlm}^T({\bf x})=T_{nl}(r)\,i{\bf L}Y_{lm}({\bf n})
   \label{B.12}
\end{equation}
with

\begin{equation}
   T_{nl}(r)=C_1(n,l)\,j_l(k_{nl}r)      \label{B.13}
\end{equation}
and $C_1(n,l)$ a dimensionless normalisation constant determined by the
general formula~(\ref{2.10}); $k_{nl}R\/$ is the $n\/$-th root of the
first equation~(\ref{B.11}) for a given $l$.

\item[{\sf ii)}] {\it Spheroidal\/} modes. These correspond to

\begin{equation}
   \det\left(\begin{array}{cc}
   \beta_2(qR)-\mbox{\large $\frac{\lambda}{2\mu}$}\,q^2R^2\,\beta_0(qR) &
   l(l+1)\,\beta_1(kR)   \\
   \beta_1(qR) & \mbox{\large $\frac{1}{2}$}\,\beta_2(kR)+
   \left\{\mbox{\large $\frac{l(l+1)}{2}$}-1\right\}\,\beta_0(kR)
   \end{array}\right) = 0      \label{B.14}
\end{equation}
and $C_1\/$\,=\,0. The frequencies of these modes {\it do\/} depend on
the Poisson ratio, and their amplitudes are

\begin{equation}
   {\bf u}_{nlm}^P({\bf x}) = A_{nl}(r)\,Y_{lm}({\bf n})\,{\bf n}
   - B_{nl}(r)\,i{\bf n}\!\times\!{\bf L}Y_{lm}({\bf n}) \label{B.15}
\end{equation}
where $A_{nl}(r)$ and $B_{nl}(r)$ have the somewhat complicated form

\begin{deqarr}
\arrlabel{B.16}
  A_{nl}(r) & = & C(n,l)\left[\beta_3(k_{nl}R)\,j_l'(q_{nl}r)
  -l(l+1)\,\frac{q_{nl}}{k_{nl}}\,\beta_1(q_{nl}R)\,
  \frac{j_l(k_{nl}r)}{k_{nl}r}\right]  \label{B.16a} \\*[0.8 em]
  B_{nl}(r) & = & C(n,l)\left[\beta_3(k_{nl}R)\,
  \frac{j_l(q_{nl}r)}{q_{nl}r}-\frac{q_{nl}}{k_{nl}}\,\beta_1(q_{nl}R)\,
  \frac{\left\{k_{nl}r\,j_l(k_{nl}r)\right\}'}{k_{nl}r}\right]
  \label{B.16b}
\end{deqarr}
with accents denoting derivatives with respect to implied (dimensionless)
arguments,

\begin{equation}
   \beta_3(z)\equiv\mbox{\large $\frac{1}{2}$}\,\beta_2(z) + 
   \left\{\mbox{\large $\frac{l(l+1)}{2}$}-1\right\}\,\beta_0(z)
   \label{B.18}
\end{equation}
and $C(n,l)$ a new normalisation constant. It is understood that
$q_{nl}\/$ and $k_{nl}\/$ are obtained after the (transcendental)
equation~(\ref{B.14}) has been solved for $\omega\/$ ---cf.
equation~(\ref{B.4}).
\end{enumerate}

In actual practice equations~(\ref{B.11}) and~(\ref{B.14}) are solved for
the {\em dimensionless\/} quantity $kR$, which will hereafter be called
the {\em eigenvalue\/}. In view of~(\ref{B.4}), the relationship between
the latter and the measurable frequencies (in Hz) is given by

\begin{equation}
   \nu\equiv\frac\omega{2\pi} = \left(\frac\mu{\varrho\,R^2}\right)^{1/2}\,
   \frac{kR}{2\pi}
   \label{B.19}
\end{equation}

It is more useful to express the frequencies in terms of the Poisson ratio,
$\sigma$, and of the speed of sound $v_{\rm s}$ in the selected material.
For this the following formulas are required ---see e.g.~\cite{ll70}:

\begin{equation}
   v_{\rm s} = \sqrt{\frac {\sf Y}\varrho}
   \label{B.20}
\end{equation}
where {\sf Y} is the {\em Young modulus\/}, related to the Lam\'e
coefficients and the Poisson ratio by

\begin{equation}
   {\sf Y} = \frac{(3\lambda + 2\mu)\,\mu}{\lambda + \mu} =
	     2\,(1+\sigma)\,\mu\ ,\qquad
   \sigma\equiv\frac\lambda{2(\lambda + \mu)}
   \label{B.21}
\end{equation}

Hence,

\begin{equation}
   \nu = \frac{(kR)}{2\pi\,\sqrt{1+\sigma}}\,\frac{v_{\rm s}}R
   \label{B.22}
\end{equation}

Equation~(\ref{B.22}) provides a suitable transformation formula from
abstract number eigenvalues $(kR)$ into physical frequencies $\nu$,
for given material's properties and sizes.

Tables~\ref{tab2} and~\ref{tab3} respectively display a set of values
of $(kR)$ for {\em toroidal\/} and {\em spheroidal\/} modes. While GWs
can only couple to quadrupole and monopole modes, it is important to have
some detailed knowledge of analytical results, as the sphere's frequency
spectrum is rather involved. It often happens, both in numerical
simulations and in experimental determinations, that it is very difficult
to disentangle the wealth of observed frequency lines, and to correctly
associate them with the corresponding eigenmode. Complications are
further enhanced by partial degeneracy lifting found in practice (due
to broken symmetries), which result in even more frequency lines in the
spectrum. Accurate analytic results should therefore be very helpful
to assist in frequency identification tasks.

\begin{table}[t]
\caption{List of a few {\it spheroidal eigenvalues\/}, ordered in
columns of ascending harmonics for each multipole value. Spheroidal
eigenvalues depend on the sphere's material Poisson ratio ---although
this dependence is weak. In this table, values are given for
$\sigma=0.33$. Note that the table contains {\it all\/} eigenvalues less
than or equal to 11.024 yet is not exhaustive for values larger than
that one; this would require to stretch the table horizontally beyond
$l\/$\,=\,10  ---see Figure I for a qualitative inspection of trends
in eigenvalue progressions.   \label{tab3}}

\begin{center}
\begin{tabular}{c|ccccccccccc}
$n$ & $l=0$   & $l=1$   & $l=2$   & $l=3$   & $l=4$   & $l=5$ &
      $l=6$   & $l=7$   & $l=8$   & $l=9$   & $l=10$ \\
\hline
 1  & 5.4322  & 3.5895  & 2.6497  & 3.9489  & 5.0662  & 6.1118
    & 7.1223  & 8.1129  & 9.0909  & 10.061  & 11.024 \\
 2  & 12.138  & 7.2306  & 5.0878  & 6.6959  & 8.2994  & 9.8529
    & 11.340  & 12.757  & 14.111  & 15.410  & 16.665 \\
 3  & 18.492  & 8.4906  & 8.6168  & 9.9720  & 11.324  & 12.686
    & 14.066  & 15.462  & 16.867  & 18.272  & 19.664 \\
 4  & 24.785  & 10.728  & 10.917  & 12.900  & 14.467  & 15.879
    & 17.243  & 18.589  & 19.930  & 21.272  & 22.619 \\
 5  & 31.055  & 13.882  & 12.280  & 14.073  & 16.125  & 18.159
    & 19.997  & 21.594  & 23.043  & 24.426  & 25.778
\end{tabular}
\end{center}
\end{table}

\begin{table}[h!]
\caption{List of a few {\it toroidal eigenvalues\/}, ordered in columns
of ascending harmonics for each multipole value. Unlike spheroidal
eigenvalues, toroidal eigenvalues are independent of the sphere's material
Poisson ratio. Note that the table contains {\it all\/} eigenvalues less
than or equal to 12.866 yet is not exhaustive for values larger than
that one; this would require to stretch the table horizontally beyond
$l\/$\,=\,11 ---see Figure II for a qualitative inspection of trends
in eigenvalue progressions.   \label{tab2}}

\begin{center}
\begin{tabular}{c|ccccccccccc}
$n$ & $l=1$   & $l=2$   & $l=3$   & $l=4$   & $l=5$   & $l=6$ &
      $l=7$   & $l=8$   & $l=9$   & $l=10$  & $l=11$ \\
\hline
 1  & 5.7635  & 2.5011  & 3.8647  & 5.0946  & 6.2658
    & 7.4026  & 8.599   & 9.6210  & 10.711  & 11.792  & 12.866 \\
 2  & 9.0950  & 7.1360  & 8.4449  & 9.7125  & 10.951
    & 12.166  & 13.365  & 14.548  & 15.720  & 16.882  & 18.035 \\
 3  & 12.323  & 10.515  & 11.882  & 13.211  & 14.511
    & 15.788  & 17.045  & 18.287  & 19.515  & 20.731  & 21.937 \\
 4  & 15.515  & 13.772  & 15.175  & 16.544  & 17.886
    & 19.204  & 20.503  & 21.786  & 23.055  & 24.310  & 25.555 \\
 5  & 18.689  & 16.983  & 18.412  & 19.809  & 21.181
    & 22.530  & 23.860  & 25.174  & 26.473  & 27.760  & 29.035
\end{tabular}
\end{center}
\end{table}

In Figures 1 and 2 a symbolic line diagramme of the two families of
frequencies of the sphere's spectrum is presented. Spheroidal eigenvalues
have been plotted for the Poisson ratio $\sigma\/$=0.33. Although only
the $l\/$=0 and $l\/$=2 {\it spheroidal\/} series couple to GW tidal
forces, the plots include other eigenvalues, as they can be useful
both in bench experiments ---cf. equation~(\ref{2.19}) above---
and for vetoing purposes in a spherical antenna.

Figures 3, 4 and 5 contain plots of the first three monopole and
quadrupole functions $T_{nl}(r)$, $A_{nl}(r)$ and $B_{nl}(r)$, always for
$\sigma\/$=0.33. $T_{n0}(r)$ and $B_{n0}(r)$ have however been omitted;
this is because they are multiplied by an identically zero angular
coefficient in the amplitude formulae~(\ref{B.12}) and~(\ref{B.15}).
Indeed, monopole vibrations are spherically symmetric, i.e., purely radial.


\newpage
\begin{center}
{\large\bf List of Figures}
\end{center}
\vspace{1.1 em}\hspace{1.7 em}

{\bf Figure I\ \ } The homogeneous sphere {\it spheroidal\/} eigenvalues
for a few {\it multipole\/} families. Only the $l\/$=0 and $l\/$=2 families
couple to metric GWs, so the rest are given for completeness and
non-directly-GW uses. Note that there are {\it fewer\/} monopole than any
other $l\/$-pole modes. The lowest frequency is the first {\it quadrupole\/}.
The diagram corresponds to a sphere with Poisson ratio $\sigma\/$=0.33.
Frequencies can be obtained from the plotted values through equation
(\protect\ref{B.22}) for any specific case.

\vspace{1.2 em}

{\bf Figure II\ \ } The homogeneous sphere {\it toroidal\/} eigenvalues.
None of these couple to GWs, but knowledge of them can be useful for
{\it vetoing\/} purposes. These eigenvalues are {\it independent\/}
of the material's Poisson ratio. To obtain actual frequencies from
plotted values, use~(\ref{B.22}). The lowest {\it toroidal\/}
eigenvalue is \mbox{$kR=2.5011$}, with $l\/$=2, and happens to be the
{\it absolute minimum\/} sphere's eigenvalue. Compared to the
{\it spheroidal\/} \mbox{$kR=2.6497$}, also with $l\/$=2, its
frequency is 5.61\%\ smaller. Note also that there are no monopole
toroidal modes.

\vspace{1.2 em}

{\bf Figure III\ \ } First three {\it spheroidal monopole\/} radial
functions $A_{n0}(r)$ ($n=1,2,3$), equation~(\ref{B.16a}).

\vspace{1.2 em}

{\bf Figure IV\ \ } First three {\it spheroidal quadrupole\/} radial
functions $A_{n2}(r)$ (continuous line) and $B_{n2}(r)$ (broken
line) ($n=1,2,3$), equations~(\ref{B.16}).

\vspace{1.2 em}

{\bf Figure V\ \ } First three {\it toroidal quadrupole\/} radial functions
$T_{n2}(r)$ ($n=1,2,3$), equation (\protect\ref{B.13}). A common feature
to these radial functions (also in the two previous Figures) is that they
present a {\it nodal\/} point at the origin ($r=0$), while the sphere's
surface ($r/R=1$) has a non-zero amplitude value, which is largest (in
absolute value) for the lowest $n\/$ in each group.

\vspace{1.2 em}

{\bf Figure VI\ \ } The {\it scalar\/} component
${\bf Z}^{(00)}({\bf x},\omega)$ of the {\it multimode\/} transfer
function, (\protect\ref{4.2a}). The diagram actually displays
$\omega^3\,{\bf Z}^{(00)}({\bf x},\omega)$, so asymptotic behaviours
are better appreciated. It is given in units of $\mu/\varrho R\/$, and
a factor $(\pi/i)\,{\bf u}_{n00}({\bf x})$, the eigenmode amplitude,
has been omitted, too. $\delta\/$-function amplitudes are symbolically
taken as 1. Note that the asymptotic regime, given by equation~(\ref{4.6}),
is quickly reached.

\vspace{1.2 em}

{\bf Figure VII\ \ } The {\it quadrupole\/} component
${\bf Z}^{(2m)}({\bf x},\omega)$ of the {\it multimode\/} transfer
function,~(\ref{4.2b}). The same prescriptions of Figure~6 apply
here; the plot is therefore {\it independent\/} of the value of $m\/$.
Note the presence of {\it three\/} sub-families of peaks; asymptotic
regimes are reached with variable speed for these sub-families, and less
rapidly than for monopole modes, anyway.


\begin{figure}
\centering
\includegraphics[width=16cm]{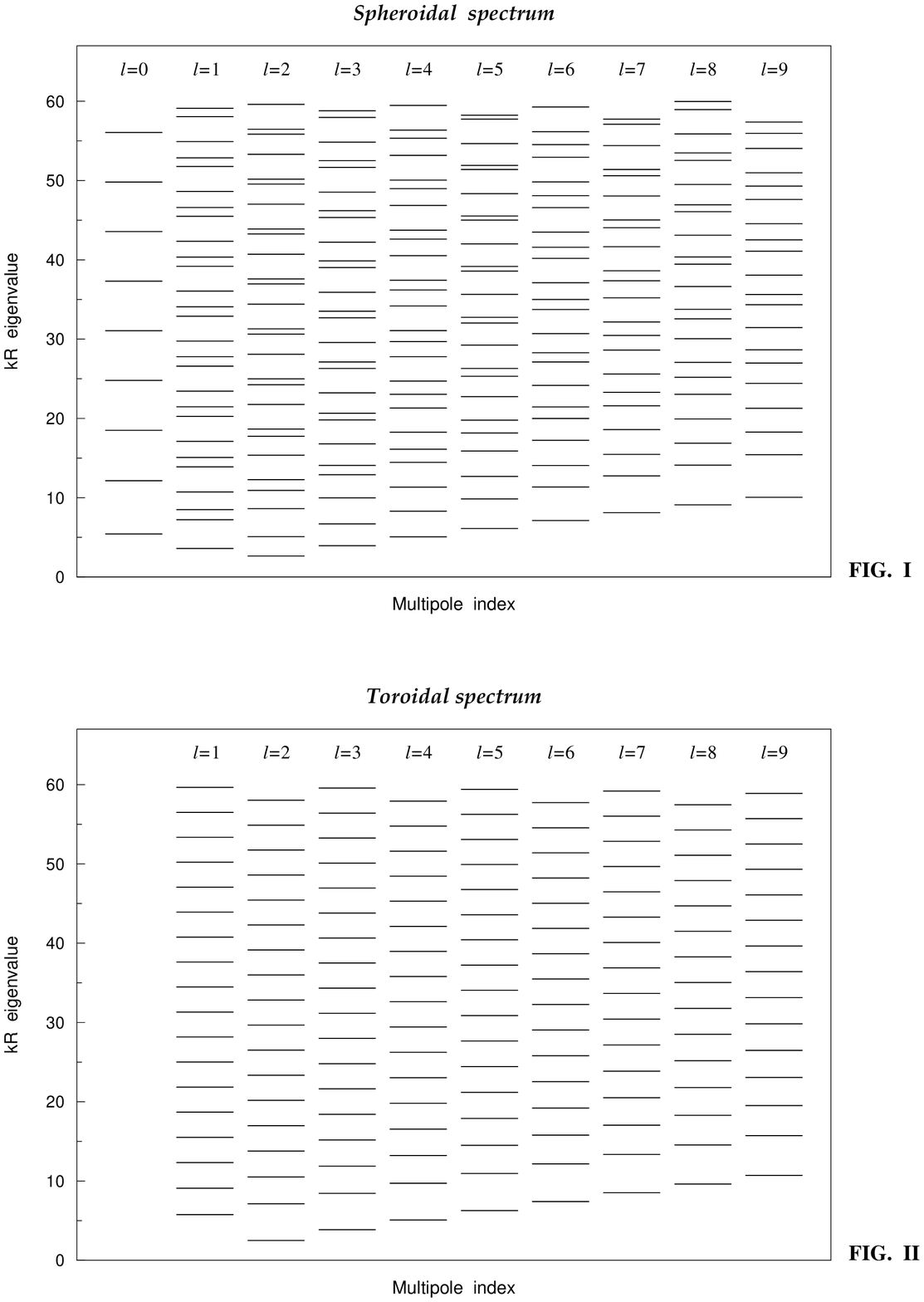}
\end{figure}

\begin{figure}
\centering
\includegraphics[width=16cm]{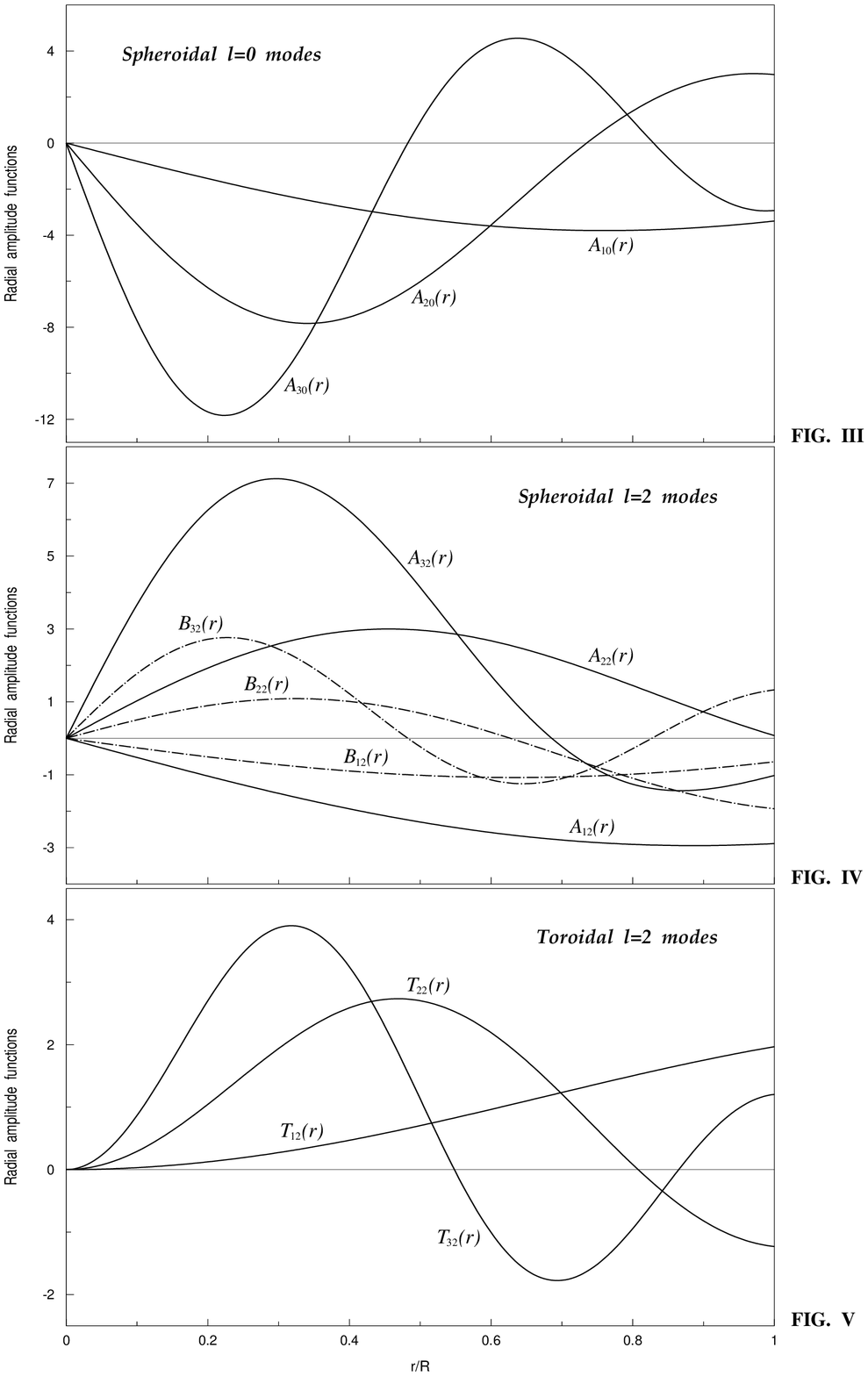}
\end{figure}

\begin{figure}
\centering
\includegraphics[width=16cm]{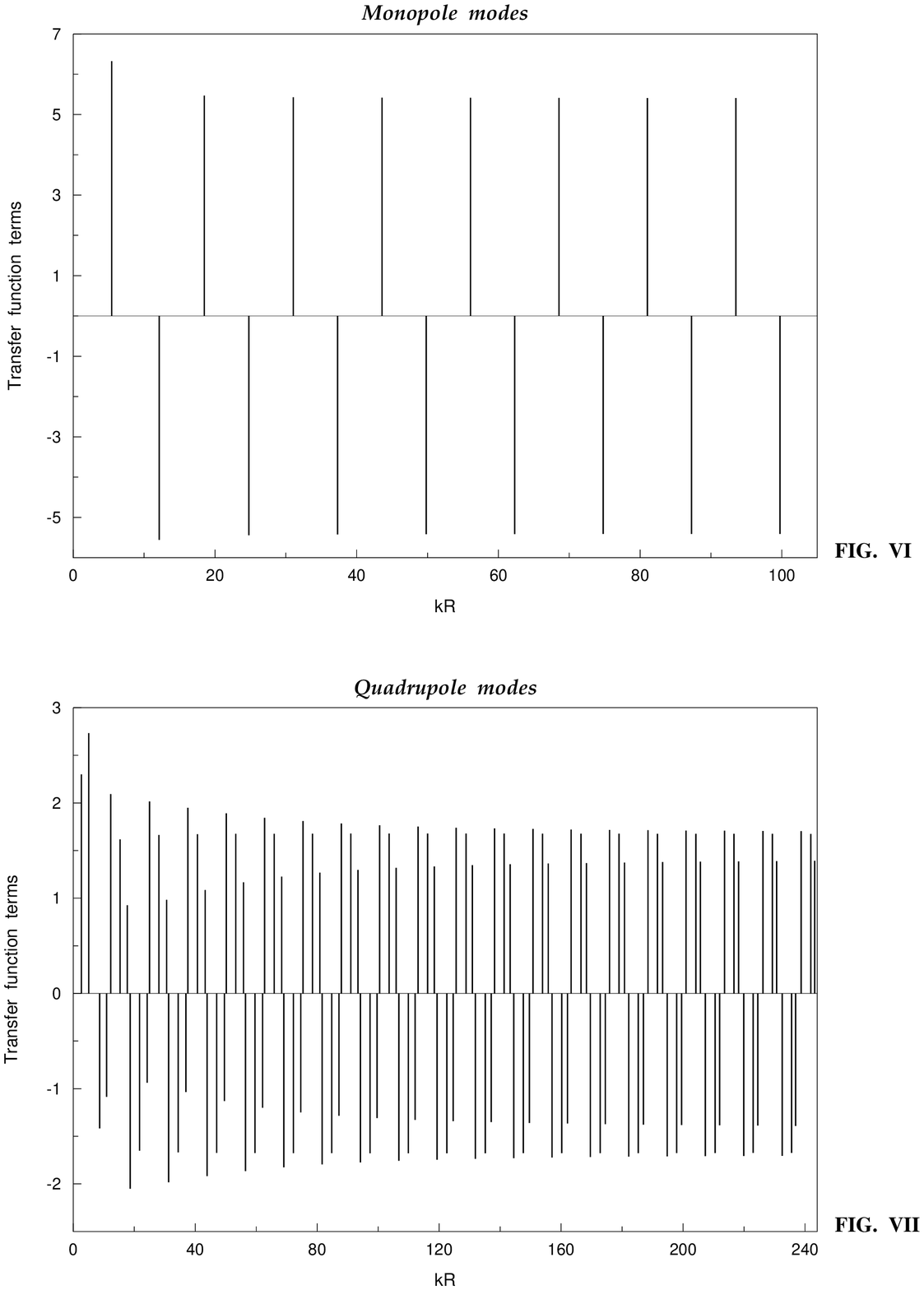}
\end{figure}

\newpage


\begin{flushleft}
{\LARGE\sf Multiple mode gravitational wave detection with a spherical
antenna}  \\[0.6 em]
MNRAS {\bf 316}, 173-194 (2000) \\[1 em]
{\large\bf Jos\'e Alberto Lobo} \\[0.5 em]
Departament de F\'\i sica Fonamental \\
Universitat de Barcelona, Spain \\
e-mail:\ {\tt lobo@hermes.ffn.ub.es}
\end{flushleft}
\vspace{2.1 em}

\begin{abstract}

Apart from omnidirectional, a solid elastic sphere is a natural multi-mode
and multi-frequency device for the detection of Gravitational Waves (GW).
Motion sensing in a spherical GW detector thus requires a {\it multiple\/}
set of transducers attached to it at suitable locations. If these are
{\it resonant\/} then they exert a significant back action on the larger
sphere and, as a consequence, the {\it joint dynamics\/} of the entire
system must be properly understood before reliable conclusions can be
drawn from its readout. In this paper, I present and develop an analytic
approach to study such dynamics which generalises currently existing  ones
and clarifies their actual range of validity. In addition, the new formalism
shows that there actually exist resonator layouts alternative to the highly
symmetric {\sl TIGA\/}, potentially having interesting properties. One of
these (I will call it {\sl PHC\/}), which only requires five resonators per
quadrupole mode sensed, and has {\it mode channels\/}, will be described in
detail. Also, the {\it perturbative\/} nature of the proposed approach makes
it very well adapted to systematically assess the consequences of realistic
mistunings in the device parameters by robust analytic methods. In order to
check the real value of the mathematical model, its predictions have been
confronted with experimental data from the {\sl LSU\/} prototype detector
{\sl TIGA\/}, and agreement between both is found to consistently reach a
satisfactory precision of {\it four\/} decimal places.

\end{abstract}

\renewcommand{\thesection}{\arabic{section}}
\setcounter{section}{0}	   

\section{Introduction}
\label{sec:intro}

The idea of using a solid elastic sphere as a gravitational wave (GW)
antenna is almost as old as that of using cylindrical bars: as far back
as 1971 Forward published a paper~\cite{fo71} in which he assessed some
of the potentialities offered by a spherical solid for that purpose. It
was however Weber's ongoing philosophy and practice of using bars which
eventually prevailed and developed up to the present date, with the highly
sophisticated and sensitive ultra-cryogenic systems currently in operation
---see~\cite{amaldi} and~\cite{gr14} for reviews and bibliography. With few
exceptions~\cite{ad75,wp77}, spherical detectors fell into oblivion for
years, but interest in them strongly re-emerged in the early 1990's, and
an important number of research articles have been published since which
address a wide variety of problems in GW spherical detector science. At
the same time, international collaboration has intensified, and prospects
for the actual construction of large spherical GW observatories (in the
range of $\sim$100 tons) are being currently considered in several countries
\footnote{
There are collaborations in Brazil, Holland, Italy and Spain.},
even in a variant {\it hollow\/} shape~\cite{vega}.

A spherical antenna is obviously omnidirectional but, most important, it
is also a natural {\it multi-mode\/} device, i.e., when suitably monitored,
it can generate information on all the GW amplitudes and incidence
direction~\cite{nadja}, a capability which possesses no other
{\it individual\/} GW detector, whether resonant or
interferometric~\cite{dt}. Furthermore, a spherical antenna could also
reveal the eventual existence of {\it monopole\/} gravitational radiation,
or set thresholds on it~\cite{maura}.

The theoretical explanation of these facts is to be found in the unique
matching between the GW amplitude structure and that of the sphere
oscillation eigenmodes~\cite{mini,lobo2}: a general {\it metric\/} GW
generates a {\it tidal\/} field of forces in an elastic body which is
given in terms of the ``electric'' components $R_{0i0j}(t)$ of the
Riemann tensor at its centre of mass by the following formula, see
equation~(\ref{3.7}) above~\cite{lobo}:

\begin{equation}
    {\bf f}_{\rm GW}({\bf x},t)\ \ \ =
    \sum_{\stackrel{\scriptstyle l=0\ {\rm and}\ 2}{m=-l,...,l}}\,
    {\bf f}^{(lm)}({\bf x})\,g^{(lm)}(t)    \label{1.1}
\end{equation}
where ${\bf f}^{(lm)}({\bf x})$ are ``tidal form factors'', while
$g^{(lm)}(t)$ are specific linear combinations of the Riemann tensor
components $R_{0i0j}(t)$ which carry all the {\it dynamical\/} information
on the GW's monopole ($l\/$\,=\,0) and quadrupole ($l\/$\,=\,2) amplitudes.
It is precisely these amplitudes, $g^{(lm)}(t)$, which a GW detector aims
to measure.

On the other hand, a free elastic sphere has two families of oscillation
eigenmodes, so called {\it toroidal\/} and {\it spheroidal\/} modes, and
modes within either family group into ascending series of $l\/$-pole
harmonics, each of whose frequencies is (2$l\/$+1)-fold degenerate
---see~\cite{lobo} for full details. It so happens that {\it only\/} monopole
and/or quadrupole spheroidal modes can possibly be excited by an incoming
{\it metric\/} GW~\cite{bian}, and their GW driven amplitudes are directly
proportional to the wave amplitudes $g^{(lm)}(t)$ of equation~(\ref{1.1}).
It is this very fact which makes of the spherical detector such a natural
one for GW observations~\cite{lobo}. In addition, a spherical antenna has
a significantly higher absorption {\it cross section\/} than a cylinder of
like fundamental frequency, and also presents good sensitivity at the
{\it second\/} quadrupole harmonic~\cite{clo}.

In order to monitor the GW induced deformations of the sphere {\it motion
sensors\/} are required. In cylindrical bars, current state of the art
technology is based upon {\it resonant transducers\/}~\cite{as93,hamil}.
A resonant transducer consists in a small (compared to the bar) mechanical
device possessing a resonance frequency accurately tuned to that of
the cylinder. This {\it frequency matching\/} causes back-and-forth
{\it resonant energy transfer\/} between the two bodies (bar and resonator),
which results in turn in {\it mechanically amplified\/} oscillations of the
smaller resonator. The philosophy of using resonators for motion sensing is
directly transplantable to a spherical detector ---only a {\it multiple\/}
set rather than a single resonator is required if its potential capabilities
as a multi-mode system are to be exploited to satisfaction.

A most immediate question in a multiple motion sensor system is:
{\it where\/} should the sensors be? The answer to this basic question
naturally depends on design and purpose criteria. Merkowitz and Johnson
(M\&J) made a very appealing proposal consisting in a set of 6 identical
resonators coupling to the {\it radial\/} motions of the sphere's surface,
and occupying the positions of the centres of the 6 non-parallel pentagonal
faces of a truncated icosahedron~\cite{jm93,jm95}. One of the most remarkable
properties of such layout is that there exist 5 linear combinations of the
resonators' readouts which are directly proportional to the 5 quadrupole
GW amplitudes $g^{(2m)}(t)$ of equation~(\ref{1.1}). M\&J call these
combinations {\it mode channels\/}, and they therefore play a fundamental
role in GW signal deconvolution in a real, {\it noisy\/}
system~\cite{m98,lms}. In addition, a reduced scale prototype antenna
---called {\sl TIGA\/}, for {\sl T\/}runcated {\sl I\/}cosahedron
{\sl G\/}ravitational {\sl A}ntenna--- was constructed at Louisiana State
University, and its working experimentally put to test~\cite{phd}. The
remarkable success of this experiment in almost every
detail~\cite{jm96,jm97,jm98} stands as a vivid proof of the practical
feasibility of a spherical GW detector~\cite{sfera,schipi}.

Despite its success, the theoretical model proposed by M\&J to describe the
system dynamics is based upon a simplifying assumption that the resonators
{\it only\/} couple to to the quadrupole vibration modes of the
sphere~\cite{jm93,jm95}. While this is seen {\it a posteriori\/} of
experimental measurements to be a very good approximation~\cite{phd,jm97},
a deeper {\it physical\/} reason which explains {\it why\/} this happens
is missing so far. The original motivation for the research I present in
this article was to develop a more general approach, based on first
principles, for the analysis of the resonator problem, very much in the
spirit of the methodology and results of reference~\cite{lobo}; this, I
thought, would not only provide the necessary tools for a rigorous
analysis of the system dynamics, but also contribute to improve our
understanding of the physics of the spherical GW detector.

Pursuing this programme, I succeeded in setting up and solving the equations
of motion for the coupled system of sphere plus resonators. The most important
characteristic of the solution is that it is expressible as a
{\it perturbative series expansion in ascending powers of the small parameter
$\eta^{1/2}$\/}, where $\eta\/$ is the ratio between the average resonator's
mass and the sphere's mass. The dominant (lowest) order terms in this
expansion appear to exactly reproduce Merkowitz and Johnson's
equations~\cite{jm95}, whence a quantitative assessment of their degree of
accuracy, as well as of the range of validity of their underlying hypotheses
obtains; if further precision is required then a well defined procedure for
going to next (higher) order terms is unambiguously prescribed by the system
equations.

Beyond this, though, the simple and elegant algebra which emerges out of the
general scheme has enabled the exploration of different resonator layouts,
alternative to the unique {\sl TIGA\/} of M\&J. In particular, I found
one~\cite{ls,lsc} requiring 5 rather than 6 resonators per quadrupole mode
sensed and possessing the remarkable property that {\it mode channels\/} can
be constructed from the system readouts, i.e., five linear combinations of the
latter which are directly proportional to the five quadrupole GW amplitudes.
I called this distribution {\sl PHC\/} ---see below for full details.

The intrinsically perturbative nature of the proposed approach makes it also
particularly well adapted to assess the consequences of small defects in the
system structure, such as for example symmetry breaking due to suspension
attachments, small resonator mistunings and mislocations, etc.\ This has
been successfully applied to account for the reported frequency measurements
of the {\sl LSU TIGA\/} prototype~\cite{phd}, which was diametrically drilled
for suspension purposes; in particular, discrepancies between measured and
calculated values (generally affecting only the {\it fourth\/} decimal place)
are precisely of the theoretically predicted order of magnitude.

The method has also been applied to analyse the stability of the spherical
detector to several mistuned parameters, with the result that it is not very
sensitive to small construction errors. This conforms again to experimental
reports~\cite{jm98}, but has the advantage that the argument depends on
{\it analytic\/} mathematical work rather than on computer simulated
procedures ---see e.g.\ \cite{jm98} or~\cite{ts}.

The paper is structured as follows: in section~\ref{sec:GE}, I present the
main physical hypotheses of the model, and the general equations of motion.
In section~\ref{sec:gff} a Green function approach to solve those equations
is set up, and in section~\ref{sec:srgw} it is used to assess the system
response to both monopole and quadrupole GW signals. In section~\ref{sec:PHC}
I describe in detail the {\sl PHC\/} layout, including its frequency spectrum
and {\it mode channels\/}. Section~\ref{sec:hs} contains a few brief
considerations on the system response to a hammer stroke calibration signal,
and finally in section~\ref{sec:symdef} I assess how the different parameter
mistunings affect the detector's behaviour. The paper closes with a summary
of conclusions, and three appendices where the heavier mathematical details
are made precise for the interested reader.

\section{General equations}  \label{sec:GE}

\begin{figure}
\label{fig1}
\centering
\includegraphics[width=4.7cm]{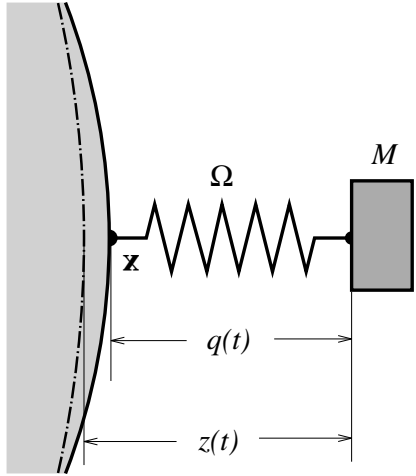}
\caption{Schematic diagram of the coupling model between a solid sphere
and a resonator. The notation is that in the text, but sub-indices have been
dropped for clarity. The dashed-dotted arc line on the left indicates the
position of the {\it undeformed\/} sphere's surface, and the solid arc its
{\it actual\/} position.}
\end{figure}

With minor improvements, I shall use the notation of references~\cite{lobo}
and~\cite{ls}, some of which is now briefly recalled. Consider a solid
sphere of mass $\cal M\/$, radius $R\/$, (uniform) density $\varrho\/$,
and elastic Lam\'e coefficients $\lambda$ and $\mu\/$, endowed with
a set of $J\/$ resonators of masses $M_a\/$ and resonance frequencies
$\Omega_a\/$ ($a\/$\,=\,1,\ldots,$J\/$), respectively. I shall model the
latter as {\it point masses\/} attached to one end of a linear spring,
whose other end is rigidly linked to the sphere at locations~${\bf x}_a\/$
---see Figure \ref{fig1}. The system degrees of freedom are given by the
{\it field\/} of elastic displacements ${\bf u}({\bf x},t)$ of the sphere
plus the {\it discrete\/} set of resonator spring deformations $z_a(t)$;
equations of motion need to be written down for them, of course, and this
is my next concern in this section.

I shall assume that the resonators only move radially, and also that
Classical Elasticity theory~\cite{ll70} is sufficiently accurate for
the present purposes\footnote{
We clearly do not expect relativistic motions in extremely small displacements
at typical frequencies in the range of 1 kHz.}.
In these circumstances we thus have

\begin{deqarr}
\arrlabel{m2.1}
    \varrho\,\frac{\partial^2 {\bf u}}{\partial t^2} & = & \mu\nabla^2 {\bf u}
    + (\lambda+\mu)\,\nabla(\nabla{\bf\cdot}{\bf u}) + {\bf f}({\bf x},t)
    \label{2.1.a}  \\*[0.7 em]
    \ddot{z}_a(t) & = & -\Omega_a^2\,
    \left[z_a(t)-u_a(t)\right]+\xi_a^{\rm external}(t)\ , \qquad a=1,\ldots,J
    \label{2.1.b}
\end{deqarr}

where ${\bf n}_a\/$\,$\equiv$\,${\bf x}_a/R\/$ is the outward pointing normal
at the the $a\/$-th resonator's attachment point, and

\begin{equation}
  u_a(t)\equiv{\bf n}_a\!\cdot\!{\bf u}({\bf x}_a,t)\ ,\qquad
  a=1,\ldots,J    \label{m3.8}
\end{equation}
is the {\it radial\/} deformation of the sphere's surface at ${\bf x}_a\/$.
A dot (\,$\dot{}$\,) is an abbreviation for time derivative. The term in
square brackets in~(\ref{2.1.b}) is thus the spring deformation ---$q(t)$ in
Figure \ref{fig1}.

${\bf f}({\bf x},t)$ in the rhs of~(\ref{2.1.a}) contains the
{\it density\/} of all {\it non-internal\/} forces acting on the sphere,
which is expediently split into a component due the resonators' {\it back
action\/} and an external action {\it proper\/}, which can be a GW signal,
a calibration signal, etc. Then

\begin{equation}
  {\bf f}({\bf x},t) = {\bf f}_{\rm resonators}({\bf x},t) +
  {\bf f}_{\rm external}({\bf x},t)    \label{m2.2}
\end{equation}

Finally, $\xi_a^{\rm external}(t)$ in the rhs of~(\ref{2.1.b}) is the
force per unit mass (acceleration) acting on the $a\/$-th resonator due
to {\it external\/} agents.

Given the hypothesis that the resonators are {\it point masses\/}, the
following holds:

\begin{equation}
    {\bf f}_{\rm resonators}({\bf x},t) =
    \sum_{a=1}^J M_a\Omega_a^2\,\left[\,z_a(t)-u_a(t)\right]\,
    \delta^{(3)}({\bf x}-{\bf x}_a)\,{\bf n}_a
    \label{2.3}
\end{equation}
where $\delta^{(3)}\/$ is the three dimensional Dirac density function.

The {\it external\/} forces I shall be considering in this paper will be
{\it gravitational wave\/} signals, and also a simple calibration signal,
a perpendicular {\it hammer stroke\/}. GW driving terms,
c.f.\ equation~(\ref{1.1}), can be written

\begin{equation}
   {\bf f}_{\rm GW}({\bf x},t) = {\bf f}^{(00)}({\bf x})\,g^{(00)}(t)\ +\ 
    \sum_{m=-2}^2\,{\bf f}^{(2m)}({\bf x})\,g^{(2m)}(t)    \label{2.4}
\end{equation}
for a general {\it metric\/} wave ---see~\cite{lobo} for explicit
formulas and technical details. While the spatial coefficients
${\bf f}^{(lm)}({\bf x})$ are pure {\it form factors\/} associated to
the {\it tidal\/} character of a GW excitation, it is the time dependent
factors $g^{(lm)}(t)$ which carry the specific information on the
incoming GW. The purpose of a GW detector is to determine the latter
coefficients on the basis of suitable measurements.

If a GW sweeps the observatory then the resonators themselves will also be
affected, of course. They will be driven, relative to the sphere's centre,
by a tidal acceleration which, since they only move radially, is given by

\begin{equation}
   \xi_a^{\rm GW}(t) = c^2\,R_{0i0j}(t)\,x_{a,i}n_{a,j}\ ,
   \qquad a=1,\ldots,J      \label{c.1}
\end{equation}
where $R_{0i0j}(t)$ are the ``electric'' components of the GW Riemann tensor
at the centre of the sphere. These can be easily manipulated to give\footnote{
$Y_{lm}({\bf n})$ are spherical harmonics \protect\cite{Ed60} ---see also the
multipole expansion of $R_{0i0j}(t)$ in reference~\cite{lobo}.}

\begin{equation}
   \xi_a^{\rm GW}(t) = R\,
   \sum_{\stackrel{\scriptstyle l=0\ {\rm and}\ 2}{m=-l,...,l}}\,
   Y_{lm}({\bf n}_a)\,g^{(lm)}(t)\ ,\qquad a=1,\ldots,J   \label{c.4}
\end{equation}
where $R\/$ is the sphere's radius.

I shall also be later considering the response of the system to a
particular {\it calibration\/} signal, consisting in a hammer stroke
with intensity ${\bf f}_0$, delivered perpendicularly to the sphere's
surface at point~${\bf x}_0$:

\begin{equation}
   {\bf f}_{\rm stroke}({\bf x},t) = {\bf f}_0\,
   \delta^{(3)}({\bf x}-{\bf x}_0)\,\delta(t)   \label{2.5}
\end{equation}
which is modeled as an impulsive force in both space and time variables.
Unlike GW tides, a hammer stroke will be applied on the sphere's surface,
so it has no {\it direct\/} effect on the resonators. In other words,

\begin{equation}
   \xi_a^{\rm stroke}(t) = 0\ ,\qquad a=1,\ldots,J    \label{c.5}
\end{equation}

The fundamental equations thus finally read:

\begin{deqarr}
\arrlabel{m2.6}
    \varrho \frac{\partial^2 {\bf u}}{\partial t^2} & = & \mu\nabla^2 {\bf u}
    + (\lambda+\mu)\,\nabla(\nabla{\bf\cdot}{\bf u}) +
    \nonumber \\
    & & \sum_{b=1}^J M_b\Omega_b^2\,\left[z_b(t)-u_b(t)\right]\,
    \delta^{(3)}({\bf x}-{\bf x}_b)\,{\bf n}_b
    + {\bf f}_{\rm external}({\bf x},t)    \label{2.6.a}  \\*[0.7 em]
    \ddot{z}_a(t) & = & -\Omega_a^2\,
    \left[z_a(t)-u_a(t)\right] + \xi_a^{\rm external}(t)\ ,
    \qquad a=1,\ldots,J    \label{2.6.b}
\end{deqarr}
where ${\bf f}_{\rm external}({\bf x},t)$ will be given by either~(\ref{2.4})
or~(\ref{2.5}), as the case may be. Likewise, $\xi_a^{\rm external}(t)$ will
be given by~(\ref{c.4}) or~(\ref{c.5}), respectively. The remainder of this
paper will be concerned with finding solutions to the system of coupled
differential equations~(\ref{m2.6}), and with their meaning and consequences.

\section{Green function formalism}  \label{sec:gff}

In order to solve equations~(\ref{m2.6}) I shall resort to Green function
formalism. The essentials of this procedure in the context of the present
problem can be found in detail in reference~\cite{lobo}; more specific
technicalities are given in appendix~\ref{app:a}.

By means of such formalism equations~(\ref{m2.6}) become the following
integro-differential system:

\begin{deqarr}
\arrlabel{m3.7}
  u_a(t) & = & u_a^{\rm external}(t) + \sum_{b=1}^J\,\eta_b\,\int_0^t
  K_{ab}(t-t')\,\left[\,z_b(t')-u_b(t')\right]\,dt'  \label{3.7.a}\\[1 ex]
  \ddot{z}_a(t) & = & \xi_a^{\rm external}(t)
  -\Omega_a^2\,\left[\,z_a(t)-u_a(t)\right]\ , \qquad a=1,\ldots,J
  \label{3.7.b}
\end{deqarr}
where $u_a^{\rm external}(t)$\,$\equiv$\,
${\bf n}_a\!\cdot\!{\bf u}^{\rm external}({\bf x}_a,t)$, and
${\bf u}^{\rm external}({\bf x},t)$ is the {\it bare\/} (i.e., without
attached resonators) sphere's response to the external forces
${\bf f}_{\rm external}({\bf x},t)$ in the rhs of~(\ref{2.6.a}).
$K_{ab}(t)$ is a {\it kernel matrix\/} defined by the following weighted
sum of diadic products of wavefunctions\footnote{
The capitalised index $N\/$ will often be used to imply the multiple index
$\{nlm\}$ which characterises the sphere's wave-functions.}:

\begin{equation}
  K_{ab}(t) = \Omega_b^2\,\sum_N\,\omega_N^{-1}\,
  \left[{\bf n}_b\!\cdot\!{\bf u}_N^*({\bf x}_b)\right]
  \left[{\bf n}_a\!\cdot\!{\bf u}_N({\bf x}_a)\right]\,\sin\omega_Nt
  \label{m3.10}
\end{equation}

Finally, the mass ratios of the resonators to the entire sphere are
defined by

\begin{equation}
  \eta_b\equiv \frac{M_b}{\cal M}\ ,\qquad b=1,\ldots,J   \label{m3.11}
\end{equation}
and will be {\it small parameters\/} in a real device.

Before proceeding further, let us briefly pause for a qualitative
inspection of equations~(\ref{m3.7}). Equation~(\ref{3.7.a}) shows
that the sphere's surface deformations $u_a(t)$ are made up of two
contributions: one due to the action of {\it external\/} agents (GWs or
other), contained in $u_a^{\rm external}(t)$, and another one due to coupling
to the resonators. The latter is commanded by the small parameters $\eta_b\/$,
and correlates to {\it all\/} of the sphere's spheroidal eigenmodes through
the kernel matrix $K_{ab}(t)$. This has consequences for GW detectors, for
even though GWs may only couple to quadrupole and monopole\footnote{
Monopole modes only exist in scalar-tensor theories of gravity, such as e.g.
Brans--Dicke \protect\cite{bd61}; General Relativity of course does not
belong in this category.}
spheroidal modes of the {\it free\/} sphere~\cite{lobo,bian},
attachment of resonators causes, as we see, coupling between these and the
{\it other\/} modes of the antenna; conversely, the latter back-act on the
former, too. As I shall shortly prove, such undesirable effects can be
minimised by suitably {\it tuning\/} the resonators' frequencies.

\subsection{Laplace transform domain equations}

A solution to equations~(\ref{m3.7}) will now be attempted.
Equation~(\ref{3.7.a}) is an integral equation belonging in
the general class of Volterra equations~\cite{tricomi}, but the usual
iterative solution to it by repeated substitution of $u_b(t)$ into the
kernel integral is not viable here due to the {\it dynamical\/} contribution
of $z_b(t)$, which is in turn governed by the {\it differential\/}
equation~(\ref{3.7.b}).

A better suited method to solve this {\it integro-differential\/} system is
to Laplace-transform it. I denote the Laplace transform of a generic function
of time $f(t)$ with a {\it caret\/} (\,$\hat{}$\,) on its symbol, e.g.,

\begin{equation}
  \hat{f}(s) \equiv \int_0^\infty f(t)\,e^{-st}\,dt	\label{m3.12}
\end{equation}
and make the assumption that the system is at rest before an instant of
time, $t\/$\,=\,0, say, or

\begin{equation}
  {\bf u}({\bf x},0)={\bf\dot u}({\bf x},0)=z_a(0)=\dot z_a(0) = 0
  \label{3.14}
\end{equation}

Equations~(\ref{m3.7}) then adopt the equivalent form

\begin{deqarr}
\arrlabel{m3.13}
    \hat u_a(s) & = & \hat u_a^{\rm external}(s)
    - \,\sum_{b=1}^J \eta_b\,\hat K_{ab}(s)\,
    \left[\hat z_b(s)-\hat u_b(s)\right]
    \label{3.13.a}   \\*[0.7 em]
    s^2\,\hat{z}_a(s) & = & \hat\xi_a^{\rm external}(s) -
    \Omega_a^2\,\left[\hat z_a(s)-\hat u_a(s)\right]\ ,\qquad a=1,\ldots,J
    \label{3.13.b}
\end{deqarr}
for which use has been made of the {\it convolution theorem\/} for Laplace
transforms\footnote{
This theorem states, it is recalled, that the Laplace transform of the
convolution product of two functions is the arithmetic product of their
respective Laplace transforms.}.
A further simplification is accomplished if we consider that we shall
in practice be only concerned with the {\it measurable\/} quantities

\begin{equation}
  q_a(t)\equiv z_a(t)-u_a(t) \ ,\qquad  a=1,\ldots,J	 \label{3.15}
\end{equation}
representing the resonators' actual elastic deformations ---cf.\ Figure
\ref{fig1}. It is readily seen that these verify the following:

\begin{equation}
  \sum_{b=1}^J \left[\delta_{ab} + \eta_b\,\frac{s^2}{s^2+\Omega_a^2}\,
  \hat K_{ab}(s)\right]\,\hat q_b(s) = -\frac{s^2}{s^2+\Omega_a^2}\,
  \hat u_a^{\rm external}(s) + \frac{\hat\xi_a^{\rm external}(s)}
  {s^2+\Omega_a^2}\ ,\qquad a=1,\ldots,J
  \label{m3.16}
\end{equation}

Equations~(\ref{m3.16}) constitute a significant simplification of the
original problem, as they are a set of just $J\/$ {\it algebraic\/} rather
than integral or differential equations. We must solve them for the unknowns
$\hat q_a(s)$, then perform {\it inverse Laplace transforms\/} to revert to
$q_a(t)$. I come to this next.

\section{System response to a Gravitational Wave}
\label{sec:srgw}

Our concern now is the actual system response when it is acted upon by
an incoming GW. It will be calculated by making a number of simplifying
assumptions, more precisely:

\begin{enumerate}
 \item[\sf i)] The detector is perfectly spherical.
 \item[\sf ii)] The resonators have identical masses and resonance frequencies.
 \item[\sf iii)] The resonators' frequency is accurately matched to one of
		 the sphere's oscillation eigenfrequencies.
\end{enumerate}

It will be shown below (section \ref{sec:symdef}) that a real system can be
appropriately treated as one which deviates by definite amounts from this
idealised construct. Therefore detailed knowledge of the ideal system
behaviour is essential for all purposes: such is the justification for the
above simplifications.

The wave-functions ${\bf u}_{nlm}({\bf x})$ of an elastic sphere can
be found in reference~\cite{lobo} in full detail, and I shall keep the
notation of that paper for them. The Laplace transform of the kernel
matrix~(\ref{m3.10}) can thus be expressed as ---see
equation~(\ref{A3.20}) in appendix~\ref{app:a}:

\begin{equation}
  \hat K_{ab}(s) = \sum_{nl}\,\frac{\Omega_b^2}{s^2+\omega_{nl}^2}\,
   \left|A_{nl}(R)\right|^2\,\frac{2l+1}{4\pi}\,
   P_l({\bf n}_a\!\cdot\!{\bf n}_b) \equiv
   \sum_{nl}\,\frac{\Omega_b^2}{s^2+\omega_{nl}^2}\,\chi_{ab}^{(nl)}
   \label{m4.2}
\end{equation}
where the last term simply {\it defines\/} the quantities $\chi_{ab}^{(nl)}$.
Note that the sums here extend over the {\it entire\/} spectrum of the
solid sphere.

The assumption that all the resonators are {\it identical\/} simply means that

\begin{equation}
  \eta_1=\,\ldots\,=\eta_J\equiv\eta\ ,\qquad
  \Omega_1=\,\ldots\,=\Omega_J\equiv\Omega
  \label{4.5}
\end{equation}

The third hypothesis makes reference to the fundamental idea behind using
resonators, which is to have them tuned to one of the frequencies of the
sphere's spectrum. This is expressed by

\begin{equation}
  \Omega = \omega_{n_0l_0}	\label{m4.6}
\end{equation}
where $\omega_{n_0l_0}$ is a specific and {\it fixed\/} frequency of the
spheroidal spectrum.

In a GW detector it will only make sense to choose $l_0$\,=\,0 or
$l_0$\,=\,2, as only monopole and quadrupole sphere modes couple to the
incoming signal; in practice, $n_0$ will refer to the first, or perhaps
second harmonic~\cite{clo}. I shall however keep the generic
expression~(\ref{m4.6}) for the time being in order to encompass
all the possibilities with a unified notation.

Based on the above hypotheses, equation~(\ref{m3.16}) can be rewritten in
the form

\begin{equation}
 \sum_{b=1}^J\,\left[\delta_{ab} + \eta\,\sum_{nl}\,
   \frac{\Omega^2s^2}{(s^2+\Omega^2)(s^2+\omega_{nl}^2)}\,\chi_{ab}^{(nl)}
   \right]\,\hat q_b(s) = -\frac{s^2}{s^2+\Omega^2}\,
   \hat u_a^{\rm GW}(s) + \frac{\hat\xi_a^{\rm GW}(s)}
   {s^2+\Omega^2}\ ,\qquad  (\Omega = \omega_{n_0l_0})
   \label{m4.8}
\end{equation}
where $\hat\xi_a^{\rm GW}(s)$ is the Laplace transform of~(\ref{c.4}), i.e.,

\begin{equation}
   \hat\xi_a^{\rm GW}(s) = R\,
   \sum_{\stackrel{\scriptstyle l=0\ {\rm and}\ 2}{m=-l,...,l}}\,
   Y_{lm}({\bf n}_a)\,\hat g^{(lm)}(s)\ ,\qquad a=1,\ldots,J
   \label{4.85}
\end{equation}

As mentioned at the end of the previous section, the matrix in the lhs
of~(\ref{m4.8}) must now be inverted; this will give us an expression for
$\hat q_a(s)$, whose {\it inverse Laplace transform\/} will take us back
to the time domain. A simple glance at the equation suffices however to
grasp the unsurmountable difficulties of accomplishing this
{\it analytically\/}.

Thankfully, though, a {\it perturbative\/} approach is applicable when
the masses of the resonators are small compared to the mass of the whole
sphere, i.e., when the inequality

\begin{equation}
  \eta\ll 1	\label{m4.10}
\end{equation}
holds. I shall henceforth assume that this is the case, as also is with
cylindrical bar resonant transducers. It is shown in appendix~\ref{app:b}
that the perturbative series happens in ascending powers of $\eta^{1/2}$,
rather than $\eta\/$ itself, and that the lowest order contribution has
the form

\begin{equation}
    \hat q_a(s) = \eta^{-1/2}\,\sum_{l,m}\,\hat\Lambda_a^{(lm)}(s;\Omega)\,
    \hat g^{(lm)}(s) + O(0) \ ,\qquad a=1,\ldots,J
    \label{6.8}
\end{equation}
where $O(0)$ stands for terms of order $\eta^0$ or smaller. Here,
$\hat\Lambda_a^{(lm)}(s;\Omega)$ is a {\it transfer function matrix\/}
which relates {\it linearly\/} the system response $\hat q_a(s)$ to the
GW amplitudes $\hat g^{(lm)}(s)$, in the usual sense that $q_a(t)$ is
given by the {\it convolution product\/} of the signal $g^{(lm)}(t)$
with the time domain expression, $\Lambda_a^{(lm)}(t;\Omega)$, of
$\hat\Lambda_a^{(lm)}(s;\Omega)$. The detector is thus seen to act as
a {\it linear filter\/} on the GW signal, whose frequency response is
characterised by the properties of $\hat\Lambda_a^{(lm)}(s;\Omega)$.
More specifically, the filter has a number of characteristic frequencies
which correspond to the {\it imaginary parts of the poles\/} of
$\hat\Lambda_a^{(lm)}(s;\Omega)$. As also shown in appendix~\ref{app:b},
these frequencies are the symmetric pairs

\begin{equation}
  \omega_{a\pm}^2 = \Omega^2\,\left(1\pm\sqrt{\frac{2l+1}{4\pi}}\,
  \left|A_{n_0l_0}(R)\right|\,\zeta_a\,\eta^{1/2}\right) + O(\eta)\ ,
  \qquad a=1,\ldots,J
  \label{5.2}
\end{equation}
where $\zeta_a^2\/$ is the $a\/$-th eigenvalue of the Legendre matrix

\begin{equation}
  P_{l_0}({\bf n}_a\!\cdot\!{\bf n}_b)\ ,\qquad a,b=1,\,\ldots,J
  \label{5.25}
\end{equation}
associated to the multipole ($l_0$) selected for tuning ---see~(\ref{m4.6}).
These frequency pairs correspond to {\it beats\/}, typical of resonantly
coupled oscillating systems ---we shall find them again in
section~\ref{sec:hs} in a particularly illuminating example.

Equation~(\ref{6.8}) neatly displays the amplification coefficient
$\eta^{-1/2}$ of the resonators' motion amplitudes, which corresponds
to the familiar resonant energy transfer in coupled systems of linear
oscillators~\cite{as93}.

The specific form of the transfer function matrix
$\hat\Lambda_a^{(lm)}(s;\Omega)$ depends both on the selected mode to
tune the resonator frequency $\Omega$ and on the resonator distribution
geometry. I now come to a discussion of these.

\subsection{Monopole gravitational radiation sensing}

General Relativity, as is well known, forbids monopole GW radiation. More
general {\it metric\/} theories, e.g. Brans-Dicke~\cite{bd61}, do however
predict this kind of radiation. It appears that a spherical antenna is
potentially sensitive to monopole waves, so it can serve the purpose of
thresholding, or eventually detecting them. It is therefore relevant to
consider the system response to scalar waves.

This clearly requires that the resonator set be tuned to a monopole
harmonic of the sphere, i.e.,

\begin{equation}
   \Omega = \omega_{n0}\ ,\qquad (l_0=0)	\label{6.9}
\end{equation}
where $n\/$ tags the chosen harmonic ---most likely the first ($n\/$\,=\,1)
in a thinkable device.

Since $P_0(z)$\,$\equiv$\,1 (for all $z\/$) the eigenvalues of
$P_0({\bf n}_a\!\cdot\!{\bf n}_b)$ are, clearly,

\begin{equation}
  \zeta_1^2=J\ ,\qquad \zeta_2^2=\,\ldots\,=\zeta_J^2=0
  \label{6.10}
\end{equation}
for {\it any resonator distribution\/}. The tuned mode frequency thus splits
into a {\it single\/} strongly coupled pair:

\begin{equation}
  \omega_\pm^2 = \Omega^2\,\left(1\pm\sqrt{\frac{J}{4\pi}}\,
  \left|A_{n0}(R)\right|\,\eta^{1/2}\right) + O(\eta)\ ,
  \qquad \Omega=\omega_{n0}
  \label{6.11}
\end{equation}

The $\Lambda$-matrix of equation~(\ref{6.8}) is seen to be in
this case

\begin{equation}
  \hat\Lambda_a^{(lm)}(s;\omega_{n0}) = (-1)^J\,\frac{a_{n0}}{\sqrt{J}}\,
  \frac{1}{2}\,\left[\left(s^2+\omega_+^2\right)^{-1} -
  \left(s^2+\omega_-^2\right)^{-1}\right]\,\delta_{l0}\,\delta_{m0}
  \label{6.12}
\end{equation}
whence the system response is

\begin{equation}
  \hat q_a(s) = \eta^{-1/2}\,\frac{(-1)^J}{\sqrt{J}}\,a_{n0}\,
  \frac{1}{2}\,\left[\left(s^2+\omega_+^2\right)^{-1} -
  \left(s^2+\omega_-^2\right)^{-1}\right]\,\hat g^{(00)}(s) + O(0)\ ,
  \qquad a=1,\ldots,J
  \label{6.13}
\end{equation}
{\it regardless of resonator positions\/}. The overlap coefficient
$a_{n0}$ is given by~(\ref{3.12a}), and can be calculated by means of
numerical computer programmes. By way of example, $a_{10}/R\/$\,=\,0.214,
and $a_{20}/R\/$\,=\,$-$0.038 for the first two harmonics.

A few interesting facts are displayed by equation~(\ref{6.13}). First,
as already stressed, it is seen that if the resonators are tuned to
a monopole {\it detector\/} frequency then only monopole {\it wave
amplitudes\/} couple strongly to the system, even if quadrupole radiation
amplitudes are significantly high at the observation frequencies
$\omega_\pm\/$. Also, the amplitudes $\hat q_a(s)$ are equal for all $a\/$,
as corresponds to the spherical symmetry of monopole sphere's oscillations,
and are proportional to $J^{-1/2}$, a factor we should indeed expect as an
indication that GW {\it energy\/} is evenly distributed amongst all the
resonators. A {\it single\/} transducer suffices to experimentally
determine the only monopole GW amplitude $\hat g^{(00)}(s)$, of course,
but~(\ref{6.13}) provides the system response if more than one sensor is
mounted on the antenna for whatever reasons.

\subsection{Quadrupole gravitational radiation sensing}

I now consider the more interesting case of quadrupole motion sensing.
The choice is now, clearly,

\begin{equation}
   \Omega = \omega_{n2}\ ,\qquad (l_0=2)	\label{6.14}
\end{equation}
where $n\/$ labels the chosen harmonic ---most likely the first
($n\/$\,=\,1) or the second ($n\/$\,=\,2) in a practical system. The
evaluation of the $\Lambda$-matrix is now considerably more
involved~\cite{serrano}, yet a remarkably elegant form is found for it:

\begin{equation}
  \hat\Lambda_a^{(lm)}(s;\omega_{n2}) = (-1)^N\,\sqrt{\frac{4\pi}{5}}\,
  a_{n2}\,\sum_{b=1}^J\,\left\{\sum_{\zeta_c\neq 0}\,\frac{1}{2}\left[
  \left(s^2+\omega_{c+}^2\right)^{-1} - \left(s^2+\omega_{c-}^2\right)^{-1}
  \right]\,\frac{v_a^{(c)}v_b^{(c)*}}{\zeta_c}\right\}\,
  Y_{2m}({\bf n}_b)\,\delta_{l2}    \label{6.15}
\end{equation}
where $v_a^{(c)}$ is the $c\/$-th normalised eigenvector of
$P_2({\bf n}_a\!\cdot\!{\bf n}_b)$, associated to the {\it non-null\/}
eigenvalue $\zeta_c^2$. Let me stress that equation~(\ref{6.15}) explicitly
shows that at most 5 pairs of modes, of frequencies $\omega_{c\pm}$, couple
strongly to quadrupole GW amplitudes, {\it no matter how many resonators in
excess of 5 are mounted on the sphere\/}. The tidal overlap coefficients
$a_{2n}\/$ can also be calculated using~(\ref{3.12b}), and give for the
first two harmonics

\begin{equation}
  \frac{a_{12}}{R} = 0.328\ ,\qquad\frac{a_{22}}{R} = 0.106   \label{6.16}
\end{equation}

The system response is thus

\begin{eqnarray}
  \hat q_a(s) & = & \eta^{-1/2}\,(-1)^J\,\sqrt{\frac{4\pi}{5}}\,a_{n2}\,
  \sum_{b=1}^J\,\left\{\sum_{\zeta_c\neq 0}\,\frac{1}{2}\left[
  \left(s^2+\omega_{c+}^2\right)^{-1} - \left(s^2+\omega_{c-}^2\right)^{-1}
  \right]\,\frac{v_a^{(c)}v_b^{(c)*}}{\zeta_c}\right\}\times
  \nonumber \\[0.5 em]  & & \hspace*{4 cm}
  \times\sum_{m=-2}^2\,Y_{2m}({\bf n}_b)\,\hat g^{(2m)}(s) + O(0)\ ,
  \qquad a=1\,\ldots,J	\label{6.17}
\end{eqnarray}

Equation~(\ref{6.17}) is {\it completely general\/}, i.e., it is valid
for any resonator configuration over the sphere's surface, and for any
number of resonators. It describes precisely how all 5 GW amplitudes
$\hat g^{(2m)}(s)$ interact with all 5 strongly coupled system modes;
like before, {\it only quadrupole wave amplitudes\/} are seen in the
detector (to leading order) when $\Omega$\,=\,$\omega_{n2}$, even if
the incoming wave carries significant monopole energy at the frequencies
$\omega_{c\pm}$.

The degree of generality and algebraic simplicity of~(\ref{6.17}) is new in
the literature. As we shall now see, it makes possible a systematic search
for different resonator distributions and their properties.

\section{The \bfsl{PHC} configuration}
\label{sec:PHC}

Merkowitz and Johnson's {\sl TIGA\/}~\cite{jm93} is highly symmetric, and is
the minimal set with maximum degeneracy, i.e., all the non-null eigenvalues
$\zeta_a\/$ are equal. To accomplish this, however, 6 rather than 5
resonators are required on the sphere's surface. Since there are just
5 quadrupole GW amplitudes one may wonder whether there are alternative
layouts with {\it only\/} 5 resonators. Equation~(\ref{6.17}) is completely
general, so it can be searched for an answer to this question. In
reference~\cite{ls} a specific proposal was made, which I now describe
in detail.

In pursuing a search for 5 resonator sets I found that distributions having
a sphere diameter as an axis of {\it pentagonal symmetry\/}\footnote{
By this I mean resonators are placed along a {\it parallel\/} of the
sphere every 72$^\circ$.}
exhibit a rather appealing structure. More specifically, let the resonators
be located at the spherical positions

\begin{equation}
   \theta_a  = \alpha \qquad ({\rm all}\,\ a)\ ,\qquad
   \varphi_a = (a-1)\,\frac{2\pi}{5}\ ,\qquad a=1,\ldots,5
\end{equation}

The eigenvalues and eigenvectors of $P_2({\bf n}_a\!\cdot\!{\bf n}_b)$ are
easily calculated:

\begin{deqarr}
\arrlabel{6.24}
   & \zeta_0^2 = \frac{5}{4}\,\left(3\,\cos^2\alpha-1\right)^2\ ,\qquad
       \zeta_1^2 = \zeta_{-1}^2 = \frac{15}{2}\,\sin^2\alpha\,\cos^2\alpha
       \ ,\qquad\zeta_2^2 = \zeta_{-2}^2 = \frac{15}{8}\,\sin^4\alpha  &
	\label{6.24.a} \\[1 em]
   & v_a^{(m)} = \sqrt{\frac{4\pi}{5}}\,\zeta_m^{-1}\,Y_{2m}({\bf n}_a)\ ,
       \qquad m=-2,\ldots,2\ ,\ \ a=1,\ldots,5	&
	\label{6.24.b}
\end{deqarr}
so the $\Lambda$-matrix is also considerably simple in structure in this
case:

\begin{equation}
  \hat\Lambda_a^{(lm)}(s;\omega_{n2}) = -\sqrt{\frac{4\pi}{5}}\,a_{n2}\,
  \zeta_m^{-1}\,\frac{1}{2}\left[\left(s^2+\omega_{m+}^2\right)^{-1} -
  \left(s^2+\omega_{m-}^2\right)^{-1}\right]\,Y_{2m}({\bf n}_a)\,\delta_{l2}
  \ ,\ \ \ \mbox{\sl PHC}     \label{6.25}
\end{equation}
where the notation

\begin{equation}
  \omega_{m\pm}^2 = \Omega^2\,\left(1\pm\sqrt{\frac{5}{4\pi}}\,
  \left|A_{n2}(R)\right|\,\zeta_m\,\eta^{1/2}\right) + O(\eta)\ ,
  \qquad m=-2,\ldots,2
  \label{6.26}
\end{equation}
has been used. As seen in these formulas, the {\it five\/} expected pairs of
frequencies actually reduce to {\it three\/}, so pentagonal distributions
keep a certain degree of degeneracy, too. The most important distinguishing
characteristic of the general {\it pentagonal\/} layout is best displayed
by the explicit system response:

\begin{eqnarray}
  \hat q_a(s) & = &-\eta^{-1/2}\,\sqrt\frac{4\pi}{5}\,a_{n2} \nonumber \\
              & \times &\left\{\,\frac{1}{2\zeta_0}\left[
  \left(s^2+\omega_{0+}^2\right)^{-1} - \left(s^2+\omega_{0-}^2\right)^{-1}
  \right]\,Y_{20}({\bf n}_a)\,\hat g^{(20)}(s)\right. \nonumber \\
  & + & \;\frac{1}{2\zeta_1}\left[
  \left(s^2+\omega_{1+}^2\right)^{-1} - \left(s^2+\omega_{1-}^2\right)^{-1}
  \right]\,\left[
     Y_{21}({\bf n}_a)\,\hat g^{(11)}(s) +
     Y_{2-1}({\bf n}_a)\,\hat g^{(1\,-1)}(s)\right] \label{6.27}  \\
  & + & \left.\frac{1}{2\zeta_2}\left[
  \left(s^2+\omega_{2+}^2\right)^{-1} - \left(s^2+\omega_{2-}^2\right)^{-1}
  \right]\,\left[
     Y_{22}({\bf n}_a)\,\hat g^{(22)}(s) +
     Y_{2-2}({\bf n}_a)\,\hat g^{(2\,-2)}(s)\right]\right\}
  \nonumber
\end{eqnarray}

This equation indicates that {\it different wave amplitudes selectively
couple to different detector frequencies\/}. This should be considered a
very remarkable fact, for it thence follows that simple inspection of the
system readout {\it spectrum\/}\footnote{
In a noiseless system, of course.}
immediately reveals whether a given wave amplitude $\hat g^{2m}(s)$ is
present in the incoming signal or not.

Pentagonal configurations also admit {\it mode channels\/}, which are
easily constructed from~(\ref{6.27}) thanks to the orthonormality property
of the eigenvectors~(\ref{6.24.b}):

\begin{equation}
  \hat y^{(m)}(s)\equiv\sum_{a=1}^5\,v_a^{(m)*}\hat q_a(s) =
  \eta^{-1/2}\,a_{n2}\,
  \frac{1}{2}\left[\left(s^2+\omega_{m+}^2\right)^{-1} -
  \left(s^2+\omega_{m-}^2\right)^{-1}\right]\,\hat g^{(2m)}(s) + O(0)
  \label{6.28}
\end{equation}

These are almost identical to the {\sl TIGA\/} mode channels~\cite{jm95},
the only difference being that each mode channel comes now at a {\it single
specific\/} frequency pair $\omega_{m\pm}$.

{\it Mode channels\/} are fundamental in signal deconvolution algorithms
in noisy systems~\cite{m98,lms}. Pentagonal resonator configurations should
thus be considered non-trivial candidates for a real GW detector.

\begin{figure}[t]
\centering
\includegraphics[width=15cm]{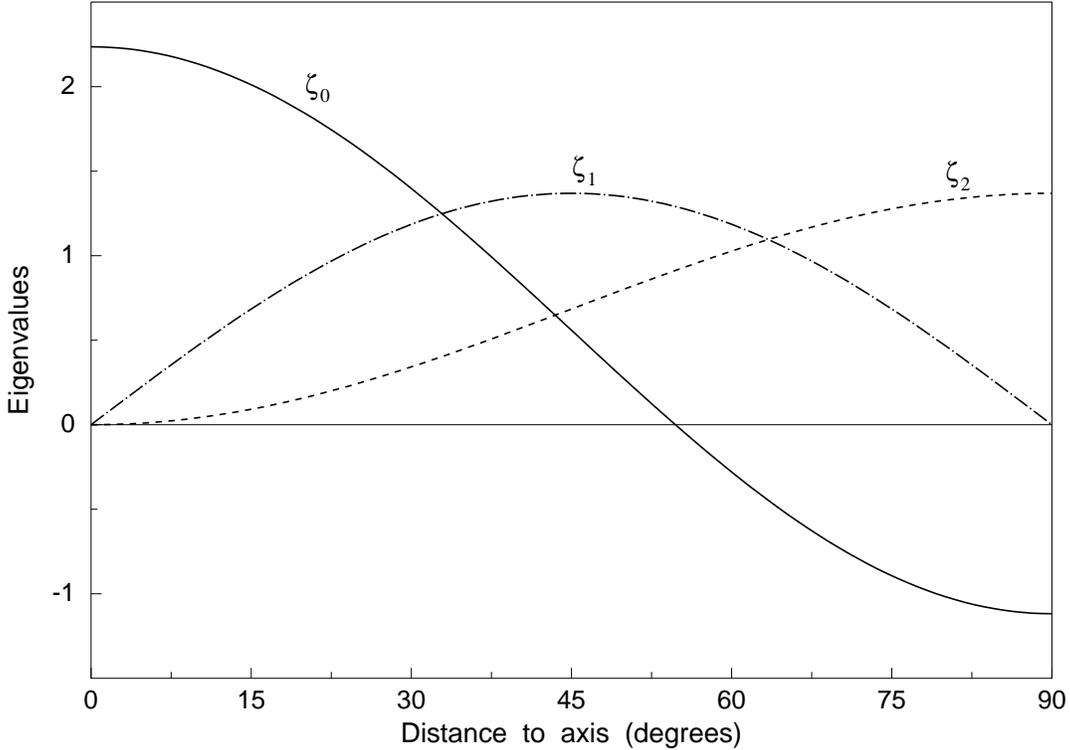}
\caption{The three distinct eigenvalues $\zeta_m\/$ ($m\/$\,=\,0,1,2) as
functions of the distance of the resonator parallel's co-latitude $\alpha\/$
relative to the axis of symmetry of the distribution, cf. equation
(\protect\ref{6.24.a}).
\label{fig3}}
\end{figure}
\begin{figure}
\centering
\includegraphics[width=8.5cm]{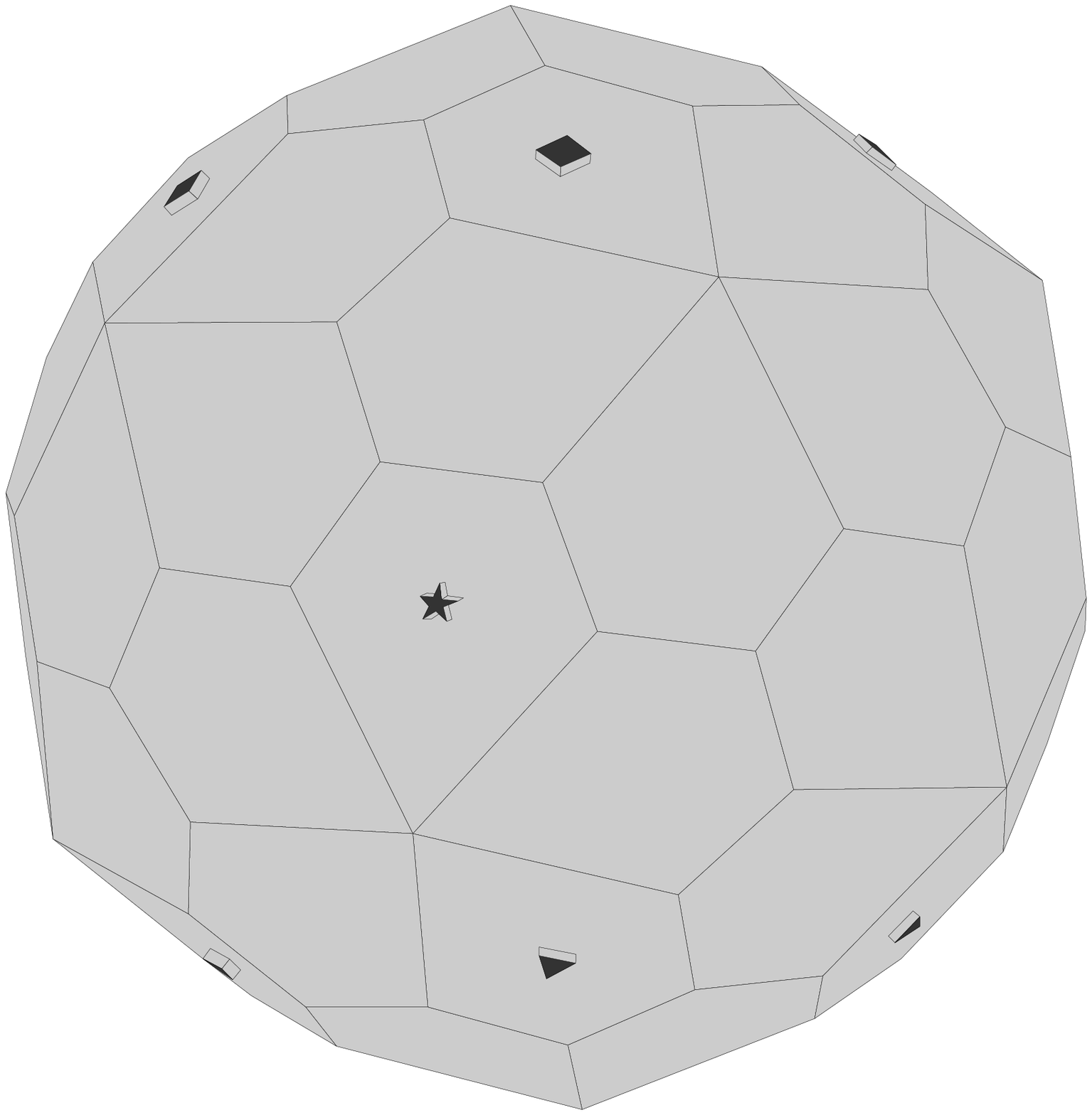} \qquad
\includegraphics[width=7.5cm]{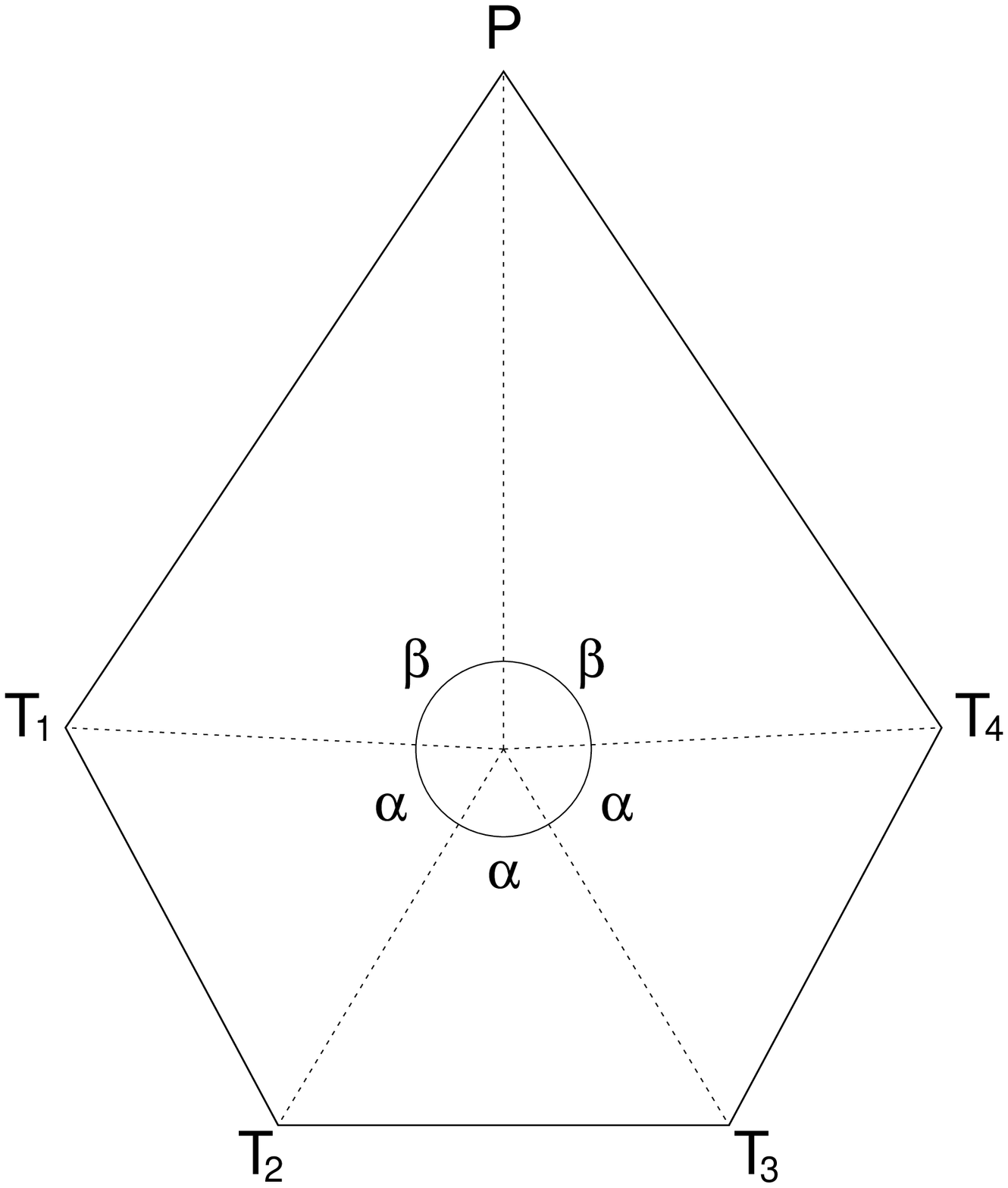}
\caption{To the left, the {\it pentagonal hexacontahedron\/} shape. Certain
faces are marked to indicate resonator positions in a specific proposal
---see text--- as follows: a {\it square\/} for resonators tuned to the
first quadrupole frequency, a {\it triangle\/} for the second, and a
{\it star\/} for the monopole. On the right we see the (pentagonal) face
of the polyhedron. A few details about it: the confluence point of the
dotted lines at the centre is the tangency point of the {\it inscribed\/}
sphere to the {\sl PHC\/}; the labeled angles have values
$\alpha\/$\,=\,61.863$^\circ$, $\beta\/$\,=\,87.205$^\circ$; the angles at
the $T\/$-vertices are all equal, and their value is 118.1366$^\circ$,
while the angle at $P\/$ is 67.4536$^\circ$; the ratio of a long edge
(e.g. $PT_1$) to a short one (e.g. $T_1T_2$) is 1.74985, and the radius of
the inscribed sphere is {\it twice\/} the long edge of the pentagon,
$R\/$\,=\,2\,$PT_1$.	\label{fig4}}
\end{figure}
\begin{figure}
\centering
\includegraphics[width=13cm]{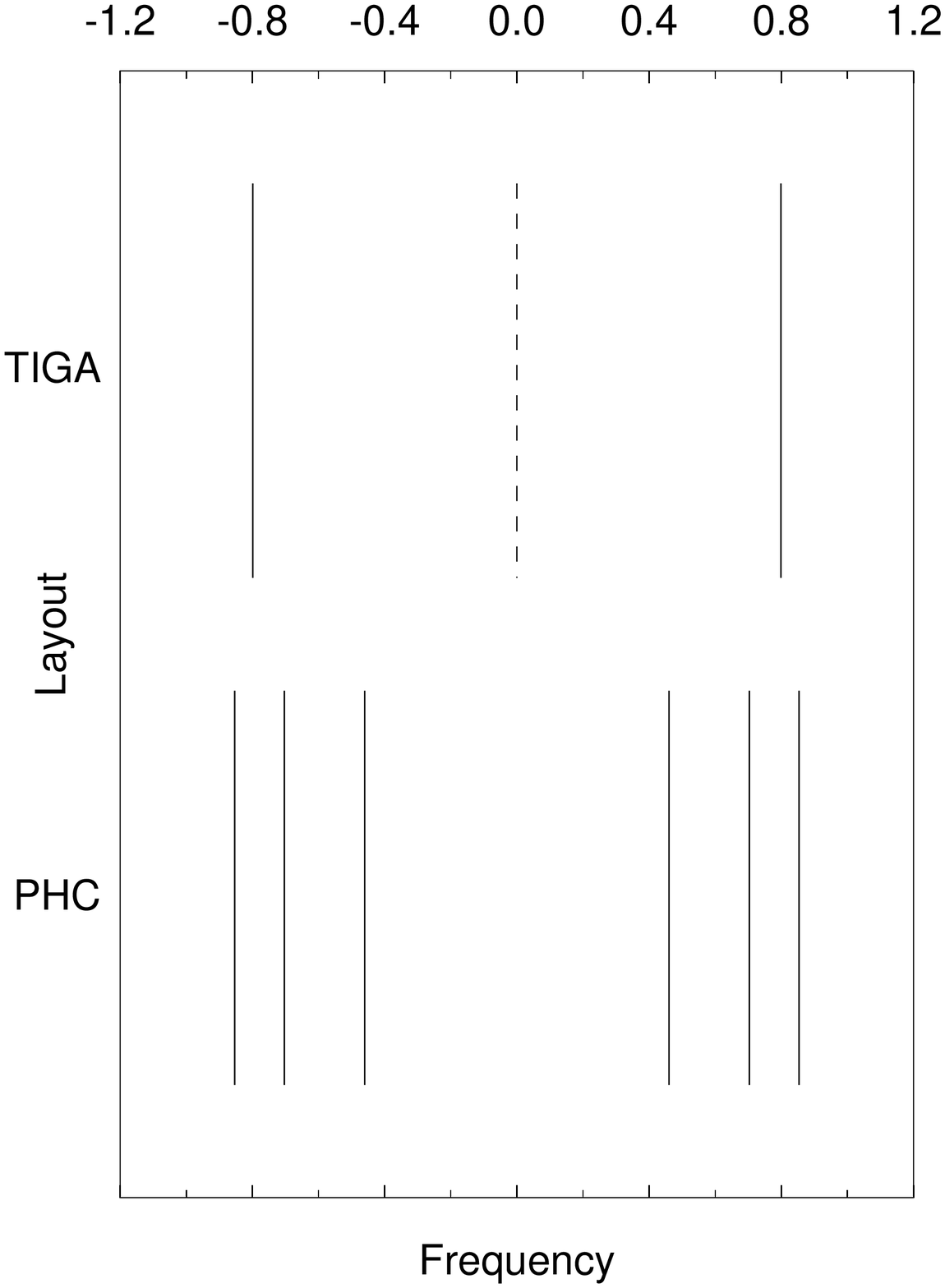}
\caption{Compared line spectrum of a coupled {\sl TIGA\/} and a {\sl PHC\/}
resonator layout in an ideally spherical system. The weakly coupled central
frequency in the {\sl TIGA\/} is drawn dashed. The frequency pair is 5-fold
degenerate for this layout, while the two outer pairs of the {\sl PHC\/}
are doubly degenerate each, and the inner pair is non-degenerate. Units in
abscissas are $\eta^{1/2}\Omega$, and the central value, labeled 0.0,
corresponds to $\Omega$.
\label{fig5}}
\end{figure}

Based on these facts one may next ask which is a suitable transducer
distribution with an axis of pentagonal symmetry. Figure~\ref{fig3} shows
a plot of the eigenvalues~(\ref{6.24.a}) as functions of $\alpha\/$,
the angular distance of the resonator set from the symmetry axis. Several
criteria may be adopted to select a specific choice in view of this graph.
An interesting one can be arrived at by the following argument. If for ease
of mounting, stability, etc., it is desirable to have the detector milled
into a close-to-spherical {\it polyhedric\/} shape\footnote{
This is the philosophy suggested and experimentally implemented by
Merkowitz and Johnson at {\sl LSU\/}.}
then polyhedra with axes of pentagonal symmetry must be searched. The
number of quasi regular {\it convex\/} polyhedra is of course finite
---there actually are only 18 of them~\cite{pacoM,tsvi}---, and I found
a particularly appealing one in the so called {\it pentagonal
hexacontahedron\/} ({\sl PHC\/}), displayed in Figure~\ref{fig4}, left
---see also~\cite{ls}. This is a 60 face polyhedron, whose faces are the
identical {\it irregular pentagons\/} of Figure~\ref{fig4}, right. The
{\sl PHC\/} admits an {\it inscribed sphere\/} which is tangent to each
face at the central point marked in the Figure. It is clearly to this point
that a resonator should be attached so as to simulate an as perfect as
possible spherical distribution.

The {\sl PHC\/} is considerably spherical: the ratio of its volume to that
of the inscribed sphere is 1.057, which quite favourably compares to the
value of 1.153 for the ratio of the circumscribed sphere to the TI volume.
If the frequency pairs $\omega_{m\pm}$ are now requested to be as
{\it evenly spaced\/} as possible, compatible with the {\sl PHC\/} face
orientations, then the choice $\alpha\/$\,=\,67.617$^\circ$ is unambiguously
singled out. Hence

\begin{equation}
 \omega_{0\pm} = \omega_{12}\,\left(1\pm 0.5756\,\eta^{1/2}\right)\ ,\ \ \ 
 \omega_{1\pm} = \omega_{12}\,\left(1\pm 0.8787\,\eta^{1/2}\right)\ ,\ \ \ 
 \omega_{2\pm} = \omega_{12}\,\left(1\pm 1.0668\,\eta^{1/2}\right)
 \label{6.29}
\end{equation}
for instance for $\Omega$\,=\,$\omega_{12}$, the first quadrupole harmonic.
Figure~\ref{fig5} shows this frequency spectrum together with the multiply
degenerate {\it TIGA\/} for comparison.

The criterion leading to the {\sl PHC\/} proposal is of course not unique,
and alternatives can be considered. For example, if the 5 faces of a
regular icosahedron are selected for sensor mounting
($\alpha\/$\,=\,63.45$^\circ$) then a four-fold degenerate pair plus a
single non-degenerate pair is obtained; if the resonator parallel is
50$^\circ$ or 22.6$^\circ$ away from the ``north pole'' then the three
frequencies $\omega_{0+}$, $\omega_{1+}$, and $\omega_{2+}$ are equally
spaced; etc. The number of choices is virtually infinite if the sphere is
not milled into a polyhedric shape~\cite{ts,grg}.

Let me finally recall that the complete {\sl PHC\/} proposal~\cite{ls} was
made with the idea of building an as complete as possible spherical GW
antenna, which amounts to making it sensitive at the first {\it two\/}
quadrupole frequencies {\it and\/} at the first monopole one. This would
take advantage of the good sphere cross section at the second quadrupole
harmonic~\cite{clo}, and would enable measuring (or thresholding)
eventual monopole GW radiation. Now, the system {\it pattern matrix\/}
$\hat\Lambda_a^{(lm)}(s;\Omega)$ has {\it identical structure\/} for all the
harmonics of a given $l\/$ series ---see~(\ref{6.12}) and~(\ref{6.15})---,
and so too identical criteria for resonator layout design apply to either
set of transducers, respectively tuned to $\omega_{12}$ and $\omega_{22}$.
The {\sl PHC\/} proposal is best described graphically in Figure \ref{fig4},
left: a {\it second\/} set of resonators, tuned to the second quadrupole
harmonic $\omega_{22}$ can be placed in an equivalent position in the
``southern hemisphere'', and an eleventh resonator tuned to the first
monopole frequency $\omega_{10}$ is added at an arbitrary position. It is not
difficult to see, by the general methods outlined earlier on in this paper,
that cross interaction between these three sets of resonators is only
{\it second order\/} in $\eta^{1/2}\/$, therefore weak.

A spherical GW detector with such a set of altogether 11 transducers would
be a very complete multi-mode multi-frequency device with an unprecedented
capacity as an individual antenna. Amongst other, it would practically enable
monitoring of coalescing binary {\it chirp\/} signals by means of a rather
robust double passage method~\cite{cf}, a prospect which was considered
so far possible only with broadband long baseline laser
interferometers~\cite{klm1,klm2}, and is almost unthinkable with currently
operating cylindrical bars.

\section{A calibration signal: hammer stroke}
\label{sec:hs}

This section is a brief digression from the main streamline of the paper.
I propose to assess now the system response to a particular, but useful,
calibration signal: a perpendicular {\it hammer stroke\/}.

Let us first go back to equation~(\ref{m3.16}) and replace
$\hat u_a^{\rm external}(s)$ in its rhs with that corresponding to a
hammer stroke, which is easily calculated ---cf.\ appendix~\ref{app:a}:

\begin{equation}
  \hat u_a^{\rm stroke}(s) = -\sum_{nl}\,\frac{f_0}{s^2+\omega_{nl}^2}\,
  \left|A_{nl}(R)\right|^2\,P_l({\bf n}_a\!\cdot\!{\bf n}_0)\ ,\qquad
  a=1,\ldots,J  \label{7.1}
\end{equation}
where ${\bf n}_0$ are the spherical coordinates of the hit point on the
sphere, and $f_0$\,$\equiv$\,${\bf n}_0\!\cdot\!{\bf f}_0/{\cal M\/}$.
Clearly, the hammer stroke excites {\it all\/} of the sphere's vibration
eigenmodes, as it has a completly flat spectrum.

The coupled system resonances are again those calculated in
appendix~\ref{app:b}. The same procedures described in section~\ref{sec:srgw}
for a GW excitation can now be pursued to obtain

\begin{eqnarray}
  \hat q_a(s) & = & \eta^{-1/2}\,(-1)^{J-1}\,\sqrt{\frac{2l+1}{4\pi}}
  \,f_0\,\left|A_{nl}(R)\right|\,\times  \nonumber \\
  & \times & \sum_{b=1}^J\,\left\{\sum_{\zeta_c\neq 0}\,\frac{1}{2}\left[
  \left(s^2+\omega_{c+}^2\right)^{-1}-\left(s^2+\omega_{c-}^2\right)^{-1}
  \right]\,\frac{v_a^{(c)}v_b^{(c)*}}{\zeta_c}\right\}\,
  P_l({\bf n}_b\!\cdot\!{\bf n}_0) + O(0)
  \label{7.2}
\end{eqnarray}
where $a=1,\ldots,J$, when the system is tuned to the $nl\/$-th spheroidal
harmonic, i.e., $\Omega$\,=\,$\omega_{nl}$. It is immediately seen from here
that the system response to this signal when the resonators are tuned to a
{\it monopole\/} frequency is given by

\begin{equation}
  \hat q_a(s) = \eta^{-1/2}\,(-1)^{J-1}\,\frac{f_0}{\sqrt{4\pi J}}\,
  \left|A_{n0}(R)\right|\,\frac{1}{2}\left[\left(
  s^2+\omega_+^2\right)^{-1}-\left(s^2+\omega_-^2\right)^{-1}\right]
  \ ,\qquad \Omega=\omega_{n0}
  \label{7.3}
\end{equation}
an expression which holds for all $a\/$, and is independent of either the
resonator layout or the hit point, which in particular prevents any
determination of the latter, as obviously expected. The frequencies
$\omega_\pm$ are those of~(\ref{6.11}), and we find here again a global
factor $J^{-1/2}$, as also expected.

Consider next the situation when quadrupole tuning is implemented,
$\Omega$\,=\,$\omega_{n2}$. Only the {\sl PHC\/} and {\sl TIGA\/}
configurations will be addressed, as more general cases are not
quite as interesting at this point.

\begin{figure}
\centering
\includegraphics[width=15.5cm]{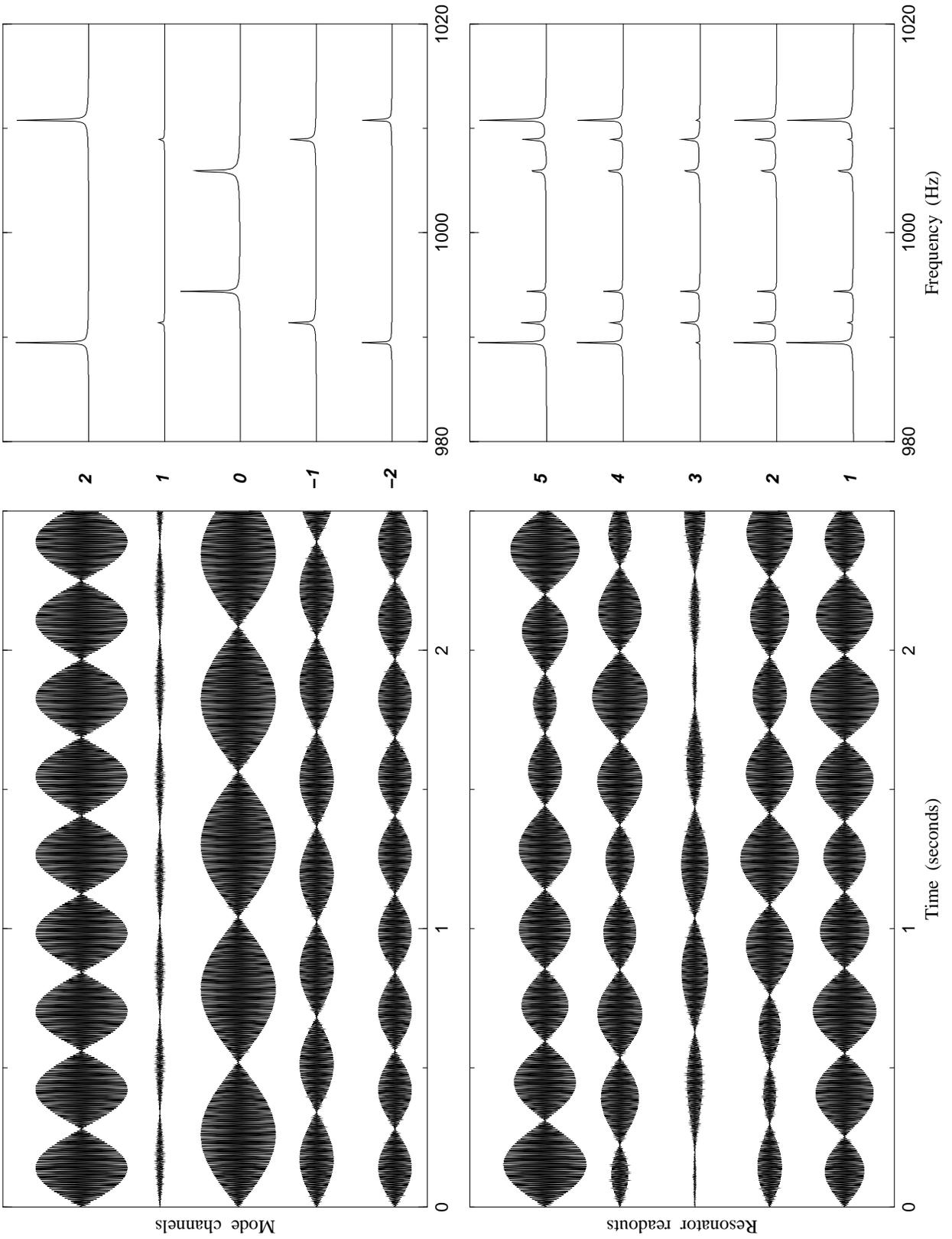}
\caption{Simulated response of a {\sl PHC\/} to a hammer stroke: the time
series and their respective spectra, both for direct resonator readouts and
mode channels. Note that while the former are {\it not\/} simple beats, the
latter are.	\label{fig7}}
\end{figure}

\subsection{{\sl PHC\/} and {\sl TIGA\/} response to a hammer stroke}

Expanding equation~(\ref{7.2}) by substitution of the eigenvalues $\zeta_m\/$
and eigenvectors $v^{(m)}_a\/$ of the {\sl PHC\/}, one readily finds that
the system response is given by

\begin{equation}
  \hat q_a(s) = \eta^{-1/2}\,f_0\,\sqrt{\frac{4\pi}{5}}\,
  \left|A_{n2}(R)\right|\,\sum_{m=-2}^2\,\frac{1}{2}\left[\left(
  s^2+\omega_{m+}^2\right)^{-1}-\left(s^2+\omega_{m-}^2\right)^{-1}\right]
  \,\zeta_m^{-1}\,Y_{2m}({\bf n}_a)\,Y_{2m}^*({\bf n}_0)
  \label{7.7}
\end{equation}
with $a=1,\ldots,5$, and the mode channels by

\begin{equation}
  \hat y^{(m)}(s) = \eta^{-1/2}\,f_0\,\left|A_{n2}(R)\right|
  \frac{1}{2}\left[\left(s^2+\omega_{m+}^2\right)^{-1} -
  \left(s^2+\omega_{m-}^2\right)^{-1}\right]\,Y_{2m}^*({\bf n}_0)
  \ ,\ \ m=-2,\ldots,2	\label{7.8}
\end{equation}

These equations indicate that the system response $q_a(t)$
is a {\it superposition of three different beats\/}\footnote{
A {\it beat\/} is a modulated oscillation of the form
$\sin\frac{1}{2}(\omega_+-\omega_-)t\,\cos\Omega t$, where $\omega_+$
and $\omega_-$ are nearby frequencies, and
$\omega_+$\,$+$\,$\omega_-$\,=\,2\,$\Omega$. The Laplace transform of such
function of time is precisely $(\Omega/2)$\,$\left[\left(
s^2+\omega_+^2\right)^{-1}-\left(s^2+\omega_-^2\right)^{-1}\right]$, up to
higher order terms in the difference $\omega_+$\,$-$\,$\omega_-$, which in
this case is proportional to $\eta^{1/2}$.}, while the mode channels are
{\it single\/} beats each, but with {\it differing modulation frequencies\/}.
This is represented graphically in Figure~\ref{fig7}, where we see the result
of a numerical simulation of the {\sl PHC\/} response to a hammer stroke,
delivered to the solid at a given location. The readouts $q_a(t)$ are
somewhat complex time series, whose frequency spectrum shows {\it three
pairs of peaks\/} ---in fact, the {\it lines\/} in the ideal spectrum of
Figure \ref{fig5}. The mode channels on the other hand are {\it pure
beats\/}, whose spectra consist of the {\it individually separate\/} pairs
of the just mentioned peaks.

The response of the {\sl TIGA\/} layout to a hammer stroke has been described
in detail by Merkowitz and Johnson ---see e.g.\ reference~\cite{jm97}. The
present formalism does of course lead to the results obtained by them; in
the notation of this paper, we have

\begin{deqarr}
\arrlabel{7.4}
  \hat q_a(s) & = & -\eta^{-1/2}\,\frac{5}{\sqrt{24\pi}}\,
  f_0\,\left|A_{n2}(R)\right|\,\frac{1}{2}\left[
  \left(s^2+\omega_+^2\right)^{-1}-\left(s^2+\omega_-^2\right)^{-1}\right]
  \,P_2({\bf n}_a\!\cdot\!{\bf n}_0)
  \label{7.4.a} \\
  \hat y^{(m)}(s) & = & -\eta^{-1/2}\,f_0\,\left|A_{n2}(R)\right|
  \frac{1}{2}\left[\left(s^2+\omega_+^2\right)^{-1} -
  \left(s^2+\omega_-^2\right)^{-1}\right]\,Y_{2m}^*({\bf n}_0)
  \ ,\ \  m=-2,...,2\qquad	\label{7.4.b}
\end{deqarr}
for the system response and the mode channels, respectively, where

\begin{equation}
  \omega_\pm^2 = \omega_{n2}^2\,\left(1\pm\sqrt{\frac{3}{2\pi}}\,
  \left|A_{n2}(R)\right|\eta^{1/2}\right) + O(\eta)\ ,
  \qquad a=1,\ldots,6
  \label{6.19}
\end{equation}
are the five-fold degenerate frequency pairs corresponding to the {\sl TIGA\/}
distribution. Comparison of the mode channels shows that they are identical
for {\sl PHC\/} and {\sl TIGA\/}, except that the former come at different
frequencies depending on the index $m\/$. One might perhaps say that the
{\sl PHC\/} gives rise to a sort of ``Zeeman splitting'' of the {\sl TIGA\/}
degenerate frequencies, which can be attributed to an {\it axial symmetry
breaking\/} of that resonator distribution: the {\sl PHC\/} mode channels
partly split up the otherwise degenerate multiplet into its components.

\section{Symmetry defects}
\label{sec:symdef}

So far we have made the assumption that the sphere is perfectly symmetric,
that the resonators are identical, that their locations on the sphere's
surface are ideally accurate, etc. This is of course unrealistic. So I
propose to address now how departures from such ideal conditions affect the
system behaviour. As we shall see, the system is rather {\it robust\/},
in a sense to be made precise shortly, against a number of small defects.

In order to {\it quantitatively\/} assess ideality failures I shall adopt
a philosophy which is naturally suggested by the results already obtained
in an ideal system. It is as follows.

As seen in previous sections, the solution to the general
equations~(\ref{m3.16}) must be given as a {\it perturbative\/} series
expansion in ascending powers of the small quantity $\eta^{1/2}$. This
is clearly a fact {\it not\/} related to the system's symmetries, so it
will survive symmetry breakings. It is therefore appropriate to
{\it parametrise\/} deviations from ideality in terms of suitable powers
of $\eta^{1/2}$, in order to address them {\it consistently with the order
of accuracy of the series solution to the equations of motion\/}. An
example will better illustrate the situation.

In a {\it perfectly ideal\/} spherical detector the system frequencies
are given by equations~(\ref{5.2}). Now, if a small departure from e.g.
spherical symmetry is present in the system then we expect that a
correspondingly small correction to those equations will be required.
Which specific correction to the formula will actually happen can be
{\it qualitatively\/} assessed by a {\it consistency\/} argument: if
symmetry defects are of order $\eta^{1/2}$ then equations~(\ref{5.2}) will
be significantly altered in their $\eta^{1/2}$ terms; if on the other hand
such defects are of order $\eta\/$ or smaller then any modifications to
equations~(\ref{5.2}) will be swallowed into the $O(0)$ terms, and the
more important $\eta^{1/2}$ terms will remain unaffected by the symmetry
failure. One can say in this case that the system is {\it robust\/}
against that symmetry breaking.

More generally, this argument can be extended to see that the only system
defects standing a chance to have any influences on lowest order ideal
system behaviour are defects of order $\eta^{1/2}$ relative to an ideal
configuration. Defects of such order are however {\it not necessarily
guaranteed\/} to be significant, and a specific analysis is required for
each concrete parameter in order to see whether or not the system response
is {\it robust\/} against the considered parameter deviations. Let us now
go into the quantitative detail.

Let $P\/$ be one of the system parameters, e.g. a sphere frequency, or a
resonator mass or location, etc. Let $P_{\rm ideal}$ be the {\it numerical
value\/} this parameter has in an ideal detector, and let $P_{\rm real}$
be its value in the real case. These two will be assumed to differ by terms
of order $\eta^{1/2}$, i.e.,

\begin{equation}
  P_{\rm real} = P_{\rm ideal}\,(1+p\,\eta^{1/2})     \label{8.1}
\end{equation}

For a given system, $p\/$ is readily determined adopting~(\ref{8.1}) as the
{\it definition\/} of $P_{\rm real}$, once a suitable {\it hypothesis\/}
has been made as to which is the value of $P_{\rm ideal}$. In order for
the following procedure to make sensible sense it is clearly required that
$p\/$ be of order 1 or, at least, appreciably larger than $\eta^{1/2}$.
Should $p\/$ thus calculated from~(\ref{8.1}) happen to be too small, i.e.,
of order $\eta^{1/2}$ itself or smaller, then the system will be considered
{\it robust\/} as regards the affected parameter.


\subsection{The suspended sphere  \label{s8.1}}

An earth based observatory obviously requires a {\it suspension
mechanism\/} for the large sphere. If a {\it nodal point\/} suspension
is e.g.\ selected then a diametral {\it bore\/} has to be drilled across
the sphere~\cite{phd}. The most immediate consequence of this is that
spherical symmetry is broken, what in turn results in {\it degeneracy
lifting\/} of the free spectral frequencies $\omega_{nl}\/$, which now
{\it split\/} up into multiplets $\omega_{nlm}\/$
($m\/$\,=\,$-l\/$,...,$l\/$). The resonators' frequency $\Omega$
{\it cannot\/} therefore be matched to {\it the\/} frequency
$\omega_{n_0l_0}$, but at most to {\it one\/} of the
$\omega_{n_0l_0m}\/$'s. In this subsection I keep the hypothesis
---to be relaxed later, see below--- that all the resonators are identical,
and assume that $\Omega$ falls {\it within\/} the span of the multiplet
of the $\omega_{n_0l_0m}\/$'s. Then

\begin{equation}
  \omega_{n_0l_0m}^2 = \Omega^2\,(1+p_m\,\eta^{1/2})\ ,\qquad
  m=-l_0,\ldots,l_0     \label{8.2}
\end{equation}

The coupled frequencies, i.e., the roots of equation~(\ref{m3.18}), will
now be searched. The kernel matrix $\hat K_{ab}(s)$ is however no longer
given by~(\ref{m4.2}), due the removed degeneracy of $\omega_{nl}\/$, and
we must stick to its general expression~(\ref{m3.17}), or

\begin{equation}
  \hat K_{ab}(s) = \sum_{nlm}\,\frac{\Omega_b^2}{s^2+\omega_{nlm}^2}\,
   \left|A_{nl}(R)\right|^2\,\frac{2l+1}{4\pi}\,
   Y_{lm}^*({\bf n}_a)\,Y_{lm}({\bf n}_b) \equiv
   \sum_{nlm}\,\frac{\Omega_b^2}{s^2+\omega_{nlm}^2}\,\chi_{ab}^{(nlm)}
   \label{8.3}
\end{equation}

Following the steps of appendix \ref{app:a} we now seek the roots of the
equation

\begin{equation}
  \det\,\left[\delta_{ab} + \eta\,\sum_{m=-l_0}^{l_0}\,
   \frac{\Omega^2s^2}{(s^2+\Omega^2)(s^2+\omega_{n_0l_0m}^2)}
   \,\chi_{ab}^{(n_0l_0m)} + \eta\,\sum_{nl\neq n_0l_0,m}\,
   \frac{\Omega^2s^2}{(s^2+\Omega^2)(s^2+\omega_{nlm}^2)}\,
   \chi_{ab}^{(nlm)}\right] = 0
  \label{8.4}
\end{equation}

Since $\Omega$ relates to $\omega_{n_0l_0m}\/$ through equation~(\ref{8.2})
we see that the roots of~(\ref{8.4}) fall again into either of the two
categories~(\ref{m4.11}) (see Appendix~\ref{app:b}), i.e., roots close
to $\pm i\Omega$ and roots close to $\pm i\omega_{nlm}\/$
($nl\/$\,$\neq$\,$n_0l_0$). I shall exclusively concentrate on the former
now. Direct substitution of the series~(\ref{4.11.a}) into~(\ref{8.4})
yields the following equation for the coefficient $\chi_{\frac{1}{2}}$:

\begin{equation}
  \det\left[\delta_{ab} - \frac{1}{\chi_\frac{1}{2}}\,\sum_{m=-l_0}^{l_0}
  \,\frac{\chi_{ab}^{(n_0l_0m)}}{\chi_\frac{1}{2}-p_m}\right] = 0
  \label{8.5}
\end{equation}

This is a variation of~(\ref{5.1}), to which it reduces when
$p_m\/$\,=\,0, i.e., when there is full degeneracy.

The solutions to~(\ref{8.5}) no longer come in symmetric pairs,
like~(\ref{5.2}). Rather, there are 2$l_0$+1+$J\/$ of them, with a
{\it maximum\/} number of 2(2$l_0$+1) non-identically zero roots if
$J\/$\,$\geq$\,2$l_0$+1\footnote{
This is a {\it mathematical fact\/}, whose proof is relatively cumbersome,
and will be omitted here; let me just mention that it has its origin in the
linear dependence of more than 2$l_0$+1 spherical harmonics of order $l_0$.}.
For example, if we choose to select the resonators' frequency close to a
quadrupole multiplet ($l_0$\,=\,2) then~(\ref{8.5}) has at most 5+$J\/$
non-null roots, {\it with a maximum ten\/} no matter how many resonators
in excess of 5 we attach to the sphere. Modes associated to null roots
of~(\ref{8.5}) can be seen to be {\it weakly coupled\/}, just like in a
free sphere, i.e., their amplitudes are smaller than those of the strongly
coupled ones by factors of order $\eta^{1/2}$.

In order to assess the reliability of this method I have applied it to
see what are its predictions for a {\it real system\/}. To this end, data
taken with the {\sl TIGA\/} prototype at {\sl LSU\/}\footnote{
These data are contained in reference~\protect\cite{phd}; I want to thank
Stephen Merkowitz for kindly handing them to me.}
were used to confront with. The {\sl TIGA\/} was drilled and suspended
from the centre, so its first quadrupole frequency split up into a
multiplet of five frequencies. Their reportedly measured values are

\begin{equation}
  \omega_{120} = 3249\ {\rm Hz}\ ,\ \ 
  \omega_{121} = 3238\ {\rm Hz}\ ,\ \ 
  \omega_{12\,-1} = 3236\ {\rm Hz}\ ,\ \ 
  \omega_{122} = 3224\ {\rm Hz}\ ,\ \ 
  \omega_{12\,-2} = 3223\ {\rm Hz}\ ,\ \ 
  \label{8.6}
\end{equation}

All 6 resonators were equal, and had the following characteristic
frequency and mass, respectively:

\begin{equation}
  \Omega = 3241\ {\rm Hz}\ ,\qquad\eta = \frac{1}{1762.45}
  \label{8.7}
\end{equation}

Substituting these values into~(\ref{8.2}) it is seen that

\begin{equation}
  p_0=0.2075\ ,\ \   p_1=-0.0777\ ,\ \   p_{-1}=-0.1036\ ,\ \ 
  p_2=-0.4393\ ,\ \  p_{-2}=-0.4650
  \label{8.8}
\end{equation}

\begin{table}
\label{t1}
\caption{Numerical values of measured and theoretically predicted
frequencies (in Hz) for the {\sl TIGA\/} prototype with varying number
of resonators. Relative errors are also shown as parts in 10$^4$. The
{\it calculated\/} values of the tuning and free multiplet frequencies
are taken {\it by definition\/} equal to the measured ones, and quoted
in brackets. In square brackets the frequency of the {\it weakly coupled\/}
sixth mode in the full, 6~resonator {\sl TIGA\/} layout. These data are
plotted in Figure~\protect\ref{fig8}.}

\begin{center}
\begin{tabular}{lccc||lccc}
Descr. & Meas. & Calc.  & \begin{tabular}{cc} Difference \\
					  (parts in 10$^4$) \end{tabular} &
Descr. & Meas. & Calc.  & \begin{tabular}{cc} Difference \\
					  (parts in 10$^4$) \end{tabular} \\
\hline Tuning & 3241 & (3241) & (0) &
4 reson. & 3159 & 3155 & $-12$ \\
No reson. & 3223 & (3223)  & (0) &
             & 3160 & 3156 & $-11$ \\
               & 3224 & (3224)  & (0) &
             & 3168 & 3165 & $-12$ \\
               & 3236 & (3236)  & (0) &
             & 3199 & 3198 & $-5$ \\
               & 3238 & (3238)  & (0) &
             & 3236 & 3236 & $ 0$ \\
               & 3249 & (3249)  & (0) &
             & 3285 & 3286 & $ 3$ \\
1 reson.     & 3167 & 3164 & $-8$ &
             & 3310 & 3310 & $ 0$ \\
             & 3223 & 3223 & $ 0$ &
             & 3311 & 3311 & $ 0$ \\
             & 3236 & 3235 & $-2$ &
             & 3319 & 3319 & $ 0$ \\
             & 3238 & 3237 & $-2$ &
5 reson.     & 3152 & 3154 & $ 8$ \\
             & 3245 & 3245 & $ 0$ &
             & 3160 & 3156 & $-14$ \\
             & 3305 & 3307 & $ 6$ &
             & 3163 & 3162 & $-3$ \\
2 reson.     & 3160 & 3156 & $-13$ &
             & 3169 & 3167 & $-8$ \\
             & 3177 & 3175 & $-7$ &
             & 3209 & 3208 & $-2$ \\
             & 3233 & 3233 & $ 0$ &
             & 3268 & 3271 & $ 10$ \\
             & 3236 & 3236 & $ 0$ &
             & 3304 & 3310 & $ 17$ \\
             & 3240 & 3240 & $ 0$ &
             & 3310 & 3311 & $ 3$ \\
             & 3302 & 3303 & $ 3$ &
             & 3313 & 3316 & $ 10$ \\
             & 3311 & 3311 & $ 0$ &
             & 3319 & 3321 & $ 6$ \\
3 reson.     & 3160 & 3155 & $-15$ &
6 reson.     & 3151 & 3154 & $ 11$ \\
             & 3160 & 3156 & $-13$ &
             & 3156 & 3155 & $-3$ \\
             & 3191 & 3190 & $-2$ &
             & 3162 & 3162 & $ 0$ \\
             & 3236 & 3235 & $-2$ &
             & 3167 & 3162 & $-14$ \\
             & 3236 & 3236 & $ 0$ &
             & 3170 & 3168 & $-7$ \\
             & 3297 & 3299 & $ 8$ &
             & [3239] & [3241] & [6] \\
             & 3310 & 3311 & $ 2$ &
             & 3302 & 3309 & $ 23$ \\
             & 3311 & 3311 & $ 0$ &
             & 3308 & 3310 & $ 6$ \\
             &  &  &  &
             & 3312 & 3316 & $ 12$ \\
             &  &  &  &
             & 3316 & 3317 & $ 2$ \\
             &  &  &  &
             & 3319 & 3322 & $ 10$
\end{tabular}
\end{center}
\end{table}

\begin{figure}
\centering
\includegraphics[width=15.3cm]{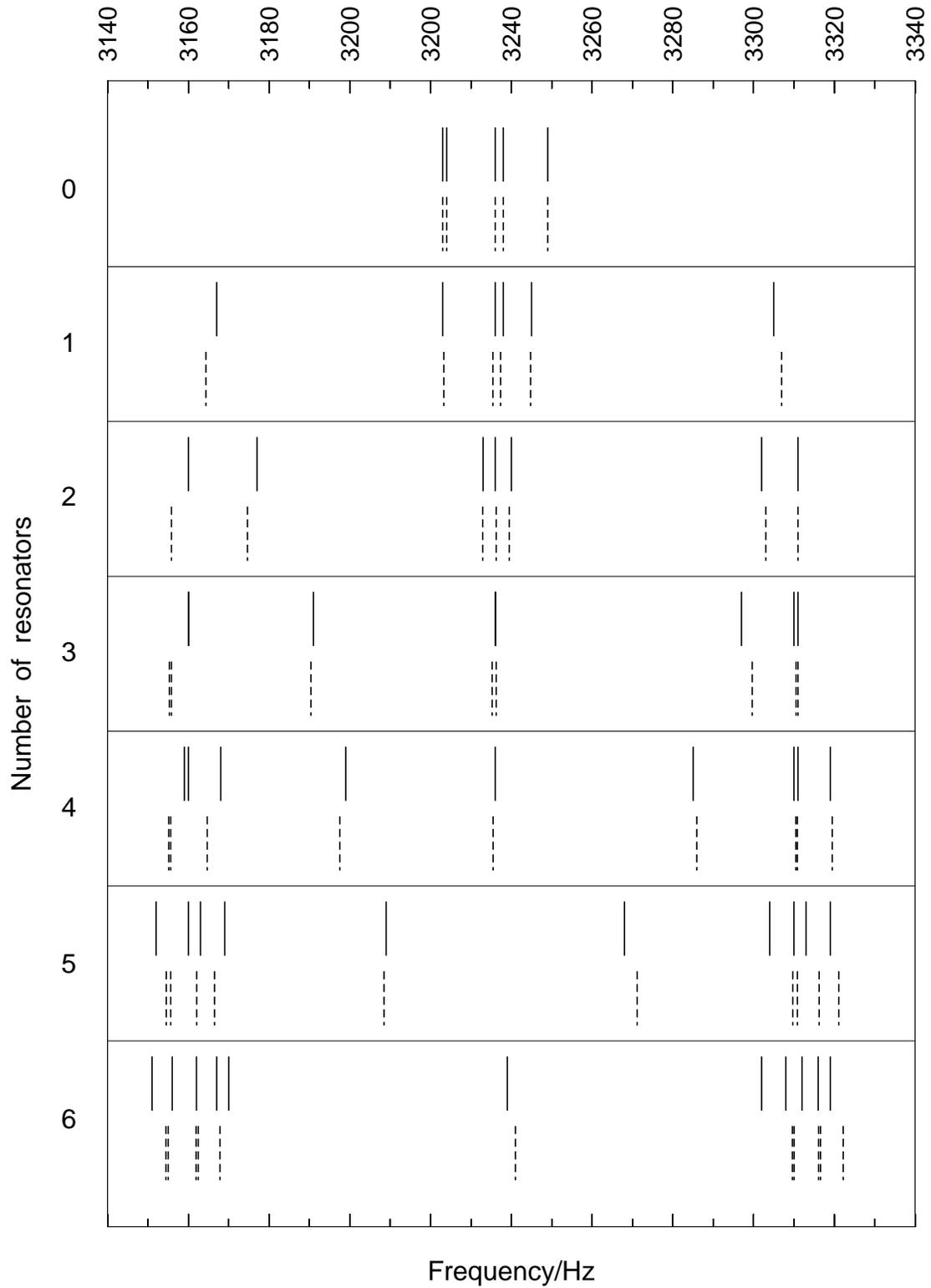}
\caption{The frequency spectrum of the {\sl TIGA\/} distribution as
resonators are progressively added from none to 6. Continuous lines
correspond to measured values, and dashed lines correspond to their
$\eta^{1/2}$ theoretical estimates with equation~(\protect\ref{8.5}).
\label{fig8}}
\end{figure}

Equation~(\ref{8.5}) can now be readily solved, once the resonator
positions are fed into the matrices $\chi_{ab}^{(12m)}$. Such positions
correspond to the pentagonal faces of a truncated icosahedron.
Merkowitz~\cite{phd} gives a complete account of all the measured system
frequencies as resonators are progressively attached to the selected faces,
beginning with one and ending with six. Figure~\ref{fig8} graphically
displays the experimentally reported frequencies along with those calculated
theoretically by solving equation~(\ref{8.5}). In Table~1 I give the
numerical values. As can be seen, coincidence between theoretical
predictions and experimental data is remarkable: the worst error is 0.2\%,
while for the most part it is below 0.1\%. This is a few parts in 10$^4$,
which is precisely the magnitude of $\eta$, as specified in
equation~(\ref{8.7}).

Therefore {\it discrepancies between theoretical predictions and experimental
data are exactly as expected\/}, i.e., of order $\eta$. In addition, it
is also reported in reference~\cite{jm97} that the 11-th, weakly coupled
mode of the {\sl TIGA\/} (enclosed in square brackets in Table 1) has a
practically zero amplitude, again in excellent agreement with the general
theoretical predictions about modes beyond the tenth ---see paragraph
after equation~(\ref{8.5}).

This is a remarkable result which encouraged a better fit by estimates
of {\it next order\/} corrections, i.e., $\chi_1$ of~(\ref{4.11.a}). As
it turned out, however, matching between theory and experiment does not
consistently improve in the next step. This is not really that surprising,
though, as M\&J explicitly state~\cite{jm97} that control of the general
experimental conditions in which data were obtained had a certain degree of
tolerance, and they actually show satisfaction that $\sim$1\% coincidence
between theory and measurement is comfortably accomplished. But 1\% is
{\it two orders of magnitude larger than $\eta\/$} ---cf. equation
(\ref{8.7})---, so failure to refine our frequency estimates to order
$\eta\/$ is again fully consistent with the accuracy of available real data.

A word on a technical issue is in order. Merkowitz and Johnson's equations
for the {\sl TIGA\/}~\cite{jm93,jm95} are identical to the equations in this
paper to lowest order in $\eta\/$. Remarkably, though, their reported
theoretical estimates of the system frequencies are not quite as accurate
as those in Table~1~\cite{jm97}. The reason is probably this: in M\&J's
model these frequencies appear within an algebraic system of 5+$J\/$ linear
equations with as many unknowns which has to be solved; here instead the
algebraic system has only $J\/$ equations and unknowns, actually
equations~(\ref{m3.16}). This is a very appreciable difference for the range
of values of $J\/$ under consideration. While the roots for the frequencies
{\it mathematically\/} coincide in both approaches, in actual practice they
are {\it estimated\/}, generally by means of computer programmes. It is here
that problems most likely arise, for the numerical reliability of an
algorithm to solve matrix equations normally decreases as the rank of the
matrix increases.

\subsection{Other mismatched parameters}

We now assess the system sensitivity to small mismatches in resonators'
masses, locations and frequencies.

\subsubsection{Resonator mass mismatches}

If the {\it masses\/} are slightly non-equal then one can write

\begin{equation}
  M_a = \eta{\cal M}\,(1+\mu_a\,\eta^{1/2})\ ,\qquad a=1,\ldots,J
  \label{8.9}
\end{equation}
where $\eta\/$ can be defined e.g. as the ratio of the {\it average\/}
resonator mass to the sphere's mass. It is immediately obvious from
equation~(\ref{8.9}) that mass non-uniformities of the resonators only
affect the main equations in {\it second order\/}, since resonator mass
non-uniformities result, as we see, in corrections of order $\eta^{1/2}$
to $\eta^{1/2}$ itself, which is the very parameter of the perturbative
expansions. The system is thus clearly {\it robust\/} to mismatches in
the resonator masses of the type~(\ref{8.9}).

\subsubsection{Errors in resonator locations}

The same happens if the {\it locations\/} of the resonators have tolerances
relative to a {\it pre-selected\/} distribution. For let ${\bf n}_a\/$ be a
set of resonator locations, for example the {\sl TIGA\/} or the {\sl PHC\/}
positions, and let ${\bf n'}_a\/$ be the real ones, close to the former:

\begin{equation}
  {\bf n'}_a = {\bf n}_a + {\bf v}_a\,\eta^{1/2}\ ,\qquad a=1,\ldots,J
  \label{8.10}
\end{equation}

The values ${\bf n}_a$ determine the eigenvalues $\zeta_a\/$ in
equation~(\ref{5.2}), and they also appear as arguments to the
spherical harmonics in the system response functions of
sections~\ref{sec:srgw}--\ref{sec:hs}. It follows
from~(\ref{8.10}) by continuity arguments that

\begin{deqarr}
\arrlabel{8.105}
  Y_{lm}({\bf n'}_a) & = & Y_{lm}({\bf n}_a) + O(\eta^{1/2})
  \label{8.105.a} \\
  \zeta'_a & = & \zeta_a + O(\eta^{1/2})
  \label{8.105.b}
\end{deqarr}

Inspection of the equations of sections~\ref{sec:srgw}--\ref{sec:hs}
shows that both $\zeta_a\/$ and $Y_{lm}({\bf n}_a)$ {\it always\/}
appear within lowest order terms, and hence that corrections to them of
the type~(\ref{8.105}) will affect those terms in {\it second order\/}
again. We thus conclude that the system is also {\it robust\/} to small
misalignments of the resonators relative to pre-established positions.

\subsubsection{Resonator frequency mistunings}

The resonator {\it frequencies\/} may also differ amongst them, so let

\begin{equation}
  \Omega_a = \Omega\,(1+\rho_a\,\eta^{1/2})\ ,\qquad a=1,\ldots,J
  \label{8.11}
\end{equation}

To assess the consequences of this, however, we must go back to equation
(\ref{3.18}) and see what the coefficients in its series solutions of the
type~(\ref{4.11.a}) are. The procedure is very similar to that of
section~\ref{s8.1}, and will not be repeated here; the lowest order
coefficient $\chi_\frac{1}{2}$ is seen to satisfy the algebraic equation

\begin{equation}
  \det\left[\delta_{ab} - \frac{1}{\chi_\frac{1}{2}}\,\sum_{c=0}^{J}\,
  \frac{\chi_{ac}^{(n_0l_0)}\,\delta_{cb}}{\chi_\frac{1}{2}-\rho_c}\right]
  = 0	  \label{8.12}
\end{equation}
which reduces to~(\ref{5.1}) when all the $\rho\/$'s vanish, as expected.
This appears to potentially have significant effects on our results to
lowest order in $\eta^{1/2}$, but a more careful consideration of the facts
shows that it is probably unrealistic to think of such large tolerances in
resonator manufacturing as implied by equation~(\ref{8.11}) in the first
place. In the {\sl TIGA\/} experiment, for example~\cite{phd}, an error of
order $\eta^{1/2}$ would amount to around 50 Hz of mistuning between
resonators, an absurd figure by all means. In a full scale sphere
($\sim$40 tons, $\sim$3 metres in diameter, $\sim$800 Hz fundamental
quadrupole frequency, $\eta\/$\,$\sim$\,10$^{-5}$) the same error would
amount to between 5 Hz and 10 Hz in resonator mistunings for the lowest
frequency. This is probably excessive for a capacitive transducer, but may
be realistic for an inductive one. With this exception, it is thus more
appropriate to consider that resonator mistunings are at least of order
$\eta\/$. If this is the case, though, we see once more that the system
is quite insensitive to such mistunings.

Summing up the results of this section, one can say that the resonator
system dynamics is quite {\it robust\/} to small (of order $\eta^{1/2}$)
changes in its various parameters. The important exception is of course
the effect of suspension drilling, which do result in significant changes
relative to the ideally perfect device, but which can be relatively easily
calculated. The theoretical picture is fully supported by experiment, as
{\it robustness\/} in the parameters here considered has been reported in
the real device~\cite{jm97}.

\section{Conclusions}

A spherical GW antenna is a natural multi-mode device with very rich
potential capabilities to detect GWs on earth. But such detector is not
just a bare sphere, it requires a set of {\it motion sensors\/} to be
practically useful. It appears that transducers of the {\it resonant\/}
type are the best suited ones for an efficient performance of the detector.
Resonators however significantly interact with the sphere, and they affect
in particular its frequency spectrum and vibration modes in a specific
fashion, which must be properly understood before reliable conclusions
can be drawn from the system readout.

The main objective of this paper has been the construction and development
of an elaborate theoretical model to describe the joint dynamics of a solid
elastic sphere and a set of {\it radial motion\/} resonators attached to
its surface at arbitrary locations, with the purpose to make predictions
of the system characteristics and response, in principle with arbitrary
mathematical precision.

The solutions to the equations of motion have been shown to be expressible
as an ascending series in powers of the small ``coupling constant''
$\eta\/$, the ratio of the average resonator mass to the mass of the larger
sphere. The {\it lowest order\/} approximation corresponds to terms of order
$\eta^{1/2}$ and, to this order, previous results~\cite{jm97,ts,grg} are
recovered. This, I hope, should contribute to clarify the nature of the
approximations inherent in earlier approaches, and to better understand
the physical reason for their remarkable accuracy~\cite{jm97}.

In addition, the methods of this paper have permitted us to discover
that there can be in fact transducer layouts alternative to the highly
symmetric {\sl TIGA\/}, and having potentially interesting practical
properties. An example is the {\sl PHC\/} distribution, which is based
on a pentagonally symmetric set of 5 rather than 6 resonators per
quadruopole mode sensed. This transducer distribution has the property
that {\it mode channels\/} can be constructed from the resonators'
readouts, much in the same way as in the {\it TIGA\/}~\cite{jm95}. In the
{\sl PHC\/} however a new and distinctive characteristic is present: different
{\it wave amplitudes\/} selectively couple to different {\it detector modes\/}
having different frequencies, so that the antenna's mode channels come at
different rather than equal frequencies. The {\sl PHC\/} philosophy can be
extended to make a {\it multi-frequency\/} system by using resonators tuned
to the first two quadrupole harmonics of the sphere {\it and\/} to the first
monopole, an altogether 11 transducer set.

The assessment of {\it symmetry failure\/} effects, as well as other
parameter departures form ideality, has also been subjected to analysis.
The general scheme is again seen to be very well suited for the purpose,
as the theory transparently shows that the system is {\it robust\/} against
relative disturbances of order $\eta\/$ or smaller in any system parameters,
also providing a systematic procedure to assess larger tolerances ---up to
order $\eta^{1/2}$. The system is shown to still be robust to tolerances of
this order in some of its parameters, whilst it is not to others. Included
in the latter group is the effect of spherical symmetry breaking due to
system suspension in the laboratory, which causes {\it degeneracy lifting\/}
of the sphere's eigenfrequencies, which split up into multiplets. A strong
point is that, by use of mostly analytic algorithms, it has been possible to
accurately reproduce the reportedly measured frequencies of the {\sl LSU\/}
prototype antenna~\cite{phd} with the predicted precision of four decimal
places. The also reported robustness of the system to resonator
mislocations~\cite{jm97} is too in satisfactory agreement with the
theoretical predictions.

The perturbative approach here adopted is naturally open to refined
analysis of the system response in higher orders in $\eta\/$. For example,
one can systematically address the weaker coupling of non-quadrupole modes,
etc. It appears however that such refinements will be largely masked by
{\it noise\/} in a real system, as shown by Merkowitz and Johnson~\cite{jm98},
and this must therefore be considered first. So the next step is to include
noise in the model and see its effect. Stevenson~\cite{ts} has already made
some progress in this direction, and partly assessed the characteristics of
{\sl TIGA\/} and {\sl PHC\/}, but more needs to be done since not too high
signal-to-noise ratios should realistically be be considered in an actual
GW detector. In particular, {\it mode channels\/} are at the basis of noise
correlations and dependencies, as well as the errors in GW parameter
estimation~\cite{lms}. I do expect the analytic tools developed in this
article to provide a powerful framework to address the fundamental problems
of {\it noise\/} in a spherical GW antenna which, to my knowledge, have not
yet received the detailed attention they require.

\section*{Acknowledgments}
\label{ack2}

I am greatly indebted with Stephen Merkowitz both for his kind supply
of the {\sl TIGA\/} prototype data, and for continued encouragement
and illuminating discussions. I am also indebted with Curt Cutler for
addressing my attention to an initial error in the general equations
of section~2, and with Eugenio Coccia for many discussions. M.A.\ Serrano
gave me valuable help in some of the calculations of Appendix~\ref{app:b}
below, and this is gratefully acknowledged, too. I have received financial
support from the Spanish Ministry of Education through contract number
PB96-0384, and from Institut d'Estudis Catalans.

\section*{Appendices}

\renewcommand{\thesection}{\Alph{section}}
\setcounter{section}{1}	   

\subsection{Green functions for the multiple resonator system}
\label{app:a}

The density of forces in he rhs of equation~(\ref{2.1.a}) happens to be
of the {\it separable\/} type

\begin{equation}
  {\bf f}({\bf x},t) = \sum_\alpha\,{\bf f}^{(\alpha)}({\bf x})\,
   g^{(\alpha)}(t)   \label{m3.1}
\end{equation}
where $\alpha\/$ is a suitable label. It is  recalled from
reference~\cite{lobo} that, in such circumstances, a formal
solution can be written down for equation~(\ref{2.1.a}) in
terms of a {\it Green function integral\/}, whereby the
following orthogonal series expansion obtains:

\begin{equation}
   {\bf u}({\bf x},t) = \sum_\alpha\sum_N\,\omega_N^{-1}\,f_N^{(\alpha)}\,
   {\bf u}_N({\bf x})\,g_N^{(\alpha)}(t)      \label{m3.2}
\end{equation}
where

\begin{deqarr}
\arrlabel{m3.3}
  f_N^{(\alpha)} & \equiv & \frac{1}{\cal M}\,\int_{\rm Sphere}
  {\bf u}_{N}^*({\bf x})\cdot{\bf f}^{(\alpha)}({\bf x})\,d^3x
  \label{3.3.a} \\[0.5 em]
  g_N^{(\alpha)}(t) & \equiv & \int_0^t g^{(\alpha)}(t')\,\sin\omega_N (t-t')
  \,dt'   \label{3.3.b}
\end{deqarr}

Here, $\omega_N\/$ and ${\bf u}_N({\bf x})$ are the eigenfrequencies and
associated normalised wave-functions of the free sphere. Also, $N\/$ is an
abbreviation for a multiple index $\{nlm\}$. The generic index $\alpha$
is a label for the different pieces of interaction happening in the system.
I quote the result of the calculations of the terms needed in this paper:

\begin{deqarr}
\arrlabel{m3.4}
 f_{{\rm resonators,} N}^{(a)} & = & \frac{M_a}{\cal M}\,\Omega_a^2\,\,
  \left[{\bf n}_a\!\cdot\!{\bf u}_N^*({\bf x}_a)\right]\ ,\qquad a=1,\ldots,J
  \label{3.4.a} \\[0.5 em]
 f_{{\rm GW,}N}^{(l'm')} & = & a_{nl}\,\delta_{ll'}\,\delta_{mm'}\ \ ,
  \qquad N\equiv\{nlm\}\ ,\ \ l'=0,2\ ,\ \ m'=-l',\ldots,l'
  \label{3.4.b} \\[0.5 em]
 f_{{\rm stroke,}N} & = &{\cal M}^{-1}\,
                        {\bf f}_0\!\cdot\!{\bf u}_N^*({\bf x}_0)
  \label{3.4.c}
\end{deqarr}
where the coefficients $a_{nl}\/$ in~(\ref{3.4.b}) are overlapping integrals
of the type~(\ref{3.3.a}), and

\begin{deqarr}
\arrlabel{m3.5}
 g_{{\rm resonators,} N}^{(a)}(t) & = & \int_0^t\left[z_a(t')-u_a(t)
   \right]\,\sin\omega_N(t-t')\,dt'
   \ \ ,\qquad a=1,\ldots,J  \label{3.5.a}  \\[0.5 em]
 g_{{\rm GW,}N}^{(lm)}(t) & = & \int_0^t g^{(lm)}(t')\,
   \sin\omega_N(t-t')\,dt'  \label{3.5.b}  \\[0.7 em]
 g_{{\rm stroke,}N}(t) & = & \sin\omega_Nt  \label{3.5.c}
\end{deqarr}

If this is replaced into~(\ref{2.1.a}) one readily finds

\begin{equation}
 {\bf u}({\bf x},t) = \sum_N\,\omega_N^{-1}\,{\bf u}_N({\bf x})\,
  \left\{\sum_{b=1}^J\,\frac{M_b}{\cal M}\,\Omega_b^2\,
  \left[{\bf n}_b\!\cdot\!{\bf u}_N^*({\bf x}_b)\right]\,
  g_{{\rm resonators,} N}^{(b)}(t) + \sum_\alpha
  f_{{\rm external,} N}^{(\alpha)}\,g_{{\rm external,} N}^{(\alpha)}(t)
  \right\}	\label{m3.6}
\end{equation}
where the label ``external'' explicitly refers to agents acting upon the
system from outside. Two kinds of such external actions are considered in
this article: those due to GWs and those due to a calibration hammer stroke
signal. Specifying {\bf x}\,=\,${\bf x}_a\/$ in the lhs of~(\ref{m3.6}) and
multiply on either side by ${\bf n}_a\/$, the following is readily found:

\begin{deqarr}
\arrlabel{A3.7}
  u_a(t) & = & u_a^{\rm external}(t) + \sum_{b=1}^J\,\eta_b\,\int_0^t
  K_{ab}(t-t')\,\left[\,z_b(t')-u_b(t')\right]\,dt'  \label{A3.7.a}\\
  \ddot{z}_a(t) & = & \xi_a^{\rm external}(t)
  -\Omega_a^2\,\left[\,z_a(t)-u_a(t)\right]\ , \qquad a=1,\ldots,J
  \label{A3.7.b}
\end{deqarr}
where $\eta_b$\,$\equiv$\,$M_b/{\cal M}$, $u_a^{\rm external}(t)$\,$\equiv$\,
${\bf n}_a\!\cdot\!{\bf u}^{\rm external}({\bf x}_a,t)$,

\begin{equation}
   {\bf u}^{\rm external}({\bf x},t) = \sum_\alpha\sum_N\,\omega_N^{-1}\,
    f_{{\rm external,}N}^{(\alpha)}\,{\bf u}_N({\bf x})\,
    g_{{\rm external,}N}^{(\alpha)}(t)      \label{m3.9}
\end{equation}
and

\begin{equation}
  K_{ab}(t) \equiv \Omega_b^2\,\sum_N\,\omega_N^{-1}\,
  \left[{\bf n}_b\!\cdot\!{\bf u}_N^*({\bf x}_b)\right]
  \left[{\bf n}_a\!\cdot\!{\bf u}_N({\bf x}_a)\right]\,\sin\omega_Nt
  \label{A3.10}
\end{equation}

The following bare sphere responses to GWs and hammer strokes
(equations!(\ref{1.1}) and~(\ref{2.5}), respectively) can be calculated
by direct substitution. The results as best presented as Laplace
transform domain functions:

\begin{deqarr}
\arrlabel{6.2}
  \hat u_a^{\rm GW}(s) & = &
   \sum_{\stackrel{\scriptstyle l=0\ {\rm and}\ 2}{m=-l,...,l}}\,\left(
   \sum_{n=1}^\infty\,\frac{a_{nl}\,A_{nl}(R)}{s^2+\omega_{nl}^2}\right)
   \,Y_{lm}({\bf n}_a)\,\hat g^{(lm)}(s)\ ,\qquad a=1,\ldots,J
   \label{6.2.a}	\\[1 em]
  \hat u_a^{\rm stroke}(s) & = & -\sum_{nl}\,\frac{f_0}{s^2+\omega_{nl}^2}\,
  \left|A_{nl}(R)\right|^2\,P_l({\bf n}_a\!\cdot\!{\bf n}_0)\ ,\qquad
   a=1,\ldots,J
  \label{6.2.b}
\end{deqarr}
where $Y_{lm}\/$ are spherical harmonics and $P_l\/$ Legendre
polynomials~\cite{Ed60}. The calculation of the Laplace transform
of the kernel matrix~(\ref{A3.10}) is likewise immediate:

\begin{equation}
  \hat K_{ab}(s) = \sum_N\,\frac{\Omega_b^2}{s^2+\omega_N^2}\,
   \left[{\bf n}_b\!\cdot\!{\bf u}_N^*({\bf x}_b)\right]
   \left[{\bf n}_a\!\cdot\!{\bf u}_N({\bf x}_a)\right]
   \label{m3.17}
\end{equation}

Given that (see~\cite{lobo} for full details)

\begin{equation}
   {\bf u}_{nlm}({\bf x}) = A_{nl}(r)\,Y_{lm}(\theta,\varphi)\,{\bf n}
   - B_{nl}(r)\,i{\bf n}\!\times\!{\bf L}Y_{lm}(\theta,\varphi)
   \label{A3.18}
\end{equation}
and that the spheroidal frequencies $\omega_{nl}\/$ are 2$l\/$+1--fold
degenerate,~(\ref{m3.17}) can be easily summed over the degeneracy index
$m\/$, to obtain

\begin{equation}
  \hat K_{ab}(s) = \sum_{nl}\,\frac{\Omega_b^2}{s^2+\omega_{nl}^2}\,
   \left|A_{nl}(R)\right|^2\,\left[\sum_{m=-l}^l\,
    Y_{lm}^*({\bf n}_b)\,Y_{lm}({\bf n}_a)\right]
   \label{A3.19}
\end{equation}
or, equivalently,

\begin{equation}
  \hat K_{ab}(s) = \sum_{nl}\,\frac{\Omega_b^2}{s^2+\omega_{nl}^2}\,
   \left|A_{nl}(R)\right|^2\,\frac{2l+1}{4\pi}\,
   P_l({\bf n}_a\!\cdot\!{\bf n}_b)
   \label{A3.20}
\end{equation}
where use has been made of the summation formula for the spherical
harmonics~\cite{Ed60}

\begin{equation}
  \sum_{m=-l}^l\,Y_{lm}^*({\bf n}_b)\,Y_{lm}({\bf n}_a) = 
  \frac{2l+1}{4\pi}\,P_l({\bf n}_a\!\cdot\!{\bf n}_b)
  \label{A3.21}
\end{equation}
and where $P_l\/$ is a Legendre polynomial:

\begin{equation}
  P_l(z) = \frac{1}{2^l\,l!}\,\frac{d^l}{dz^l}\,(z^2-1)^l
  \label{A3.22}
\end{equation}

\subsection{System response algebra}
\label{app:b}

From equation~(\ref{m4.8}), i.e.,

\begin{equation}
 \sum_{b=1}^J\,\left[\delta_{ab} + \eta\,\sum_{nl}\,
   \frac{\Omega^2s^2}{(s^2+\Omega^2)(s^2+\omega_{nl}^2)}\,\chi_{ab}^{(nl)}
   \right]\,\hat q_b(s) = -\frac{s^2}{s^2+\Omega^2}\,
   \hat u_a^{\rm GW}(s) + \frac{\hat\xi_a^{\rm GW}(s)}
   {s^2+\Omega^2}\ ,\qquad  (\Omega = \omega_{n_0l_0})
   \label{A2.1}
\end{equation}
we must first isolate $\hat q_b(s)$, then find inverse Laplace transforms
to revert to time domain quantities. Substituting the values of
$\hat u_a^{\rm GW}(s)$ and $\hat\xi_a^{\rm GW}(s)$ from~(\ref{6.2.a})
and~(\ref{4.85}) into~(\ref{A2.1}) we find

\begin{equation}
   \hat q_a(s) = \sum_{\mbox{\scriptsize $\begin{array}{c}
    l=0\ \mbox{and}\ 2 \\ m=-l,...,l \end{array}$}}\hat\Phi_a^{(lm)}(s)\,
    \hat g^{(lm)}(s)\ ,\qquad a=1,\ldots,J
    \label{6.3}
\end{equation}
where

\begin{equation}
   \hat\Phi_a^{(lm)}(s) = -\frac{s^2}{s^2+\Omega^2}\,\left(-\frac{R}{s^2} +
   \sum_{n=1}^\infty\,\frac{a_{nl}\,A_{nl}(R)}{s^2+\omega_{nl}^2}\right)\,
   \sum_{b=1}^J\,\left[\delta_{ab} +
   \eta\,\sum_{nl}\,\frac{\Omega^2s^2}{(s^2+\Omega^2)(s^2+\omega_{nl})^2}\,
   \chi_{ab}^{(nl)}\right]^{-1}\,Y_{lm}({\bf n}_b)
   \label{6.4}
\end{equation}

Now, using the convolution theorem of Laplace transforms, we see that the
time domain version of equation~(\ref{6.3}) is

\begin{equation}
   q_a(t) = \sum_{lm}\,\int_0^t\,\Phi_a^{(lm)}(t-t')\,g^{(lm)}(t')\,dt'
   \ ,\qquad a=1,\ldots,J
   \label{6.5}
\end{equation}
where $\Phi_a^{(lm)}(t)$ is the {\it inverse Laplace transform\/}
of~(\ref{6.4}). The inverse Laplace transform of $\hat\Phi_a^{(lm)}(s)$
can be expediently calculated by the {\it residue theorem\/} through
the formula~\cite{he68}

\begin{equation}
   \Phi_a^{(lm)}(t) = 2\pi i\,\sum\,\left\{ {\rm residues\ of}\ \ 
   \left[\hat\Phi_a^{(lm)}(s)\,e^{st}\right]\ \ {\rm at\ its\ poles\ 
   in\ complex\ {\it s\/}-plane}\right\}
   \label{6.6}
\end{equation}

Clearly thus, the {\it poles\/} of $\hat\Phi_a^{(lm)}(s)$ must be determined
in the first place. It is immediately clear from equation~(\ref{6.4}) that
there are no poles at either $s\/$\,=\,0, or $s\/$\,=\,$\pm i\Omega$, or
$s\/$\,=\,$\pm i\omega_{nl}$, for there are exactly compensated infinities
at these locations. The only possible poles lie at those values of $s\/$
for which the matrix in square brackets in~(\ref{6.4}) is not invertible,
and these of course correspond to the zeroes of its determinant, i.e.,

\begin{equation}
  \Delta(s)\equiv\det\left[\delta_{ab} + \eta\,
  \frac{s^2}{s^2+\Omega^2}\,\hat K_{ab}(s)\right]= 0\ ,
  \qquad {\rm poles}	\label{m3.18}
\end{equation}

There are infinitely many roots for equation~(\ref{m3.18}), but
{\it analytic\/} expressions cannot be found for them.
{\it Perturbative\/} approximations in terms of the small parameter
$\eta\/$ will thus be applied instead. It is assumed that

\begin{equation}
  \Omega = \omega_{n_0l_0}
\end{equation}
for a {\it fixed\/} multipole harmonic $\{n_0l_0\}$. Equation~(\ref{m3.18})
can then be recast in the more convenient form

\begin{equation}
  \Delta(s) \equiv \det\,\left[\delta_{ab} + \eta\,\frac{\Omega^2s^2}
   {(s^2+\Omega^2)^2}\,\chi_{ab}^{(n_0l_0)} + \eta\,\sum_{nl\neq n_0l_0}\,
   \frac{\Omega^2s^2}{(s^2+\Omega^2)(s^2+\omega_{nl}^2)}\,\chi_{ab}^{(nl)}
   \right] = 0
   \label{m4.9}
\end{equation}

Since $\eta\/$ is a small parameter, the {\it denominators\/} of the
fractions in the different terms in square brackets in~(\ref{m4.9}) must be
{\it quantities of order $\eta\/$} at the root locations for the determinant
to vanish at them. A distinction however arises depending on whether $s^2$ is
close to $-\Omega^2$ or to the other $-\omega_{nl}^2$. There are accordingly
two categories of roots, more precisely:

\begin{deqarr}
\arrlabel{m4.11}
    s_0^2 & = & -\Omega^2\,\left(1 + \chi_\frac{1}{2}\,\eta^{1/2}
    + \chi_1\,\eta + \ldots\right)  \qquad (\Omega=\omega_{n_0l_0})
    \label{4.11.a}   \\
    s_{nl}^2 & =& -\omega_{nl}^2\,\left(1 + b_1^{(nl)}\,\eta +
	b_2^{(nl)}\,\eta^2 + \ldots\right) \qquad ({nl\neq n_0l_0})
    \label{4.11.b}
\end{deqarr}

\begin{figure}
\centering
\includegraphics[width=13cm]{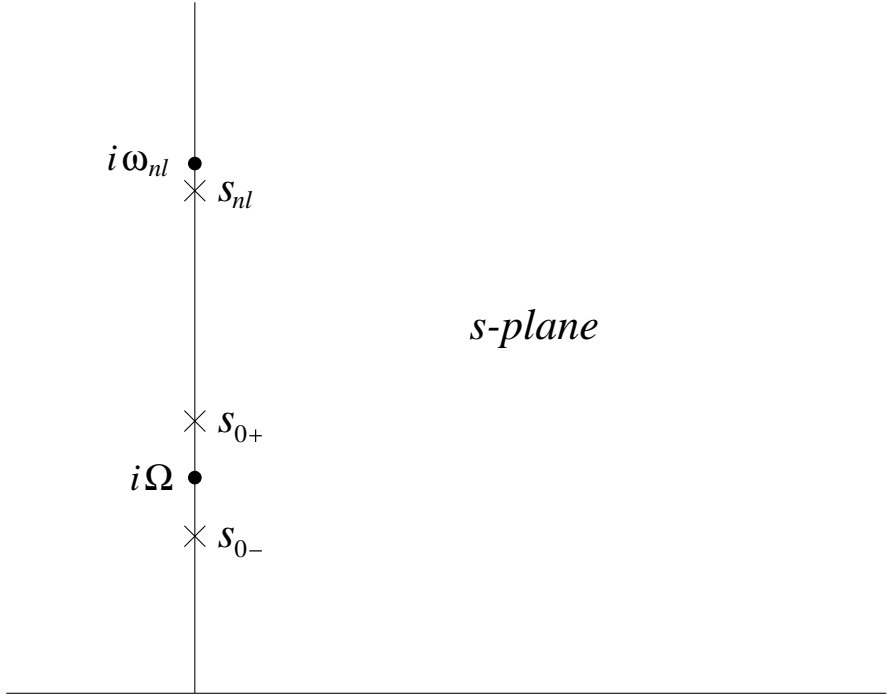}
\caption{Position of the roots (\protect\ref{m4.11}) in the complex
$s\/$--plane for $\Omega$ tuned to the lowest frequency of the sphere's
spectrum. Note that $s_{nl}\/$ is closer to $i\omega_{nl}\/$ than
$s_{0\pm}\/$ are to $i\Omega$; this is a consequence of the
differences being proportional to $\eta\/$ and $\eta^{1/2}$,
respectively, as seen in equations (\protect\ref{m4.11}).
\label{fig9}}
\end{figure}

Figure \ref{fig9} provides a visual characterisation of these roots.
The coefficients $\chi_\frac{1}{2}$, $\chi_1$,... and $b_1^{(nl)}$,
$b_1^{(nl)}$,... can be calculated recursively, starting form the
first, by substitution of the corresponding series expansions into
equation~(\ref{m4.9}). The lowest order terms are easily seen to be
given by

\begin{equation}
  \det\left[\delta_{ab} - \frac{1}{\chi_\frac{1}{2}^2}\,
  \chi_{ab}^{(n_0l_0)}\right] = 0
  \label{5.1}
\end{equation}
and

\begin{equation}
  \det\left[\frac{\Omega^2-\omega_{nl}^2}{\omega_{nl}^2}\,b_1^{(nl)}\,
  \delta_{ab} - \chi_{ab}^{(nl)}\right] = 0
  \label{5.4}
\end{equation}
respectively. Both equations~(\ref{5.1}) and~(\ref{5.4}) are algebraic
eigenvalue equations. As shown in appendix~\ref{app:c}, the matrix
$\chi_{ab}^{(nl)}$ has at most $(2l+1)$ non-null positive eigenvalues
---all the rest up to $J\/$ are identically zero.

As a final step we must evaluate~(\ref{6.6}). This is accomplished by
standard textbook techniques (see e.g.\ \cite{porter}); the algebra is
quite straightforward but rather lengthy, and I shall not delve into its
details here, but quote only the most interesting results. It appears that
the {\it dominant\/} contribution to $\Phi_a^{(lm)}(t)$ comes from the poles
at the locations~(\ref{4.11.a}), whereas all other poles only contribute as
higher order corrections; generically, $\Phi_a^{(lm)}(t)$ is seen to have
the form

\begin{equation}
   \Phi_a^{(lm)}(t)\propto\eta^{-1/2}\,\sum_{\zeta_c\neq 0}\,
   \left(\sin\omega_{c+}t - \sin\omega_{c-}t\right)\,\delta_{ll_0}
   + O(0)	\label{6.7}
\end{equation}
where

\begin{equation}
  \omega_{a\pm}^2 = \Omega^2\,\left(1\pm\sqrt{\frac{2l+1}{4\pi}}\,
  \left|A_{n_0l_0}(R)\right|\,\zeta_a\,\eta^{1/2}\right) + O(\eta)\ ,
  \qquad a=1,\ldots,J
\end{equation}

In Laplace domain one has,

\begin{equation}
   \hat\Phi_a^{(lm)}(s)\propto\eta^{-1/2}\,\sum_{\zeta_c\neq 0}\,
   \left[\left(s^2+\omega_{c+}^2\right)^{-1} -
   \left(s^2+\omega_{c-}^2\right)^{-1}\right]\,\delta_{ll_0}
   \label{6.78}
\end{equation}

Detailed calculation of the residues~\cite{serrano} yield
equation~(\ref{6.8}), which must be evaluated for each particular
tuning and resonator distribution, as described in section~\ref{sec:srgw}.

\subsection{Eigenvalue properties}
\label{app:c}

This Appendix presents a few important properties of the matrix
$P_l({\bf n}_a\!\cdot\!{\bf n}_b)$ for arbitrary $l\/$ and resonator
locations ${\bf n}_a\/$ ($a\/$=1,\ldots,$J\/$) which are useful for
detailed system resonance characterisation.

Recall the {\it summation formula\/} for spherical harmonics~\cite{Ed60}:

\begin{equation}
  \sum_{m=-l}^l\,Y_{lm}^*({\bf n}_a)\,Y_{lm}({\bf n}_b) = 
  \frac{2l+1}{4\pi}\,P_l({\bf n}_a\!\cdot\!{\bf n}_b)\ ,\qquad
  a,b=1,\ldots,J        \label{mA.1}
\end{equation}
where $P_l\/$ is a Legendre polynomial

\begin{equation}
  P_l(z) = \frac{1}{2^l\,l!}\,\frac{d^l}{dz^l}\,(z^2-1)^l   \label{A.15}
\end{equation}

To ease the notation I shall use the symbol ${\cal P}_l\/$ to mean the
entire $J\/$$\times$$J\/$ matrix $P_l({\bf n}_a\!\cdot\!{\bf n}_b)$, and
introduce Dirac {\it kets\/} $|m\rangle$ for the column $J\/$-vectors

\begin{equation}
  |m\rangle\equiv\sqrt{\frac{4\pi}{2l+1}}\left(\begin{array}{c}
    Y_{lm}({\bf n}_1) \\ \vdots  \\ Y_{lm}({\bf n}_J)
  \end{array}\right)\ \ ,\qquad m=-l,\ldots,l
  \label{mA.2}
\end{equation}

These kets are {\it not\/} normalised; in terms of them equation~(\ref{mA.1})
can be rewritten in the more compact form

\begin{equation}
  {\cal P}_l = \sum_{m=-l}^l\,|m\rangle\langle m|     \label{mA.3}
\end{equation}

Equation~(\ref{mA.3}) indicates that the {\it rank\/} of the matrix
${\cal P}_l\/$ cannot exceed $(2l+1)$, as there are only $(2l+1)$ kets
$|m\rangle$. So, if $J\/$\,$>$\,$(2l+1)$ then it has at least
$(J-2l-1)$ identically null eigenvalues ---there can be more if some of
the ${\bf n}_a\/$'s are parallel, as this causes rows (or columns) of
${\cal P}_l\/$ to be repeated.

We now prove that the non-null eigenvalues of ${\cal P}_l\/$ are
{\it positive\/}. Clearly, a regular eigenvector, $|\phi\rangle$, say,
of ${\cal P}_l\/$ will be a linear combination of the kets $|m\rangle$:

\begin{equation}
  {\cal P}_l \,|\phi\rangle = \zeta^2\,|\phi\rangle\ ,\qquad
  |\phi\rangle = \sum_{m=-l}^l\,\phi_m\,|m\rangle
  \label{mA.4}
\end{equation}
where $\zeta^2$ is the corresponding eigenvalue, having a positive value,
as we now prove. If the second~(\ref{mA.4}) is substituted into the first
then it is immediately seen that

\begin{equation}
  \sum_{m'=-l}^l\,\left(\zeta^2\,\delta_{mm'} -
  \langle m|m'\rangle\right)\,\phi_{m'} = 0    \label{A.5}
\end{equation}
which admits non-trivial solutions if and only if

\begin{equation}
  \det\,\left(\zeta^2\,\delta_{mm'} -
  \langle m|m'\rangle\right) = 0    \label{A.6}
\end{equation}

In other words, $\zeta^2$ are the eigenvalues of the
$(2l+1)$$\times$$(2l+1)$ matrix $\langle m|m'\rangle$, which is positive
definite because so is the ``scalar product'' $\langle\phi|\phi'\rangle$.
All of them are therefore strictly positive.

Finally, since the {\it trace\/} is an invariant property of a matrix, and

\begin{equation}
  {\rm trace}({\cal P}_l) \equiv \sum_{a=1}^J\,
  P_l({\bf n}_a\!\cdot\!{\bf n}_a) = \sum_{a=1}^J\,1 = J
  \label{A.7}
\end{equation}
we see that the eigenvalues $\zeta_a^2$ add up to $J\/$:

\begin{equation}
  {\rm trace}({\cal P}_l) =
  \sum_{a=1}^J\,\zeta_a^2\equiv\sum_{\zeta_a\neq 0}\,\zeta_a^2 = J
\end{equation}


\begin{thebibliography}{99}

\bibitem{ab72} Abramowitz M. and Stegun I.A., {\it Handbook of Mathematical
	Functions}, Dover 1972.

\bibitem{ad75} Ashby N. and Dreitlein J., {\it Phys. Rev.\/}
	{\bf D12}, 336 (1975).

\bibitem{as91} Astone P. {\em et al.}, {\it Europhys. Lett.\/}
	{\bf 16}, 231 (1991).

\bibitem{as93} Astone P. {\em et al}, {\it Phys. Rev.\/}
	{\bf D47}, 2  (1993).

\bibitem{als} Astone P., Lobo J.A. and Schutz B.F., {\it Class Quan Grav\/}
	{\bf 11}, 2093 (1994).

\bibitem{sfera} Astone P. et al., ``{\sl SFERA\/}: Proposal for
	a spherical GW detector'', Roma 1997.

\bibitem{Barut} Barut A.O., {\it Electrodynamics and Classical Theory of
	Fields and Particles}, Dover 1980.

\bibitem{maf} Bassan M., Cosmelli C., Frasca S., Papa M.A., Puppo P.,
	Rapagnani P. and  Ricci F., {\it High Frequency Gravitational
	Array}, ``La Sapienza'' preprint, Roma 1994.

\bibitem{bian} Bianchi M., Coccia E, Colacino C. N., Fafone V., Fucito F.,
	{\it Class Quan Grav\/} {\bf 13}, 2865 (1996).

\bibitem{maura} Bianchi M., Brunetti M., Coccia E., Fucito F, Lobo J. A.,
	{\it Phys. Rev.\/} {\bf D57}, 4525 (1998).

\bibitem{bd61} Brans C. and Dicke R. H., {\it Phys. Rev.\/}
	{\bf 124}, 925 (1961)

\bibitem{eupc} Eugenio Coccia, personal communication.

\bibitem{clo} Coccia E., Lobo J.A. and Ortega J.A., {\it Phys. Rev.\/}
	{\bf D52}, 3735 (1995).

\bibitem{amaldi} Coccia E., Pizzella G., Ronga F., eds., Proceedings
	of the First Edoardo Amaldi Conference, World Scientific,
	Singapore 1995.

\bibitem{cf} Coccia E. and Fafone V., {\it Phys Lett A\/} {\bf 213},
	16 (1996).

\bibitem{gr14} Coccia E., in Francaviglia M., Longhi G., Lusanna L.,
	and Sorace E., eds., Proceedings of the GR-14 Conference, World
	Scientific, Singapore 1997.

\bibitem{vega} Coccia E., Fafone V., Frossati G., Lobo J. A. and Ortega
	J. A., {\it Phys. Rev.\/} {\bf D57}, 2051 (1998).

\bibitem{klm1} Dhurandhar S. V., Krolak A., Lobo J. A., {\it Mon Not of Roy
	Ast Soc\/}, {\bf 237}, 333 (1989), and {\bf 238}, 1407 (1989).

\bibitem{dt} Dhurandhar S. V. and Tinto M., {\it Mon Not of the Royal
	Ast Soc\/}, {\bf 236}, 621 (1989).

\bibitem{el73} Eardley D.M., Lee D.L. and Lightman A.P., {\it Phys Rev\/}
	{\bf D8}, 3308 (1973).

\bibitem{Ed60} Edmonds A.R., {\it Angular Momentum in Quantum Mechanics},
	Princeton Univ.\ Press 1960.

\bibitem{fo71} Forward R., {\it Gen Rel and Grav\/}
	{\bf 2}, 149 (1971).

\bibitem{ko92} See e.g. Hamilton O.W. in Gleiser R.J., Kozameh C.N.
	and Moreschi O.M., {\sl Proceedings of GR13, C\'ordoba,
	Argentina}, IOP 1993. Also, Hamilton O.W. in
	reference~\protect\cite{amaldi} above.

\bibitem{hamil} Hamilton W. O., Johnson W. W., Xu B. X., Solomonson N.,
	Aguiar O. D., {\it Phys Rev\/} {\bf D40}, 1741 (1989).

\bibitem{tsvi} Har'El Z., {\it Geometri\ae\ Dedicata\/} {\bf 47}, 57 (1993).

\bibitem{he68} Helstrom C.W., {\it Statistical Theory of Signal Detection\/},
	Pergamon Press 1968.

\bibitem{ric} Hier R.G. and Rasband S.N., {\it Ap Jour\/} {\bf 195},
	507 (1975).

\bibitem{pacoM} Holden A., {\sl Formes, espace et sym\'etries\/},
	CEDIC, Paris 1977.

\bibitem{wjpc} Warren Johnson, private communication during the First
	Edoardo Amaldi Meeting in Frascati, June 1994.

\bibitem{jm93} Johnson W. and Merkowitz S. M.,{\it Phys. Rev. Lett.\/}
	{\bf 70}, 2367 (1993).

\bibitem{klm2} Krolak A., Lobo J. A., Meers B. J., {\it Phys Rev\/}
	{\bf D43}, 2470 (1991).

\bibitem{ll70} Landau L.D. and Lifshitz E.M., {\it Theory of Elasticity\/},
	Pergamon Press 1970.

\bibitem{ll85} Landau L.D. and Lifshitz E.M., {\it The Classical Theory
	of Fields\/}, Pergamon Press 1985.

\bibitem{ol} Lobo J.A. and Ortega J.A., in Coccia E., Pizzella G.,
	Ronga F., eds., {\sl Proceedings of the First Edoardo Amaldi
	Conference on Gravitational Waves\/} (Frascati 1994), pag 449,
	World Scientific, Singapore (1995).

\bibitem{mzno} I take this denomination from reference~\cite{MZ94}.
	It is, apparently, less common in the classical literature
	than the term {\it toroidal\/} applied to the other modes.

\bibitem{lobonotes} Lobo J.A., unpublished.

\bibitem{lobo} Lobo J. A., {\it Phys Rev\/} {\bf D52}, 591 (1995). This
	paper constitutes the first part of the present file, {\sl What
	can we learn about GW Physics with an elastic spherical antenna?}.

\bibitem{ls} Lobo J. A. and Serrano M. A., {Europhys Lett\/}, {\bf 35},
	253 (1996).

\bibitem{mini} Lobo J.A., in A. Kr\'olak, ed, {\sl Mathematics of
	Gravitation\/}, Banach Center Publications, vol 41, part II,
	pag 163-178, Warszawa (1997).

\bibitem{lsc} Lobo J. A. and Serrano M. A., {\it Class Quan Grav\/}
	{\bf 14}, 1495 (1997).

\bibitem{lobo2} Lobo J.A., in E. Coccia, G. Pizzella, G. Veneziano, eds,
	{\sl Proceedings of the Second Edoardo Amaldi Conference on
	Gravitational Waves\/} ({\sl CERN\/}, Geneva 1997), pag 168-179,
	World Scientific Singapore (1998).

\bibitem{lo44} Love A.E.H., {\it A Treatise on the Mathematical Theory
	of Elasticity}, Dover 1944.

\bibitem{nadja} Magalh\~aes N. S., Johnson W.W., Frajuca C, Aguiar O.,
	{\it Mon Not of the Royal Ast Soc\/}, {\bf 274}, 670 (1995).

\bibitem{grg} Magalh\~aes N. S., Aguiar O. D., Johnson W. W., Frajuca C.,
	{\it Gen Rel and Grav\/} 29, 1509 (1997).

\bibitem{phd} Merkowitz S. M., PhD Thesis Memoir, Louisiana State
	University (1995).

\bibitem{jm95} Merkowitz S. M. and Johnson W. W., {\it Phys Rev\/} {\bf D51},
	2546 (1995).

\bibitem{jm96} Merkowitz S. M. and Johnson W. W., {\it Phys Rev\/} {\bf D53},
	5377 (1996).

\bibitem{jm97} Merkowitz S. M. and Johnson W. W., {\it Phys Rev\/} {\bf D56},
	7513 (1997).

\bibitem{m98} Merkowitz S. M., {\it Phys Rev\/} {\bf D58}, 062002 (1998).

\bibitem{jm98} Merkowitz S. M. and Johnson W. W., {\it Europhys Lett\/},
	{\bf 41}, 355 (1998).

\bibitem{lms} Merkowitz S. M., Lobo J. A., Serrano M. A., {\it Class Quan
	Grav\/} {\bf 16}, 3035 (1999).

\bibitem{schipi} Merkowitz S. M., Coccia E., Fafone V., Raffone G.,
	Schipilliti M., and Visco M., {\it  Rev Sci Instr\/}, {\bf 70},
	1553 (1999).

\bibitem{MZ94} Michelson P.F. and Zhou C.Z., {\it Phys. Rev.\/}
	{\bf D51}, 2517 (1995).

\bibitem{mtw} Misner, Thorne, Wheeler, {\sl Gravitation\/}, Freeman 1973.

\bibitem{porter} Porter D. and Stirling D. S. G., 1990, Integral equations:
	a practical treatment from spectral theory to applications,
	Cambridge University Press

\bibitem{Ras} Rasband S.N., {\it J Acous Soc Am}
	{\bf 57}, 899 (1975).

\bibitem{serrano} Serrano M. A., PhD Thesis, University of Barcelona (1999).

\bibitem{ts} Stevenson T. R., {\it Phys. Rev.\/} {\bf D56}, 564 (1997).

\bibitem{tricomi} Tricomi F. G., {\sl Integral equations\/}, Interscience
	Publishers 1957.

\bibitem{tm87} Tym Myint--U, {\sl Partial Differential Equations for
	Scientists and Engineers}, North Holland 1987.

\bibitem{wp77} Wagoner R. V. and Paik H. J. in {\it Experimental
	Gravitation}, Proceedings of the Pavia International Symposium,
	Acad. Naz. dei Lincei 1977.

\bibitem{we72} Weinberg S., {\sl Gravitation and Cosmology}, Wiley 1972.

\end{thebibliography}
\end{document}